\documentclass[a4paper,11pt]{article}
\usepackage{jheppub}\usepackage{lineno}
\usepackage{amsthm,amsmath,amssymb,amsfonts,empheq}
\usepackage{mathrsfs,bm, comment}
\usepackage{slashed,braket,tensor,cancel}
\usepackage[dvipsnames]{xcolor}
\usepackage{tikz,tikz-cd,tkz-euclide}
\usepackage[compat=1.1.0]{tikz-feynman}
\usetikzlibrary{cd}\usetikzlibrary{calc}\usetikzlibrary{patterns}
\usetikzlibrary{feynman}\tikzfeynmanset{warn luatex=false}\usetikzlibrary{arrows.meta}
\allowdisplaybreaks[4]
\def\eq{\eqref}

\def\0{{(0)}}\def\1{{(1)}}\def\2{{(2)}}\def\3{{(3)}}\def\4{{(4)}}
\def\ab{{\alpha\beta}}\def\mn{{\mu\nu}}\def\rs{{\rho\sigma}}
\newcommand{\e}{\mathrm{e}}\newcommand{\p}{\partial} 
\newcommand{\Rn}[1]{{\rm\uppercase\expandafter{\romannumeral#1}}}
 
\def\nn{\nonumber}

\def\qaq{\quad\text{and}\quad}

\def\wh{\widehat}
\def\c.c.{\mathrm{c.c.}}

\newcommand{\abs}[1]{\left\lvert #1\right\rvert} 

\def\g{\gamma}

\def\ep{\epsilon}\def\ve{\varepsilon}
\def\th{\theta}
\def\l{\lambda}
\def\m{\mu}\def\n{\nu}

\def\Th{\Theta}

\def\cl{\mathcal{L}}
\def\cm{\mathcal{M}}

\def\cw{\mathcal{W}}

\def\C{\mathbb{C}}
\def\R{\mathbb{R}}

\author{Geoffrey Compère$^a$, Dima Fontaine$^a$, Wen-Bin Liu$^{a,b}$, and Kevin Nguyen$^a$.}
\affiliation{$^a$Université Libre de Bruxelles, BLU-ULB Brussels Laboratory of the Universe, International Solvay Institutes, CP 231, B-1050 Brussels, Belgium}
\affiliation{$^b$Huazhong University of Science and Technology, School of Physics, \\ Luoyu Road 1037, Wuhan, Hubei, 430074, China}
\emailAdd{geoffrey.compere@ulb.be, dima.fontaine@ulb.be, wenbin.liu@ulb.be, kevin.nguyen2@ulb.be}
\definecolor{dima}{RGB}{255, 164, 213}
\newcommand{\scri}{\mathscr{I}}

\abstract{
We derive the classical logarithmic soft graviton theorem for the scattering of massive particles in four-dimensional asymptotically flat spacetime using a position-space analysis of the Weyl tensor. Starting from the Iyer–Damour multipole solution in harmonic gauge, we obtain an infinite tower of antipodal matching relations across spatial infinity for all five Newman–Penrose Weyl scalars, at linear order in $G$ and for the leading matter-induced logarithms at order $G^2$. We connect the relation relevant to the logarithmic soft theorem to the radiative-gauge formulation of Boschetti–Campiglia and Compère–Robert. We then compute the required asymptotic fields directly from scattering data, including matter contributions and nonlinear effects responsible for graviton drag. Combining these results with the Newman–Penrose evolution equations yields position-space proofs of the classical leading and logarithmic soft graviton theorems, independently confirming the frequency-space derivation of the latter originally performed by Laddha and Sen.
}

\begin{document}
\title{Gravitational multipoles, antipodal matching relations, and the logarithmic soft graviton theorem}
\maketitle


\section{Introduction}

Important developments in establishing and refining soft graviton theorems have occurred in recent years, going beyond Weinberg's seminal soft graviton theorem \cite{Weinberg:1965nx}. In 2014, Cachazo and Strominger provided strong evidence for the existence of a subleading soft graviton theorem at tree level \cite{Cachazo:2014fwa} (see also earlier work \cite{1968PhRv..166.1287G,White:2011yy}). Beyond tree level and in four dimensions, infrared divergences occur which give divergent corrections to this subleading soft graviton theorem \cite{Bern:2014oka,He:2014bga,Broedel:2014fsa}. Yet, the divergences themselves admit a universal structure which is one-loop exact and can therefore be expressed in a quite simple form. At the same time, one-loop diagrams generate a logarithmic soft factor \cite{Sahoo:2018lxl} whose classical counterpart was originally established by Laddha and Sen in 2018 \cite{Laddha:2018myi}. Later, it was recognized that these two seemingly distinct one-loop corrections appear with the exact same coefficients which indicate that they share a common origin \cite{Donnay:2022hkf,Agrawal:2023zea}. The classical logarithmic soft theorem can also be understood as originating from a specific subleading antipodal map at spatial infinity \cite{Compere:2026jmk,Boschetti:2026ogm} once a dictionary between the scattering data and the asymptotic fields at spatial infinity has been established \cite{Boschetti:2026gfd}.  The derivations of the logarithmic soft theorem originally presented in \cite{Sahoo:2018lxl,Laddha:2018myi} are performed in Fourier space and crucially involve selecting parts of the integration domain of various integrals to isolate the contributions to the logarithmic branch in frequency of the gravitational waveform. The aim of this paper is to develop a position-space approach and independently confirm the classical version of the logarithmic soft theorem.  \\

\noindent In asymptotically flat gravity, two subleading quantities in the radial expansion at null and spatial infinities stand up as relevant to the logarithmic soft graviton theorem. First, as the Fourier transform of a logarithm in frequency space involves inverse times, the logarithmic soft graviton theorem is related to the so-called $1/u$ tails of gravitational waveforms in the early or late limits $u \to \pm \infty$. Second, as $O(G^2)$ interactions (or one-loop effects in field theory language) are determinant, the violation of peeling of the asymptotic Weyl tensor is also relevant as it generically occurs in gravitational scattering at $O(G^2)$ and leads to logarithmic divergences \cite{PhysRevD.19.3495,Damour:1985cm,1985FoPh...15..605W,christodoulou2002global,Kehrberger:2021uvf,Kehrberger:2021vhp,Sahoo:2021ctw}. As found out by Boschetti and Campiglia \cite{Boschetti:2026gfd}, a particular linear combination of the two main asymptotic quantities sourcing these two phenomena is precisely the relevant quantity that allows us to rewrite the logarithmic soft theorem as an antipodal map at spatial infinity once all quantities at spatial infinity have been recognized in terms of scattering data. This antipodal identity has now been formally proven under a polyhomogeneity hypothesis \cite{Compere:2026jmk,Boschetti:2026ogm}. In the rewriting of Boschetti-Campiglia \cite{Boschetti:2026gfd}, however, the results of Laddha-Saha-Sahoo-Sen \cite{Laddha:2018myi,Sahoo:2018lxl,Saha:2019tub,Sen:2024qzb} were used as input in order to evaluate the relevant asymptotic fields at spatial infinity. In this work, we will take one step back, and derive the relevant fields at spatial infinity from first principles, assuming only that a gravitational scattering event is taking place, which will allow us to rederive the logarithmic soft theorem without input from these prior works. 
\\

\noindent The antipodal map instrumental in proving the logarithmic soft graviton theorem is only part of an infinite tower whose structure has been progressively understood \cite{Strominger:2017zoo,McLoughlin:2022ljp,Sen:2024qzb}. In this paper we take some steps in the derivation of all-order antipodal map relationships by working in the post-Minkowskian multipolar formalism \cite{Blanchet:2013haa}. This approach was successful at reproducing the subleading soft photon theorem and discovering an infinite set of subleading antipodal relations in electrodynamics \cite{Compere:2025tzr} (see also earlier work \cite{Campiglia:2018dyi,AtulBhatkar:2020hqz}). In this paper, we generalize this approach to the gravitational case, based on the multipolar expansion of the general retarded linear metric obtained by Iyer and Damour \cite{Damour:1990gj} in the tradition of  \cite{PhysRevD.15.2156,Thorne:1980ru,Blanchet:1985sp,1989AIHPA..50..377B}. We will limit ourselves to $O(G^2)$ in this paper. \\

\noindent The paper is organised as follows.
Section~\ref{sec:conventions} sets out our conventions and notations.
Section~\ref{Section:AntipodalMatchings} develops the multipole expansions
of the Weyl scalars, derives the antipodal matching relations at linear
order and for the leading matter-induced logarithms at order $G^2$,
and relates the matching condition relevant to the logarithmic soft
theorem to its radiative-gauge formulation.
Section~\ref{Section:GravitationalScattering} computes the asymptotic
Weyl coefficients for the scattering of massive particles, including
the nonlinear contribution responsible for graviton drag, and discusses
the transformation from harmonic to radiative gauge.
Section~\ref{Section:SoftTheorems} combines these results with the
Newman--Penrose evolution equations to derive the classical leading
and logarithmic soft graviton theorems.
Appendix~\ref{App:Multipole} supplies the detailed multipole calculations,
Appendix~\ref{app: soft factor identities} establishes the differential
identities for the soft factors, and
Appendix~\ref{Appendix:GravitationalTail} derives the logarithmic Weyl
coefficients generated by graviton drag.

\section{Conventions and notations}
\label{sec:conventions}

\paragraph{Coordinates and measures.}
The spacetime metric $g_{\mu\nu}(x^\mu)$ in coordinates ${x^\mu \equiv (t, x^i)}$ with $x^i \equiv r n^i$ is defined as a perturbation of the Minkowski metric ${\eta_{\mu \nu}\equiv\mathrm{diag}(-1,1,1,1)}$ where  $r\equiv|\vec{x}|$ is the coordinate radius and $n^i=(\sin\theta\cos\phi,\sin\theta\sin\phi,\cos\th)$ the unit-length directional spatial vector. Capital-indexed coordinates refer to coordinates on the sphere $x^A=(\theta, \phi)$ while repeated $L$ indices denote summation over multipole indices $i_1...i_\ell$ such as $n_L \equiv n_{i_1}... n_{i_\ell}$.  Moreover, $\wh n_L$ denotes the tracefree part of $n_L$.\\

\noindent The coordinate basis on the unit sphere $e_A\equiv\p_A$ has components $e_A^i=\p_An^i$ when embedded in $\R^3$. We have 
\begin{align}
e^i_Ae^j_B\delta_{ij}=\g_{AB}, \qquad \perp^{ij} \equiv e^i_Ae^j_B\g^{AB}=\delta^{ij}-n^in^j,
\end{align}
where $\gamma_{AB}\equiv\mathrm{diag}(1, \sin^2 \theta)$ is the metric on the unit sphere. We define the standard measure over the sphere $\epsilon^{AB}$ with the orientation convention $\epsilon^{\theta\phi}=\csc \theta$. We also have
\begin{align}
n_ie^i_A=0,\qquad D_Ae^i_B=-\g_{AB}n^i,
\end{align}
where $D_A$ is the covariant derivative of the Christoffel connection of $\g_{AB}$. The Levi-Civita tensor of $\R^3$ is denoted as $\ve_{ijk}$ which is related to $\ep_{AB}$ through
\begin{align}
\epsilon_{AB}=e^i_Ae^j_Bn^k\ve_{kij} \qaq \ep_{AB}e_j^Ae_k^B=n^i\ve_{ijk},
\end{align}
which implies
\begin{align}
\ve_{ijk}e^j_Ae^k_B=n_i\epsilon_{AB} \qaq \ve_{ijk}n^ie_j^A=\epsilon^A{}_Be_k^B.
\end{align}
The measure over the sphere is defined as
\begin{align}
\oint d\Omega\,\equiv \int_{0}^\pi d\th\,\sin\th\int_0^{2\pi} d\phi.
\end{align}
The antipodal map $\Upsilon: S^2 \to S^2$ defined as $\Upsilon(\theta,\phi)=(\pi - \theta,\phi+\pi)$ maps $\vec n \mapsto - \vec n$ and $n_L \mapsto (-1)^\ell n_L$. It preserves the metric $\Upsilon^*\gamma_{AB}=\gamma_{AB}$ but reverses the orientation $\Upsilon^*\epsilon_{AB}= -\epsilon_{AB}$. It acts on scalars as $(\Upsilon^* f)(x)=f(\Upsilon x)$ and on one-forms as $(\Upsilon^* V)_A(x)=J_A^{\;\; B}V_B(\Upsilon x)$ where $J_A^{\;\; B}=\text{diag}(-1,1)$ is the Jacobian.

\paragraph{Null tetrad and Newman-Penrose scalars.}
We introduce the complex dyad on the sphere\footnote{We warn the reader that this dyad is not universal in the Newman-Penrose literature but it is used e.g. in \cite{Geiller:2024ryw}, which we follow here. }
\begin{align}
\th^A\equiv\left(1, \frac{-i}{\sin\theta}\right)\,.  \end{align}
The dyad and its conjugate $\bar\th^A$ satisfy $\g^{AB}=\th^{(A}\bar\th^{B)}$ and $\epsilon^{AB}=-i \theta^{[A}\bar\theta^{B]}$, which also imply $\theta^A \bar \theta_A=2$ and $\theta^A=-i \theta^B \epsilon\indices{_B^A}$. Under the antipodal map we have 
\begin{align}\label{thetaUps}
  \Upsilon^* \theta^A=-\bar \theta^A  .
\end{align}
We also introduce the Newman-Penrose (NP) tetrad of the background Minkowski spacetime,
\begin{equation}
\begin{aligned}
l^\m&\equiv\frac{1}{\sqrt{2}}(1,n^i)\equiv\frac{1}{\sqrt{2}}q^\mu ,\qquad\quad n^\m\equiv\frac{1}{\sqrt{2}}(1,-n^i)\equiv\frac{1}{\sqrt{2}}\tilde{q}^\mu, \\
m^\m&\equiv\dfrac{1}{\sqrt2}(0,\th^Ae^i_A),\qquad \qquad  \quad \hspace{3 pt}  \bar m^\m\equiv\dfrac{1}{\sqrt2}(0,\bar\th^A e^i_A).\label{tetradlnm}
\end{aligned}
\end{equation}
We use the ``democratic'' convention between $l^\mu$ and $n^\mu$, which allows to treat the limits to $\scri^\pm$ symmetrically. The tetrad elements are normalised such that 
\begin{align}
l^\m n_\m=-m^\m\bar m_\m=-1 \qaq \eta_\mn=-2l_{(\m}n_{\n)}+2m_{(\m}\bar m_{\n)}.
\end{align}
We define the spacetime orientation for $\epsilon_{\mn\ab}$ as $\epsilon_{0123}=1$ (hence $\epsilon^{0123}=-1$). The NP tetrad orientation is given by 
\begin{equation}
	\epsilon_{\mu\nu\alpha\beta} l^{\mu} n^{\nu} m^{\alpha} \bar{m}^{\beta} = \epsilon_{0123}  \det(l, n, m, \bar{m}) = -i.
\end{equation}
The five complex NP Weyl scalars are defined as
\begin{align}
\hspace{-8 pt}\Psi_0\equiv-C_{lmlm},\quad \Psi_1\equiv-C_{lnlm},\quad \Psi_2\equiv-C_{lm\bar m n},\quad \Psi_3\equiv-C_{ln\bar mn},\quad \Psi_4\equiv-C_{n\bar mn\bar m},
\end{align}
where  $C_{lmlm}\equiv l^\alpha m^\beta l^\rho m^\delta C_{\alpha \beta \rho \delta}$ and other contractions are defined similarly. The electro-magnetic dual Weyl tensor is $\mbox{}^*C_{\alpha\beta\gamma\delta}\equiv \frac{1}{2}\epsilon_{\alpha\beta \rho\sigma} C^{\rho\sigma}_{\;\;\;\;\; \gamma\delta}$. The dual Weyl scalars are defined as $\mbox{}^*\Psi_0=-\mbox{}^*C_{lmlm}$ in terms of the dual Weyl tensor and similarly for $\mbox{}^*\Psi_I$, $I=1,\dots , 4$. We have the identities
\begin{equation}
	\begin{aligned}
i \epsilon_\mn{}^\ab l^\m m^\n&= l^{\alpha} m^{\beta}-m^{\alpha} l^{\beta},\\
i\epsilon_{\mu\nu}{}^{\alpha\beta} l^{\mu} n^{\nu} &= \bar{m}^{\alpha} m^{\beta}-m^{\alpha} \bar{m}^{\beta},\\
i\epsilon_{\mu\nu}{}^{\alpha\beta} n^{\mu} \bar{m}^{\nu} &=  n^{\alpha} \bar{m}^{\beta}-\bar{m}^{\alpha} n^{\beta}\,,
\end{aligned}
\end{equation}
which imply 
\begin{equation}
i \, {}^*\Psi_I = \Psi_I\,, \qquad  I=0,1,2,3,4\,.
\end{equation}

\paragraph{Spin-weighted derivatives.} We define the spin weight of $\theta^A$ (and therefore $m^A$) as $+1$ and of $\bar\theta^A$ (and therefore $\bar m^A$) as $-1$. Then, the NP scalars $\Psi_I$ have spin weight $s=2-I$. We define the $\eth$ and $\bar \eth$ derivatives acting on a scalar of spin weight $s$ as 
\begin{equation}
\label{eth definition}
\eth F_s\equiv\left(\theta^A \partial_A-s \cot \theta \right) F_s\,\,, \qquad \bar \eth F_s\equiv\left(\bar \theta^A \partial_A+s \cot \theta \right) F_s\,.
\end{equation}
For a generic vector $V_A$, we have 
\begin{equation}
\label{eth identity}
\begin{split}
\eth (\bar \theta^A V_A)&=(\theta^A \partial_A+\cot \theta) (\bar \theta^B V_B)
=(D^A V_A +i  \epsilon^{AB}D_A V_B)\,,\\
\eth (\theta^A V_A) &=(\theta^A \partial_A-\cot \theta) (\theta^B V_B)= \theta^A \theta^B D_A V_B. 
\end{split}
\end{equation}
It will be useful to compare our conventions with those used in \cite{Geiller:2022vto,Geiller:2024ryw}. That work uses the tetrads $(l^*,n^*,m^*)$ with $l^*=\sqrt{2}l^{\text{rad}+}$, $n^*=\frac{1}{\sqrt{2}}n^{\text{rad}+}$, $m^*=m^{\text{rad}+}$ as compared to our outgoing radiation tetrad defined in Eq. \eqref{tetrad:rad}. Our $\theta^A$ corresponds to $\sqrt{2}\, m_1^a$ as defined in Eq. (2.12) of \cite{Geiller:2024ryw} and $\sqrt{2}\, m_0^a$ as given in Eq. (2.54b) of \cite{Geiller:2022vto}. Our spin-weighted derivative differs by the same factor, $\eth=\sqrt{2}\, \eth^*$. The change of normalization for the NP scalars is as follows:  $\Psi_0^*=2 \Psi_0\,, \Psi_1^*=\sqrt{2} \Psi_1\,, \Psi_2^*=\Psi_2\,, \Psi_3^*= \Psi_3/\sqrt{2}\,, \Psi_4^*=\Psi_4/2.$

\paragraph{Scattering conventions.}
We are considering the scattering of incoming massive bodies to outgoing massive bodies. We denote each body by a letter $a,b,c,\dots $ and we state $a \in \text{in}$ or $a \in -$ if the body is incoming while $a \in \text{out}$ or $a \in +$ if the body is outgoing. The momentum of the body $a$ of mass $m_a$ and velocity $v_a^\mu$ normalized as $v_a^\mu v_{a \mu}=-1$ is ${p}^\mu_{a} = \eta_a m_a {v}_{a}^{\mu}$ where $\eta_a=1$ if it is outgoing and $\eta_a=-1$ if it is incoming. The total momentum is 
\begin{align}
\label{total P}
    P^\mu \equiv \sum_{a \in \text{out}}p_a^\mu = - \sum_{a \in \text{in}}{p}_{a}^{\mu}\,,
\end{align}
such that momentum conservation can be expressed as 
\begin{equation}
  \sum_{a \in \text{out}} m_a {v}_{a}^{\mu}   - \sum_{a \in \text{in}} m_a {v}_{a}^{\mu} = \sum_{a \in \pm } p^{\mu}_a=0\,. \label{Eq:MomentumConservation}
\end{equation}
Here  $a \in \pm$ denotes the sum over both incoming and outgoing bodies. 


\section{Antipodal matchings from multipole expansions}
\label{Section:AntipodalMatchings}

In this section, we will derive features of the gravitational field at order $G^2$ around spatial infinity in the multipolar formalism, which will allow us to relate the asymptotic behavior of the Weyl tensor in the limit $u \to -\infty$ of future null infinity $\scri^+_-$ with the behavior of the Weyl tensor in the limit $v \to +\infty$ of past null infinity $\scri^-_+$.

\subsection{Elements of the post-Minkowskian formalism}

We use the notations of Blanchet \cite{Blanchet:2013haa} and define the gothic metric  deviation $\mathfrak{h}$ such that 
\begin{equation}
    \mathfrak{h}^{\alpha \beta} \equiv \sqrt{-g} g^{\alpha \beta} - \eta^{\alpha \beta},
\end{equation}
with $g^{\alpha \beta}$ the full inverse spacetime metric, such that $g^{\alpha \beta}g_{\beta \gamma}= \delta^\alpha_\gamma$ and $\eta^{\alpha \beta}=\mathrm{diag}(-1,1,1,1)$. In the harmonic gauge, the Einstein equations are 
\begin{equation}
    \Box \mathfrak{h}^{\alpha \beta}= 16 \pi G\, \tau^{\alpha \beta}\,,
\end{equation}
with $\Box$ the Minkowski d'Alembertian and 
\begin{equation}
    \tau^{\alpha \beta} \equiv \abs{g} T^{\alpha \beta} + \dfrac{1}{16 \pi G}\, \Lambda^{\alpha \beta}\,,
\end{equation}
where $T^{\alpha \beta}$ is the matter stress-energy tensor and $\Lambda^{\alpha \beta}$ is the gravitational source term. Hence, up to first non-linear order, 
\begin{align}
  \mathfrak{h}^{\alpha \beta}&\equiv G\mathfrak{h}^{\alpha \beta}_{(1)}+G^2\mathfrak{h}^{\alpha \beta}_{(2)}+ O(G^3)\,,\\
  T^{\alpha \beta}&\equiv T^{\alpha \beta}_{(0)} + G\, T^{\alpha \beta}_{(1)} + O(G^2)\,,
\end{align}
and since 
\begin{equation}
    \abs{g}= - \mathrm{det} \left(\eta^{\alpha \beta} + \mathfrak{h}^{\alpha \beta} \right)= 1 + G\, \eta_{\alpha \beta} \mathfrak{h}^{\alpha \beta}_{(1)} + O(G^2)\,,
\end{equation}
the Einstein equations form the following hierarchy: 
\begin{empheq}[left={\empheqlbrace}]{align}
    &\Box \mathfrak{h}^{\alpha \beta}_{(1)} =  16 \pi T^{\alpha \beta}_{(0)}, \label{Eq:Hierarchy1}\\
    & \Box \mathfrak{h}^{\alpha \beta}_{(2)} =  16 \pi\left( T^{\alpha \beta}_{(1)} + \ \mathfrak{h}_{(1)}T^{\alpha \beta}_{(0)}  \right) + N^{\alpha \beta}\left[\mathfrak{h}_{(1)}\right],\label{Eq:Hierarchy2} 
\end{empheq}
with 
\begin{equation}
    \begin{aligned}
        N^{\alpha\beta}
        \!\left[
        \mathfrak h_{(1)}
        \right]
        \equiv{}&
        - \mathfrak h^{\mu\nu}_{(1)}
        \partial^2_{\mu\nu} \mathfrak h^{\alpha\beta}_{(1)}
        + \frac{1}{2}
        \partial^\alpha \mathfrak h^{(1)}_{\mu\nu}
        \partial^\beta \mathfrak h^{\mu\nu}_{(1)}
        - \frac{1}{4}
        \partial^\alpha \mathfrak h_{(1)}
        \partial^\beta \mathfrak h_{(1)}
        \\
        &+\partial_\nu \mathfrak h^{\alpha\mu}_{(1)}\left(\partial^\nu \mathfrak h^\beta{}_{\mu(1)}+\partial_\mu \mathfrak h^{\beta\nu}_{(1)}\right)-2\partial^{(\alpha}\mathfrak h^{(1)}_{\mu\nu}\partial^\mu\mathfrak h^{\beta)\nu}_{(1)}
        \\
        &+\eta^{\alpha\beta}\left[-\frac{1}{4}\partial_\tau \mathfrak h^{(1)}_{\mu\nu}\partial^\tau \mathfrak h^{\mu\nu}_{(1)}+\frac{1}{8}\partial_\mu \mathfrak h_{(1)}\partial^\mu \mathfrak h_{(1)}+\frac{1}{2}\partial_\mu\mathfrak h^{(1)}_{\nu\tau}\partial^\nu \mathfrak h^{\mu\tau}_{(1)}\right] . 
    \end{aligned}\label{Eq:SourceFormula}
\end{equation}
and where $\mathfrak h_{(1)}\equiv \eta_{\mu\nu}\mathfrak h^{\mu\nu}_{(1)}$ and $ \mathfrak h^{(1)}_{\mu\nu} \equiv \eta_{\mu\rho}\eta_{\nu\sigma} \mathfrak h^{\rho\sigma}_{(1)}.$ Let us furthermore split the second-order perturbation in a matter part and a gravitational part: 
\begin{empheq}[left={\empheqlbrace}]{align}
    &\Box \mathfrak{h}^{\alpha \beta}_{(2), \text{matter}} =  16 \pi\left( T^{\alpha \beta}_{(1)} + \ \mathfrak{h}_{(1)}T^{\alpha \beta}_{(0)}  \right),\label{Eq:SourceMatterTerm} \\
    &\Box \mathfrak{h}^{\alpha \beta}_{(2), \text{grav.}}= N^{\alpha \beta}\left[\mathfrak{h}_{(1)}\right]. \label{Eq:SourceTerm}
\end{empheq}

\subsection{Multipole expansion of gravitational perturbations}

At first post-Minkowskian order, the most general solution to the vacuum Einstein equations with no incoming radiation in canonical harmonic gauge can be expressed in terms of two sets of symmetric and trace-free (STF) multipole moments  \cite{Damour:1990gj, Thorne:1980ru, Blanchet:2020ngx} as 
\begin{subequations}\label{multi}
	\begin{align}
		\mathfrak{h}^{00} &=-4\sum_{\ell=0}^{\infty}\frac{(-1)^{\ell}}{\ell!}\p_L\Big(\frac{M_{L}(u)}{r} \Big),\\
		\mathfrak{h}^{0j} &=4\sum_{\ell=1}^{\infty} \frac{(-1)^{\ell}}{\ell!}\left[\p_{L-1}\Big(\frac{M^{(1)}_{j L-1}(u)}{r} \Big) + \frac{\ell}{\ell+1}\p_{p L-1}\Big(\frac{\ve_{jpq}S_{q L-1}(u)}{r} \Big)\right],\\
		\mathfrak{h}^{jk}&=-4\sum_{\ell=2}^{\infty} \frac{(-1)^{\ell}}{\ell!}\left[\p_{L-2}\Big(\frac{M^{(2)}_{jk L-2}(u)}{r} \Big) +  \frac{2\ell}{\ell+1}\p_{p L-2}\Big(\frac{\ve_{pq(j}S^{(1)}_{k)q L-2}(u)}{r} \Big)\right].
	\end{align}
\end{subequations}
Here $M_L(u)$ and $S_L(u)$ are the STF canonical \textit{mass} and \textit{current multipole moments}. In this linear solution, these multipole moments are linear in $G$. We recall the conservation laws of the lowest multipole moments at linear order 
\begin{align}
\label{conservation laws}
\dot M=0,\qquad \ddot M_i=0,\qquad \dot S_i=0.
\end{align}
At second post-Minkowskian order, matter and quadratic gravitational terms source the metric perturbation, see Eqs. \eqref{Eq:SourceMatterTerm}--\eqref{Eq:SourceTerm}. Since the matter stress tensor vanishes near spatial infinity, the perturbation $\mathfrak{h}^{\alpha \beta}_{(2), \text{matter}}$ that it generates is homogeneous there and admits the same STF expansion \eqref{multi} (up to a shift by a pure gauge homogeneous solution that depends upon 4 gauge moments \cite{Damour:1990gj} and which do not contribute to the Weyl scalars at order $G^2$), 
thereby contributing \(O(G^2)\) corrections to \(M_L(u)\) and \(S_L(u)\). On the other hand, the quadratic gravitational perturbations yield various effects including tails, violations of peeling and memory effects. All these effects can be modeled around $\scri^+_-$ in the limit $r \to \infty$ at fixed $u$ followed by $u \to -\infty$ using a double polyhomogeneous expansion in $(\log r,r)$ and $(\log |u|,u)$ \cite{Chrusciel:1993hx,Hintz:2017xxu,Kadar:2025xmo}. The linear conservation laws \eqref{conservation laws} become flux-balance laws of Poincar\'e charges at quadratic order in $G$ \cite{Thorne:1980ru, Blanchet:2018yqa,Compere:2019gft}. As a consequence, the metric perturbation  $\mathfrak{h}^{\alpha \beta}_{(2), \text{matter}}$ given in the form \eqref{multi} alone violates the harmonic gauge condition and requires a completion with the gravitational part. \\

\noindent Here, we will only consider the second-order contribution sourced by matter, such that the resulting antipodal relations are a priori incomplete. However, in Section~\ref{sec: 3.6} we present an argument showing the validity of one such antipodal map derived in this way, including all possible $O(G^2)$ effects, which we use in Section~\ref{Section:SoftTheorems} to provide an independent derivation of the classical logarithmic soft graviton theorem.

\subsection{Gravitational electric-magnetic duality}

Before describing the post-Minkowskian metric around spatial infinity, we first make a useful detour to the gravitational electric-magnetic duality, which will help us generate the magnetic sector from the electric sector.\\

\noindent In our conventions, the linear gravitational electric-magnetic duality relation restricted to the  retarded fields in the multipolar expansion is given by \cite{Compere:2022zdz} 
\begin{equation}
\label{duality relation}
	{}^*R_{\mn\rs}(M_L^+,M_L^-)= R_{\mn\rs}(-M_L^-,M_L^+),
\end{equation}
where the dual linearized Riemann tensor is defined as ${}^* R_{\mn\rs}=\frac{1}{2}\epsilon_\mn{}^\ab  R_{\ab\rs}$ and the multipole moments $M_L^\pm$ are defined as
\begin{equation}
	M_L^+(u)\equiv M_L(u), \qquad M_L^-(u)\equiv \frac{2\ell}{\ell+1}S_L(u)\,.
\end{equation}
where $M_L^-$ is defined for $\ell \geq 1$. The duality relation \eqref{duality relation} holds locally in vacuum regions, i.e., away from matter sources, and only at the linearized level. Note that the electric-magnetic duality is explicitly broken in our parameter space as the NUT parameter is set to zero, $M^-_{L=0}=0$. The duality map is however useful to formally generate the current sector from the mass sector as a solution generating technique.  We define the mass and current components of the Weyl scalars, 
\begin{equation}
	\Psi^M_I\equiv \Psi_I(M^+_L,0), \qquad \Psi^S_I\equiv \Psi_I(0, M^-_L).
\end{equation}
The gravitational electric-magnetic duality \eqref{duality relation} then implies that 
\begin{equation}
	{}^* \Psi^M_I=-i\Psi^M_I=\Psi^S_I\left(\frac{2\ell}{\ell+1}S_L\to M_L\right)\,,
\end{equation}
and
\begin{equation}	{}^*\Psi^S_I=-i\Psi^S_I=\Psi^M_I\left(M_L\to -\frac{2\ell}{\ell+1}S_L\right),
\end{equation}
which ensures that the study of the mass multipole moments suffices to infer general properties of the Weyl perturbations in the wave zone. Moreover, by linearity, the Weyl scalars can be simply expressed as
\begin{equation}
	 \Psi_I(M_L,S_L) = \Psi_I^M\left(M_L - i \frac{2\ell}{\ell+1}S_L\right). 
\end{equation}
This motivates the introduction of complex multipole moments
\begin{equation}
	\cm_L\equiv M_L -i \frac{2\ell}{\ell+1}S_L\,,
\end{equation}
which have the property 
\begin{equation}
\quad{}^* \Psi_I(M_L,S_L)=\Psi_I(-{i}\cm_L).
\end{equation}
The electric-magnetic duality $\Psi_I(\cm_L)\mapsto {}^* \Psi_I(\cm_L)$ amounts to a clockwise rotation \(-i=\e^{-i\pi/2}\) in the \((M_L,S_L)\) space
\begin{align}
 \mathcal M_L \mapsto - i \mathcal M_L. \label{EMduality}   
\end{align}
Let us offer another perspective on the electric-magnetic duality. By expanding the definition of the Weyl scalars using our explicit tetrad, we have
\begin{subequations}
	\begin{align}
	\Psi_0&=-\frac{1}{4}\th^A\th^B( C_{0i0j}-2 C_{0ijp}n^p+ C_{ipjq}n^pn^q)e^i_Ae^j_B\,,\\
		\Psi_1&=\frac{1}{2}\th^A( C_{0i0j}n^ie^j_A+ C_{0ijp}n^in^je^p_A)\,,\\
		\Psi_2&=\frac{1}{4}\th^A\bar\th^B( C_{0i0j}+2 C_{0[ij]p}n^p- C_{ipjq}n^pn^q)e^i_Ae^j_B\,,\\
		\Psi_3&=-\frac{1}{2}\bar\th^A( C_{0i0j}n^ie^j_A- C_{0ijp}n^in^je^p_A)\,,\\
		\Psi_4&=-\frac{1}{4}\bar\th^A\bar\th^B( C_{0i0j}+2 C_{0ijp}n^p+ C_{ipjq}n^pn^q)e^i_Ae^j_B\,.
	\end{align}
\end{subequations}
Introducing now the electric and magnetic fields
\begin{align}
	E_{ij} \equiv C_{0i0j}\,, \qaq B_{ij}\equiv\frac{1}{2}\ep\indices{_i^{pq}} C_{0jpq}\,,
\end{align}
the definition of the magnetic field can be inverted to find
\begin{align}
	C_{0ijp}=\epsilon\indices{_{jp}^q}B_{qi}\,.
\end{align}
Tracelessness of the Weyl tensor implies that $E_{ij}$ and $B_{ij}$ are STF tensors. The Weyl scalars can all be expressed solely in terms of these fields. With the help of 
\begin{equation}\label{LeviCivitaId}
\epsilon_{ipq} n^p m^q = i m_i\,, \qquad \epsilon_{ipq} n^p \bar{m}^q =-i \bar{m}_i\,,
\end{equation}
(here $n_i$ is the spatial vector, not the spatial component of $n^\mu$ which differs by a factor $-\sqrt{2}$) and
\begin{equation}
	\begin{aligned}
		\epsilon_{ipm} \epsilon_{jqn} n^p n^q E^{mn} m^i m^j &= -m^i m^j E_{ij}\,,\\
		\epsilon_{ipm} \epsilon_{jqn} n^p n^q E^{mn} m^i \bar{m}^j &= m^i \bar{m}^j E_{ij}\,, 
	\end{aligned}
\end{equation}
we can visualize all the Weyl scalars as the various projections of a complex tensor 
\begin{equation}
	\mathcal{W}_{ij} \equiv E_{ij} + i B_{ij}\,.
\end{equation}
Specifically, we have
\begin{subequations}
	\begin{align}
	\Psi_0 &=- m^i m^j \mathcal{W}_{ij}\,,\label{Eq:Psi0EM}\\
		\Psi_1 &= \frac{1}{\sqrt{2}} m^i n^j  \mathcal{W}_{ij}\,,\label{Eq:Psi1EM}\\
		\Psi_2 &=- \frac{1}{2} n^in^j\mathcal{W}_{ij}\,,\label{Eq:Psi2EM}\\
		\Psi_3 &= -\frac{1}{\sqrt{2}} \bar{m}^i n^j \mathcal{W}_{ij}\,,\label{Eq:Psi3EM}\\
		\Psi_4 &= -\bar{m}^i \bar{m}^j\mathcal{W}_{ij}\label{Eq:Psi4EM}\,.
	\end{align}
\end{subequations}
Conversely, we may write
\begin{align}
	\mathcal{W}_{ij} = -\Psi_{0} \bar{m}_{i}\bar{m}_{j} - \Psi_{4} m_{i}m_{j} + 2\sqrt{2} \Psi_{1} n_{(i}\bar{m}_{j)} - 2\sqrt{2} \Psi_{3} n_{(i}m_{j)} -  \Psi_{2} (3n_{i}n_{j} - \delta_{ij})\,.
\end{align}
Using $ \l\cw_{ij}(\cm_L)=\cw_{ij}( \l\cm_L)$ with $\l\in\C$, the electric-magnetic duality \eq{EMduality} can then be neatly stated as
\begin{align}
\cw_{ij}(\cm_L)\mapsto -i\cw_{ij}(\cm_L).
\end{align}
which is immediate since the Weyl scalars and thus the complex tensor $\cw_{ij}$ are linear in the multipole moments.

\subsection{Antipodal matchings at $O(G)$}

We assume that the linear solution is non-radiative in a neighborhood of spatial infinity. More precisely, we assume that the linearized Bondi news locally vanishes in a neighborhood of spatial infinity. The news is the time derivative of the shear which is proportional, at order $G$, to a linear sum of $\ell$ derivatives of multipoles $M_L(u)$, $S_L(u)$, see e.g.~Eq.~(192) of \cite{Blanchet:2013haa}. Therefore, requiring a vanishing local news implies that the canonical multipole moments $M_L(u)$, $S_L(u)$ are polynomials of degree at most $\ell$ as $u \to -\infty$: 
\begin{equation}
  M_L(u) = \sum_{k=0}^\ell M_{L,k}\, u^k+ O(G^2)\,, \qquad  S_L(u) = \sum_{k=0}^\ell S_{L,k}\, u^k+ O(G^2)\,,  \label{Eq:GMultipole}
\end{equation}
with $M_{L,k}$, $S_{L,k}$ some undetermined STF tensor coefficients. Contributions of order $G^2$ require additional functional dependence in $u$ as we will discuss in Section~\ref{subsec: 3.5}. Hence, the results of this subsection hold to first order in $G$. The $p$-th derivative of the mass multipoles are thus given by 
\begin{equation}
    M_L^{(p)}(u)=   \sum_{k=p}^\ell M_{L,k} u^{k-p} \dfrac{k!}{(k-p)!} + O(G^2)\,. \label{Eq:DerivativeMultipoleG}
\end{equation}
Substituting such expansions in the gothic metric perturbation \eqref{multi} is a straightforward exercise and results in a double homogeneous expansion in $r$ and $u$. One can then compute the Weyl scalars and obtain expansions of the form 
\begin{equation}
    \begin{aligned}
         \Psi_0 &=\sum_{n=0}^\infty \dfrac{1}{r^{n+2}}\,  \overset{n}{\Psi}_{0}(u), \qquad \Psi_1 =\sum_{n=0}^\infty \dfrac{1}{r^{n+2}}\,  \overset{n}{\Psi}_{1}(u), \qquad  \Psi_2 =\sum_{n=0}^\infty \dfrac{1}{r^{n+1}}\,  \overset{n}{\Psi}_{2}(u), \\
         & \qquad  \Psi_3 =\sum_{n=0}^\infty \dfrac{1}{r^{n+1}}\,  \overset{n}{\Psi}_{3}(u), \qquad \qquad  \Psi_4=\sum_{n=0}^\infty \dfrac{1}{r^{n+1}}\,  \overset{n}{\Psi}_{4}(u).
    \end{aligned}
\end{equation}
This computation is explicitly carried through in Appendix \ref{App:Multipole}, to which we refer the reader for the details. At first, these radial expansions seem to violate peeling, which is not expected at linear order in $G$. However, Eqs. \eqref{Eq:Kernel0}-\eqref{Eq:Kernel4} demonstrate that the following quantities vanish for any mass or current multipole moment:
\begin{equation}
    \overset{0}{\Psi}_0=  \overset{1}{\Psi}_0=  \overset{2}{\Psi}_0= \overset{0}{\Psi}_1= \overset{1}{\Psi}_1 =  \overset{0}{\Psi}_2=\overset{1}{\Psi}_2= \overset{0}{\Psi}_3=0.
\end{equation}
In what follows, we will however keep these vanishing fictitious orders in our expansions as they help treat expansions on either side of spatial infinity democratically and lighten notations. For a generic multipole moment $M_L(u)$, by Eq. \eqref{Eq:MultipoleShortExpansion}, we have
\begin{equation}\label{Eq:ShortMultipoleInText}
   \left\langle (\overset{n}{\Psi}_I)_{(s)}, \widehat{n}_L^{(s)}\right \rangle  \equiv \oint d\Omega\,(\overset{n}{\Psi}_I)_{(s)} \widehat{n}_L^{(s)}= \dfrac{C_\ell^{(s)}}{\ell!} M_{L}^{(\ell-n+1+\Theta_{I-2})} R_I(n,\ell),
\end{equation}
where $\Th_n$ is the discrete Heaviside step function defined as $\Th_n = 1$ for $n \geq 0$ and $0$ otherwise. Here $\Psi_I^{(s)}$ with $s=|I-2|$ denotes the tensor, vector or scalar dyad decomposition on the sphere of the Weyl scalars, namely
\begin{equation}
    \begin{aligned}
        \Psi_0\equiv \theta^A \theta^B (\Psi_0)_{AB} \quad &\text{and} \quad \Psi_4\equiv \bar{\theta}^A \bar{\theta}^B (\Psi_4)_{AB} \qquad \text{for } s=2, \\
        \Psi_1\equiv \theta^A (\Psi_1)_A \quad &\text{and} \quad  \Psi_3\equiv \bar{\theta}^A (\Psi_3)_A \qquad \!\qquad\text{for } s=1,
    \end{aligned}
\end{equation}
and $\Psi_2$ remaining a true scalar for $s=0$, the coefficients $R_I(n,\ell)$ given by Eqs. \eqref{Eq:Kernel0}-\eqref{Eq:Kernel4}, $\widehat{n}_L^{(s)}$
denoting tensor, vector or scalar STF harmonics, respectively, and the coefficients $C_\ell^{(s)}$ arising from harmonics orthogonality relations, see e.g. Eqs. \eqref{TensorProjection}, \eqref{VectorProjection} and \eqref{ScalarProjection}. We omit writing the explicit expressions of these coefficients as they do not explicitly affect the antipodal matching relations which we are about to derive. Now inserting the linear multipoles \eqref{Eq:GMultipole} in \eqref{Eq:ShortMultipoleInText} and writing 
\begin{equation}
    \Psi_I= \sum_{n=0}^\infty \sum_{k=0}^n \dfrac{u^{n-k}}{r^{n+2- \Theta_{I-2}}}\overset{n,k}{\Psi}_{\!I} + O(G^2),
\end{equation}
we find 
\begin{equation}
   \left\langle (\overset{n,k}{\Psi}_{\!I})_{(s)}, \widehat{n}_L^{(s)}\right\rangle= \dfrac{C_\ell^{(s)}}{\ell!} \dfrac{(\ell-k+1 + \Theta_{I-2})!}{(n-k)!} M_{L,\ell-k+1 + \Theta_{I-2}}R_I(n,\ell). \label{Eq:GProjectionAllScalars}
\end{equation}
Upon changing coordinates from the retarded time $u$ to the advanced time $u=v-2r$, with $v\equiv t+r$, it is straightforward but cumbersome to demonstrate that the Weyl scalars admit an analogous expansion at past null infinity,
\begin{equation}
    \Psi_I =  \sum_{n=0}^\infty \sum_{k=0}^n \dfrac{v^{n-k}}{r^{n+2-\Theta_{I-2}}}  \left.\overset{n,k}{\Psi}_{\hspace{-2 pt}I}\right|_{\scri^-_+} + O(G^2)\,,
\end{equation}
where
\begin{equation}\label{psim}
   \left.\overset{n,k}{\Psi}_{\hspace{-2 pt}I}\right|_{\scri^-_+}= \sum_{j=0}^\infty(-2)^{j}\binom{n+j-k}{j}\left.\overset{n+j,k}{\Psi}_{\hspace{-7 pt}I}\,\,\right|_{\scri^+_-}\,.
\end{equation}
The proof exactly follows the one presented in \cite{Compere:2025tzr}. 
Although seemingly spanning all positive integers, this sum is truncated by the definition of the coefficients $R_I(n,\ell)$ given by Eqs. \eqref{Eq:Kernel0}-\eqref{Eq:Kernel4} which are zero outside of their defined range. Now we consider the example of $\Psi_3$ to construct the antipodal matching relations. Under the parity transformation, we have
\begin{align}
\Upsilon^*(\bar\theta^A V_A)=-\theta^A\Upsilon^*V_A \qaq \Upsilon^*(D_A\wh n_L)=(-1)^\ell D_A\wh n_L,
\end{align}
where $V_A$ is a generic vector. Using Eqs. \eqref{Eq:GProjectionAllScalars} and \eqref{psim}, we find 
 \begin{equation}
     \left\langle\;\; (\hspace{-9 pt} \overset{n+1, k+1}{\Psi}_{\hspace{-12 pt}3})_{B}\Big|_{\scri^-_+}, D^B \wh n_L \right\rangle = (-1)^{\ell+n} \left\langle (\overset{n,k}{\Psi}_{\!1})_{B}\Big|_{\scri^+_-}, D^B \wh n_L \right\rangle .
 \end{equation}
Since $(\Psi_{3})_{A}$ has vanishing projection on odd-parity vector harmonics and given the following expansion for such one-forms on the sphere:
\begin{align}
U_A=\sum_{\ell \geq 1} \dfrac{1}{{\ell}C_\ell^{(1)}}\left\langle U_{B}, D^B \wh n_L \right\rangle D_A\wh n_L,
\end{align}
we can derive
\begin{align}
\begin{aligned}
\Upsilon^*(\hspace{-9 pt} \overset{n+1, k+1}{\Psi}_{\hspace{-12 pt}3})_A\Big|_{\scri^-_+}
&=\sum_{\ell\ge1}\frac{1}{\ell C_\ell^{(1)}}\left\langle\;\; (\hspace{-9 pt} \overset{n+1, k+1}{\Psi}_{\hspace{-12 pt}3})_B\Big|_{\scri^-_+},D^B\widehat n_L\right\rangle (-1)^\ell D_A\widehat n_L\\
&=\sum_{\ell\ge1}\frac{(-1)^\ell(-1)^{\ell+n}}{\ell C_\ell^{(1)}}\left\langle (\overset{n,k}{\Psi}_{\!1})_B\Big|_{\scri^+_-},D^B\widehat n_L\right\rangle D_A\widehat n_L\\
&=(-1)^n(\overset{n,k}{\Psi}_{\!1})_A\Big|_{\scri^+_-}.
\end{aligned}
\end{align}
Repeating the analysis of \cite{Compere:2025tzr} for the other components, we find the following antipodal matching relations for the Weyl scalar dyad components:
\begin{equation}
 \boxed{\hspace{5 pt}\begin{aligned}\left.(\overset{n,k}{\Psi}_{\hspace{-2 pt}0})_{AB}\right|_{\scri^+_-} &=\left.(-1)^{n+1} \Upsilon^* \hspace{6 pt}(\hspace{-13 pt}\overset{n+1,k+1}{\Psi}_{\hspace{-12 pt}4})_{AB}\right|_{\scri^-_+}, { \qquad n\geq0,\ 0\leq k\leq n,}\\ 
 \left.(\hspace{-13 pt}\overset{n+1,k+1}{\Psi}_{\hspace{-12 pt}4})_{AB}\right|_{\scri^+_-} &=\left.(-1)^{n+1}  \Upsilon^*(\hspace{-2 pt}\overset{n,k}{\Psi}_{\hspace{-1 pt}0})_{AB}\right|_{\scri^-_+}\,, {\qquad n\geq0,\ 0\leq k\leq n,}\end{aligned}}
\end{equation}
\begin{equation}
 \boxed{\hspace{5 pt}\begin{aligned}\left.(\overset{n,k}{\Psi}_{\hspace{-2 pt}1})_{A}\right|_{\scri^+_-} &=(-1)^{n}  \left.\Upsilon^* \; \;(\hspace{-13 pt}\overset{n+1,k+1}{\Psi}_{\hspace{-12 pt}3})_{A}\right|_{\scri^-_+},{\qquad n\geq0,\ 0\leq k\leq n,}\\  
 (\hspace{-13 pt} \left.\overset{n+1,k+1}{\Psi}_{\hspace{-12 pt}3})_{A}\right|_{\scri^+_-}&=(-1)^{n}  \Upsilon^*\left.(\hspace{-2 pt}\overset{n,k}{\Psi}_{\hspace{-1 pt}1})_{A}\right|_{\scri^-_+}\,,{\qquad n\geq0,\ 0\leq k\leq  n}\end{aligned}}
\end{equation}
and 
\begin{equation}
\label{antipodal matching Psi2}
 \boxed{\left.\overset{n,k}{\Psi}_{\hspace{-2 pt}2}\right|_{\scri^+_-} =(-1)^{n}\Upsilon^*\left.\overset{n,k}{\Psi}_{\hspace{-2 pt}2}\right|_{\scri^-_+}\,,{\qquad n\geq0,\ 0\leq k\leq n.}}
\end{equation}

\subsection{Antipodal matchings at $O(G^2)$: matter-induced leading logarithms}
\label{subsec: 3.5}
Let us now compute part of the $O(G^2)$ perturbations: those generated by matter, see Eq.~\eqref{Eq:SourceMatterTerm}, while discarding non-linear gravitational perturbations coming from Eq.~\eqref{Eq:SourceTerm}. More precisely, we are interested only in the leading logarithmic terms (in $r$ or $u$) of the second order metric in the large radius expansion. Long-range gravitational interactions generate logarithmic corrections to the asymptotic particle trajectories \cite{Blanchet:2013haa,Sahoo:2018lxl,Saha:2019tub}. In the Newtonian limit, inserting these trajectories into the mass multipole moments produces a leading logarithmic term proportional to \(u^{\ell-1}\log|u|\). Motivated by this behavior, we consider the following asymptotic expansion
\begin{equation}
    M_L(u)= \sum_{k=0}^\ell M_{L,k} u^k + \left( M_{L,\ell-1}^{\text{log.}} \log\abs{u} \,  u^{\ell-1} \Theta_{\ell-1} + O(u^{\ell-1}) \right) + O(G^3). \label{Eq:G2Multipole}
\end{equation}
Contributions of order $O(G^3)$ may require polylogarithmic dependence, which we do not consider in the present work.
 Let us focus solely on this leading logarithmic contribution to the multipole moments. We then have for $p \leq \ell-1$, 
\begin{equation}
    M_L^{(p)}= \dfrac{(\ell-1)!}{(\ell-1-p)!}u^{\ell-1-p} M_{L,\ell-1}^{\text{log.}} \log \abs{u} + \text{(non-log)} + O(G^3). 
\end{equation}
As in the electromagnetic case \cite{Compere:2025tzr}, we show that the existence of the  antipodal matching relations at order $G$ immediately implies the existence of the same matchings for the leading logarithmic piece at order $G^2$. Writing
\begin{equation}
    \Psi_I= \sum_{n=0}^\infty \sum_{k=0}^n \dfrac{u^{n-k}}{r^{n+2- \Theta_{I-2}}}\overset{n,k}{\Psi}_{\!I} + \left(\sum_{n=0}^\infty \dfrac{u^n \log \vert u\vert }{r^{n+4}}\, \overset{w_I, \text{log.}}{\Psi}_{\hspace{-9 pt}I}+(\text{non-log})\right)+ O(G^3),
\end{equation}
with $w_I = n+2+\Theta_{I-2}$, we find, by plugging \eqref{Eq:G2Multipole} in \eqref{Eq:ShortMultipoleInText},
\begin{equation}
   \left\langle \, (\hspace{-6 pt}\overset{n,\text{log.}}{\Psi}_{\hspace{-7 pt}I})_{(s)}, \wh n_L^{(s)}\right \rangle= \dfrac{C_\ell^{(s)}}{\ell!} \dfrac{(\ell-1)!}{(n-2-\Theta_{I-2})!} M^\text{log.}_{L,\ell-1} \Theta_{\ell-1}R_I(n,\ell).
\end{equation}
Up to the STF coefficients $M_{L, \ell-1}^\text{log.}$ that do not affect antipodal matchings, this is exactly Eq. \eqref{Eq:GProjectionAllScalars} with $k=2+ \Theta_{I-2}$. Since the antipodal matching conditions at order $G$ are valid for all $0\leq k \leq n$, they ensure that the logarithmic coefficients also match antipodally on either side of spatial infinity. Through the change of coordinates to advanced time $v=t+r$, we get 
\begin{equation}
    \sum_{n=0}^\infty  \dfrac{u^{n} \log |u|}{r^{n+4}}\,  \left.\overset{w_I,\text{log.}}{\Psi}_{\hspace{-9pt}I}\;\right|_{\scri^+_-} \!= \sum_{n=0}^\infty  \dfrac{v^{n} \log r}{r^{n+4}}\, \left.\overset{w_I,\text{log.}}{\Psi}_{\hspace{-9 pt}I}\;\right|_{\scri^-_+} +(\text{non-log})\,,\label{eq3.49}
\end{equation}
with 
\begin{equation}
    \left.\overset{n,\text{log.}}{\Psi}_{\hspace{-6 pt}I}\;\right|_{\scri^-_+}= \sum_{j=0}^\infty(-2)^{j}\binom{n+j-2-\Theta_{I-2}}{j} \left.\overset{n+j,\text{log.}}{\Psi}_{\hspace{-12 pt}I}\: \:\right|_{\scri^+_-}\,.
\end{equation}
Thus, we find the following antipodal matching conditions at order $O(G^2)$: 
\begin{equation}
     \boxed{\hspace{5 pt}\begin{aligned}(\hspace{-5 pt}\left.\overset{n,\text{log.}}{\Psi}_{\hspace{-6 pt}0})_{AB} \:\right|_{\scri^+_-}&=(-1)^{n+1} \Upsilon^* \;(\hspace{-12 pt}\left.\overset{n+1,\text{log.}}{\Psi}_{\hspace{-12 pt}4})_{AB}\:\right|_{\scri^-_+}\,,{ \qquad n\ge0,}\\ \; (\hspace{-12 pt}\left.\overset{n+1,\text{log.}}{\Psi}_{\hspace{-12 pt}4})_{AB}\:\right|_{\scri^+_-} &=(-1)^{n+1} \Upsilon^* (\hspace{-5 pt}\left.\overset{n,\text{log.}}{\Psi}_{\hspace{-6 pt}0})_{AB}\:\right|_{\scri^-_+},\,{ \qquad n\ge0},\end{aligned}}
 \end{equation}
\begin{equation}
     \boxed{\hspace{2 pt}\begin{aligned}(\hspace{-5 pt}\left.\overset{n,\text{log.}}{\Psi}_{\hspace{-6 pt}1})_{A}\:\right|_{\scri^+_-}&=(-1)^{n}\hspace{2 pt}  \Upsilon^*\;(\hspace{-12 pt}\left.\overset{n+1,\text{log.}}{\Psi}_{\hspace{-12 pt}3})_{A}\:\right|_{\scri^-_+}, {\qquad n\ge0,}\\
     \left.(\hspace{-12 pt}\overset{n+1,\text{log.}}{\Psi}_{\hspace{-12 pt}3})_{A}\:\right|_{\scri^+_-} &=(-1)^{n} \Upsilon^* (\hspace{-5 pt}\left.\overset{n,\text{log.}}{\Psi}_{\hspace{-6 pt}1})_{A}\:\right|_{\scri^-_+},{\qquad n\ge0},\hspace{-12 pt}\end{aligned}} \label{Eq:PsiAMatchingLog}
 \end{equation}
and
\begin{equation}
 \boxed{\hspace{2 pt}\begin{aligned}\left.\hspace{-2 pt}\overset{n,\text{log.}}{\Psi}_{\hspace{-7 pt}2}\:\right|_{\scri^+_-} &=(-1)^{n}  \Upsilon^*\!\left.\overset{n,\text{log.}}{\Psi}_{\hspace{-6 pt}2}\:\right|_{\scri^-_+}, { \qquad n\ge0}.\hspace{-8 pt}\end{aligned}}
\end{equation}
These matching conditions are provisional, since they might admit corrections due to \textit{non-linear} gravitational interactions at the same order $O(G^2)$. However, for the matching condition underlying the logarithmic soft graviton theorem, we will argue that no correction is needed.

\subsection{Comparison between one particular antipodal relation and the literature}
\label{sec: 3.6}

In the previous sections, we demonstrated that up to $O(G^3)$ corrections and non-linear gravitational $G^2$ corrections, the matching conditions \eqref{Eq:PsiAMatchingLog} hold. In particular, for $n=2$,
\begin{equation}
     (\hspace{-6 pt}\left.\overset{2, \text{log.}}\Psi_{\hspace{-5 pt}1})_A\:\right|_{\scri^+_-}\hspace{-8 pt}= \Upsilon^*\left.(\hspace{-6 pt}\overset{3, \text{log.}}\Psi_{\hspace{-5 pt}3})_A\:\right|_{\scri^-_+}\hspace{-8 pt}\, \quad\Longleftrightarrow \quad   (\hspace{-6 pt}\left.\overset{2, \text{log.}}\Psi_{\hspace{-5 pt}1})_A \theta^A\:\right|_{\scri^+_-}\hspace{-8 pt}=- \Upsilon^* \left[ \left.(\hspace{-6 pt}\overset{3, \text{log.}}\Psi_{\hspace{-5 pt}3})_A\bar \theta^A\:\right|_{\scri^-_+}\right].\label{log: starting point0}
\end{equation}
Now, the classical logarithmic soft graviton theorem has been demonstrated to follow from one particular antipodal map at spatial infinity that  has been proven to hold in General Relativity under the assumption that a polyhomogeneous Bondi-Sachs expansion exists and that the news falls off at spatial infinity as $1/u^2$ as $u \to -\infty$ \cite{Agrawal:2023zea,Choi:2024ajz,Compere:2026jmk,Boschetti:2026gfd}. Let us now connect this non-perturbative result with our antipodal map \eqref{log: starting point0} obtained in perturbation theory under restricted assumptions. We first summarize the result of \cite{Compere:2026jmk,Boschetti:2026gfd}. We define the outgoing radiative tetrad $(l^{\text{rad}+},n^{\text{rad}+},m^{\text{rad}+})$ in the non-linear theory as 
\begin{equation}
\label{tetrad:rad}
\begin{split}
l^{\text{rad}+}&\equiv \frac{1}{\sqrt{2}}(\partial_R)_U =\frac{1}{\sqrt{2}} (1,n^i(X^A)) \,, \\
n^{\text{rad}+}& \equiv
\sqrt{2} (\partial_U - \frac{1}{2}(\partial_R)_U ) +O(1/R)=\frac{1}{\sqrt{2}} (1,-n^i(X^A)) + O(1/R)\,,\\
m^{\text{rad}+}&=\frac{1}{\sqrt{2}}(0,\theta^A e^i_A) +O(1/R)\,,
\end{split}
\end{equation}
where $(U,R,X^A)$ are the outgoing Bondi-Sachs coordinates. The vectors are expressed in components in $(T,X_i)$ coordinates where $X_i \equiv R\,  n_i(X^A)$, $T \equiv U +R$. This tetrad differs from the background outgoing radiative tetrad \eqref{tetradlnm} by subleading radial corrections that do not matter to derive leading order quantities in the large radius expansion. The normalisation of the tetrad treats advanced and retarded coordinates on a democratic setting.\\

\noindent The two matching conditions derived by Comp\`ere and Robert in \cite{Compere:2026jmk} can be defined in terms of the NP scalars defined from this outgoing radiative tetrad in our conventions as
\begin{align}
-\Upsilon^* \left( \Psi^{\text{rad}+}_1 \vert_{\scri^+_-, \frac{\log R}{R^4}}+  \Psi_1^{\text{rad}+} \vert_{\scri^+_-, \frac{\log (-U)}{R^4}} \right)&=  \Psi^{\text{rad}+}_3 \vert_{\scri^-_+, \frac{\log R^-}{(R^-)^4}}+ \Psi^{\text{rad}+}_3 \vert_{\scri^-_+, \frac{\log V}{(R^-)^4}}\nonumber \\ 
&\hspace{-1cm}-2G\, \bar \theta^A\left(3D^B (\tilde q \cdot P)+ (\tilde q \cdot P)D^B \right)\mathcal M^{-(0)}_{AB}(\vec{n}) . \label{DAmatch} 
\end{align}
This equality is a rewriting of Eq. (147) of  \cite{Compere:2026mdj} after using their definition (58) and after expressing all quantities in terms of the asymptotic Weyl scalar using their Eqs. (24)-(34).\\

\noindent The matching condition \eqref{DAmatch} was shown to be consistent with the results of Boschetti and Campiglia \cite{Boschetti:2026ogm} and was demonstrated to yield the classical logarithmic soft graviton theorem \cite{Boschetti:2026gfd} after assuming some of the results of Laddha, Saha, Sahoo and Sen \cite{Laddha:2018myi,Laddha:2018vbn,Sahoo:2018lxl,Saha:2019tub,Sahoo:2021ctw}. In this section, we prove that Eq.~\eqref{DAmatch} at order $O(G^2)$ is equivalent to our much simpler map \eqref{log: starting point0}  when expressed in the harmonic tetrad and when the asymptotic coefficients are identified in harmonic coordinates. This establishes at the same time that there are no $O(G^2)$ corrections to Eq.~\eqref{log: starting point0} due to non-linear interactions, and identifies Eq.~\eqref{DAmatch} as its non-perturbative completion. Moreover, this suggests that harmonic coordinates are best suited to express antipodal map, as the harmonic expression is much simpler than the radiative expression. Let us now demonstrate this mapping.\\

\noindent We recall that our result \eqref{log: starting point0} is written in the harmonic background tetrad
\begin{align}
l^{\mathrm{harm}} \equiv \frac{1}{\sqrt{2}}\partial_r,\qquad n^{\mathrm{harm}} \equiv \sqrt{2}\left(\partial_u-\frac{1}{2}\partial_r\right),\qquad m^{\mathrm{harm}} \equiv \frac{\theta^A}{\sqrt{2}r}\partial_A, 
\end{align}
where radial derivatives are taken at fixed $u$. The non-linear metric in harmonic coordinates has the form $\eta_{\mu\nu}+O(r^{-1})$. We can therefore promote this harmonic background tetrad as the harmonic tetrad to the non-linear theory up to corrections proportional to $O(r^{-1})\partial_t$, $O(r^{-1})\partial_r$ and $O(r^{-2})\partial_A$. In what follows, we will consider the leading contributions to the Weyl scalar including subleading $\log r/r$ corrections. The $1/r$ subleading corrections can be safely ignored for our arguments. The change of coordinates from harmonic coordinates to outgoing radiative coordinates is the inverse of Eq. \eqref{mapcoord}, namely  
\begin{subequations}
\begin{align}
u &=U-2G(q\cdot P)\log (R/b)+O(R^0),\label{uU}\\ 
r &=R+2Gn^iP_i\log (R/b)+O(R^0),\\
x^A &=X^A+\frac{2G}{R}e_i^AP^i\log (R/b)+O(R^{-1}),
\end{align}
\end{subequations}
where the subleading terms will not enter our arguments. Applying this change of coordinates combined with a type III rotation to align the phase of $m^{\text{rad}+}$ to $m^\text{harm}$, we obtain
\begin{subequations}\label{tetrad trans}
\begin{align}
l^{\mathrm{rad}+}&=l^{\mathrm{harm}}+\bar a  m^{\mathrm{harm}}+a \bar m^{\mathrm{harm}}+O(a^2),\\
n^{\mathrm{rad}+}&=n^{\mathrm{harm}}-\bar am^{\mathrm{harm}}-a\bar m^{\mathrm{harm}}+O(a^2),\\
m^{\mathrm{rad}+}&=m^{\mathrm{harm}}+a\left(n^{\mathrm{harm}}-l^{\mathrm{harm}}\right)+O(a^2),
\end{align}
\end{subequations}
where
\begin{align}
a\equiv -\frac{G\log(R/b)}{R}\theta^AD_A(q\cdot P)+O(1/R).
\end{align}
We can neglect $O(a^2)$ terms as they are $O(G^2)$ and will therefore affect the Weyl scalars at order $O(G^3)$. The harmonic and outgoing radiation tetrad are therefore related by a combined type I and type II rotation. The Weyl scalars are related as
\begin{subequations} \label{changeWeyl}
\begin{align}
\Psi_1^{\mathrm{rad}+} &=\Psi_1^{\mathrm{harm}}-\bar a\Psi_0^{\mathrm{harm}}+3a\Psi_2^{\mathrm{harm}}+O(a^2),\\
\Psi_3^{\mathrm{rad}+} &=\Psi_3^{\mathrm{harm}}-3\bar a\Psi_2^{\mathrm{harm}} +a\Psi_4^{\mathrm{harm}}+O(a^2),
\end{align}    
\end{subequations}
In the absence of matter at spatial infinity, 
\begin{align}
\Psi_0^{\mathrm{harm}}\vert_{\scri^+_-}=O(GR^{-4}), \qquad \Psi_2^{\mathrm{harm}}\vert_{\scri^+_-}=O(GR^{-3}), 
\end{align}
see Eqs. (24)-(28) of \cite{Compere:2026mdj}. Therefore, only the term $+3a\Psi_2^{\mathrm{harm}}\vert_{\scri^+_-}$ contributes at order \(G^2\log R/R^4\) to $\Psi_1^{\mathrm{rad}+}$ at $\scri^+_-$.  \\

\noindent Let us now look at the same outgoing tetrad but in a past radiative coordinate chart adapted to the description of past null infinity. 
 In the ingoing Bondi-Sachs coordinate chart $(V,R^-,X^{-A})$, one can rewrite the outgoing radiative tetrad as 
\begin{subequations}
\begin{align}\label{tetrad:radminus}
l^{\text{rad}+}&= \sqrt{2}( \partial_V + \frac{1}{2}(\partial_{R^-})_V )  +O(1/R^-)=\frac{1}{\sqrt{2}} (1,n^i(X^{-A})) +O(1/R^-) , \\
n^{\text{rad}+}& = - \frac{1}{\sqrt{2}} (\partial_{R^-})\vert_{V} =\frac{1}{\sqrt{2}} (1,-n^i(X^{-A})) ,\\
m^{\text{rad}+}&=\frac{1}{\sqrt{2}}(0,\theta^A e^i_A) +O(1/R^-). 
\end{align}
\end{subequations}
Its components written in the Cartesian frame $X^{-\mu} \equiv (T^-,X_i^-)$ where $T^-=V-R^-$, $X_i^-=R^- n_i(X^{-A})$ are identical to the ones in the $(T,X_i)$ frame described in Eq. \eqref{tetrad:rad} at leading order in the large radius expansion.  The change of coordinates from harmonic coordinates to incoming radiative coordinates is $X^{-\mu} =x^\mu + 2G P^\mu \log (r/b)+\dots$ which leads to $v=V+2G(\tilde q \cdot P)\log (R^-/b)+\dots$\footnote{This change of coordinates can be formally obtained from the change of coordinates from harmonic coordinates to outgoing radiative coordinates as follows. Following the procedure outlined in Section 2.5. of \cite{Compere:2023qoa}, one performs an active time reversal and then a passive coordinate transformation. In practice, it amounts to the substitution $u \to -v$, $U \to -V$ and $R \to R^-$ without antipodal map over the sphere. Since incoming and outgoing momenta have opposite signs in our conventions and the final $v$ increases towards the future after the active time reversal, one needs to also replace $P^i \to -P^i$ when keeping $P^0$ unchanged.}. Applying this change of coordinates leads to a frame rotation identical to Eq. \eqref{changeWeyl} but with parameter $a$ replaced by $a^- \equiv -a$. \\

\noindent We deduce from Eqs. (30)-(34) of \cite{Compere:2026mdj} that in the absence of matter at spatial infinity, 
\begin{align}
\Psi_4^{\mathrm{harm}}\vert_{\scri^-_+}=O(G(R^-)^{-4}),\qquad \Psi_2^{\mathrm{harm}}\vert_{\scri^-_+}=O(G(R^-)^{-3}).
\end{align}
Therefore, only the term $-3\bar a^-\Psi_2^{\mathrm{harm}}\vert_{\scri^-_+}$ contributes at order \(G^2\log R^-/(R^-)^4\) at $\scri^-_+$. Using Eqs. (24)-(34) and Eq. (38c) of \cite{Compere:2026mdj} we have 
\begin{subequations}
\begin{align}
\theta_A\Psi^{\text{rad}+}_2\Big|_{\scri^+_-, \frac{U^0}{R^3}}=\theta_A\Psi^{\text{harm}+}_2\Big|_{\scri^+_-, \frac{u^0}{r^3}}=\theta^B \mathcal M^{+(0)}_{BA}(\vec{n}), \\   
\bar \theta_A\Psi^{\text{rad}+}_2\Big|_{\scri^-_+, \frac{V^0}{(R^-)^3}}=\bar \theta_A\Psi^{\text{harm}+}_2\Big|_{\scri^-_+, \frac{v^0}{r^3}}=\bar \theta^B \mathcal M^{-(0)}_{BA}(\vec{n}),
\end{align}
\end{subequations}
where one has the antipodal relationship $\mathcal M_{AB}^{-(0)}=\Upsilon^*\mathcal M_{AB}^{+(0)}$. The shift of $\Psi_1^{\text{rad}+}\vert_{\scri^+_-, \frac{\log R}{R^4}}$ due to the change of tetrad is therefore 
\begin{align}
\delta_{\mathrm{tetrad}}\Psi_1^{\text{rad}+}\vert_{\scri^+_-, \frac{\log R}{R^4}}=-3 G\, \theta^AD^B(q\cdot P)\mathcal M_{AB}^{+(0)}(\vec{n}), 
\end{align}
or, equivalently, after using Eq. \eqref{thetaUps}, 
\begin{align}
\delta_{\mathrm{tetrad}}\Upsilon^*\Psi_1^{\text{rad}+}\vert_{\scri^+_-, \frac{\log R}{R^4}}=+3 G\, \bar \theta^AD^B(\tilde{q}\cdot P)\mathcal M_{AB}^{-(0)}(\vec{n}). \label{tetrad variation 1}
\end{align}
Similarly, we have 
\begin{equation}
\delta_{\mathrm{tetrad}}\Psi_3^{\text{rad}+}\vert_{\scri^-_+, \frac{\log R^-}{(R^-)^4}}=3 G\, \bar \theta^AD^B(\tilde{q}\cdot P)\mathcal M_{AB}^{-(0)}(\vec{n})
\end{equation}
after using $a^- = - a$ and switching $D^B(q\cdot P)=-D^B(\tilde{q}\cdot P)$, so the two contributions add up in Eq. \eqref{DAmatch}.
In addition to this frame rotation, a second consideration matters when comparing Eqs. \eqref{DAmatch} and \eqref{log: starting point0}. The $\log(-U)/R^4$ term of $\Psi_1^{\text{rad}+}$ at $\scri^+_-$ (resp. the $\log V/(R^-)^4$ term of $\Psi_3^{\text{rad}+}$ at $\scri^-_+$) is not the leading term in the $-U \to \infty$ (resp. $V \to \infty$) expansion after taking large $R/R^-$. Instead, the leading term is 
\begin{align}
\Psi^{\text{rad}+}_1 \vert_{\scri^+_-, \frac{U}{R^4}}=\frac{1}{2}D^B \mathcal M_{AB}^{+(0)}\theta^A, \quad \Psi^{\text{rad}+}_3 \vert_{\scri^-_+, \frac{V}{(R^-)^4}}=\frac{1}{2}D^B \mathcal M_{AB}^{-(0)}\bar \theta^A.  \label{eq3.60}
\end{align}
The shift of retarded time \eqref{uU} then brings a shift 
\begin{align}
\delta_{\mathrm{time}}\Psi_1^{\text{rad}+}\vert_{\scri^+_-, \frac{\log R}{R^4}}=- G\, \theta^A (q \cdot P)D^B \mathcal M_{AB}^{+(0)}(\vec{n}).  
\end{align}
Similarly, 
\begin{equation}\delta_{\mathrm{time}}\Psi_3^{\text{rad}+}\vert_{\scri^-_+, \frac{\log R^-}{(R^-)^4}}=+ G\, \bar \theta^A (\tilde{q} \cdot P)D^B \mathcal M_{AB}^{-(0)}(\vec{n}).\end{equation}
Linearly combining the frame shifts and the time shifts at both $\scri^+$ and $\scri^-$ maps Eq. \eqref{DAmatch} to 
\begin{align}
-\Upsilon^* \left( \Psi^{\text{harm}}_1 \vert_{\scri^+_-, \frac{\log r}{r^4}}+  \Psi_1^{\text{harm}} \vert_{\scri^+_-, \frac{\log |u|}{r^4}} \right)&=  \Psi^{\text{harm}}_3 \vert_{\scri^-_+, \frac{\log r}{r^4}}+ \Psi^{\text{harm}}_3 \vert_{\scri^-_+, \frac{\log v}{r^4}}.     
\end{align}
As there is no incoming radiation in our scattering scheme, the $\log r/r^4$ branch of $\Psi^{\text{harm}}_1\vert_{\scri^+_-}$ and the $\log v/r^4$ branch of $\Psi^{\text{harm}}_3 \vert_{\scri^-_+}$ are vanishing, see Eqs. \eqref{eq:421}, \eqref{Eq:Psi1insoft}, \eqref{logrbranch} demonstrated below.  We have therefore proven that in the context of scattering, Eq. \eqref{log: starting point0} is the order-\(G^2\) form of the radiative matching law \eqref{DAmatch} valid in the full non-linear theory, under the polyhomogeneous hypotheses described in \cite{Compere:2026jmk,Boschetti:2026ogm}. In particular, no additional order-\(G^2\) term is required in Eq. \eqref{log: starting point0}. The radiative matching law \eqref{DAmatch} provides a nonperturbative completion, but the present coordinate and tetrad calculation does not establish the uncorrected harmonic formula beyond this perturbative order.

\section{Asymptotic Weyl scalars from classical gravitational scattering}
\label{Section:GravitationalScattering}

The previous section was established for generic multipole moments and made no specific reference to particle scattering. In this section, we explicitly evaluate the asymptotic NP Weyl scalars resulting from the scattering of massive particles. First, this allows us to perform explicit checks of the antipodal relations derived in Section~\ref{Section:AntipodalMatchings}. Second, the explicit form of these NP Weyl scalars will be needed for the new derivation of the leading and logarithmic soft graviton theorems presented in Section~\ref{Section:SoftTheorems}.

\subsection{Hierarchy of equations and stress-energy tensor}

\paragraph{Matter stress tensor and Liénard-Wiechert solution.}
We now consider a set of massive particles following trajectories $X_a^\mu(s_a)$, which, asymptotically, are either all incoming or all outgoing. We denote this by the summation index $a \in \text{in/out}$. The corresponding matter stress-tensor is given by 
\begin{equation}
    T^{\alpha \beta}(x)= \sum_{a \in \text{in/out}} \frac{m_a}{\sqrt{-g}} \int ds_a\, \dfrac{dX_a^\alpha}{ds_a}\dfrac{dX_a^\beta}{ds_a} \delta^{(4)}(x- X_a(s))\,. \label{Eq:MatterStressTensor}
\end{equation}
We further specify the asymptotic trajectories to be those of particles subject to acceleration due to the long-range nature of the gravitational force:
\begin{equation}
    X_a^\mu(s_a) =y_a^\mu + v_a^\mu s_a + G\,  \left( c_a^\mu \log\abs{s_a} + O\left(s_a^{-1}\right)  \right)+ O(G^2), \label{Eq:Trajectory}
\end{equation}
with $c_a^\mu$ given by \cite{Sahoo:2018lxl}
\begin{equation}
\label{deviation vector}
    c^\mu_a = \eta_a\sum_{\substack{b \neq a \\ \eta_b=\eta_a}}\dfrac{ m_b}{((v_a \cdot v_b)^2-1)^{3/2}} \left( v_a^\mu - v_b^\mu (2 (v_a\cdot v_b)^3 - 3 (v_a \cdot v_b) )\right),
\end{equation}
where the sum is over either the outgoing or incoming particles distinct from $a$. The only logarithmic contribution to the $O(G)$ trajectory deviation is the leading term proportional to $c_a^\mu$. Plugged in the matter stress tensor \eqref{Eq:MatterStressTensor}, these trajectories contribute to both equations \eqref{Eq:Hierarchy1}-\eqref{Eq:Hierarchy2}. We have 
\begin{equation}
    \begin{aligned}
        \abs{g}T^{\alpha \beta}(x)&=  \sum_{a \in \text{in/out}} \! \!m_a \! \int \! \!ds_a  \left\{v_a^\alpha v_a^\beta - G \left[ v_a^\alpha v_a^\beta c_a^\lambda \, \log \abs{s_a} \partial_\lambda + \text{(non-log)}  \right] \right\}\\
        & \hspace{280 pt}\delta^{(4)}(x-v_a s_a).
    \end{aligned}
\end{equation}
In the rest of this work, we will be interested in the leading logarithmic contributions (in $u$ and $r$ at $\scri^+$, in $v$ and $r$ at $\scri^-$)  to the Newman-Penrose Weyl scalars at $O(G^2)$. Using the retarded Green function 
\begin{equation}
    G_\text{ret}(x,x')\equiv -\dfrac{1}{2\pi} \Theta(x^0-x'^{0}) \delta\left((x-x')^2 \right), \label{Eq:GreenRet}
\end{equation}
we get the ingoing and outgoing waveforms
\begin{equation}
    \mathfrak{h}_{(1)}^{\alpha \beta}(x) = - \sum_{a \in \text{in/out}} \dfrac{4 m_a v_a^\alpha v_a^\beta }{\sqrt{(v_a \cdot (x-y_a))^2 + (x-y_a)^2}} \equiv - \sum_{a \in \text{in/out}}\dfrac{4 m_a v_a^\alpha v_a^\beta }{\rho_a} \label{Eq:h1matter}
\end{equation}
and
\begin{equation}
    \mathfrak{h}_{(2),\text{matter}}^{\alpha \beta}(x) =  \sum_{a \in \text{in/out}} 4 m_a v_a^\alpha v_a^\beta c_a^\gamma \, \partial_\gamma \left( \dfrac{\log|s_{a,\text{ret}}|}{\rho_a}\right) +\, (\text{non-log}),\label{Eq:h2matter} 
\end{equation}
where $s_{a,\text{ret}}$ is obtained by solving the retarded lightcone equation
\begin{equation}
    \left(x-X_a(s_{a,\text{ret}})\right)^2=0, \qquad x^0 \geq X_a^0(s_{a,\text{ret}}). 
\end{equation}

\paragraph{Gravitational stress tensor.}
Unlike \eqref{Eq:h2matter} which relies on the asymptotic trajectories, the first-order perturbation \eqref{Eq:h1matter} is valid everywhere outside of the scattering region and can be plugged in \eqref{Eq:SourceTerm} to determine the source of the second-order perturbation $\mathfrak{h}_{(2),\text{grav.}}^{\alpha \beta}$. With 
\begin{equation}
    \begin{aligned}
        \partial^\mu \rho_a = \dfrac{(x-y_a)^\mu + v_{a}^{\mu}\, v_a \cdot (x-y_a) }{\rho_a} \equiv \dfrac{Z_{a}^{\mu}}{\rho_a}, \qquad \partial^\nu Z_{a}^{\mu} = v_{a}^{\mu} v_{a}^{\nu}  + \eta^{\mu \nu} \equiv \Pi_a^{\mu \nu} \label{Eq:SourceObjectsDefinitions}
    \end{aligned}
\end{equation}
and the self-particle identities
\begin{equation}
    v_a^\mu Z_{a \mu}=0, \quad Z_a^\mu Z_{a \mu}= \rho_a^2, \quad (x-y_a)^\mu Z_{a\mu}= \rho_a^2,
\end{equation}
the source \eqref{Eq:SourceTerm} can be written as 
\begin{equation}
    \begin{aligned}
        &N^{\alpha \beta}[\mathfrak{h}_{(1)},\mathfrak{h}_{(1)}]
        =  \sum_{a \in \text{in/out}} m_a^2
        \left[4\frac{Z_a^\alpha Z_a^\beta}{\rho_a^6}-16\frac{v_a^\alpha v_a^\beta}{\rho_a^4}-2\frac{\eta^{\alpha\beta}}{\rho_a^4}\right]\\
        &+ \sum_{a\in\text{in/out}}\sum_{b\neq a}m_a m_b\,\left[ 16\frac{v_b^\alpha v_b^\beta}{\rho_a}\left(\frac{(v_a\cdot v_b)^2-1}{\rho_b^3}-\frac{3(v_a\cdot Z_b)^2}{\rho_b^5}\right)  \right.\\
        &+ \left(8(v_a\cdot v_b)^2-4\right)
        \frac{Z_a^\alpha Z_b^\beta}{\rho_a^3\rho_b^3}+16\frac{v_a^\alpha v_b^\beta}{\rho_a^3\rho_b^3}\left[(v_a\cdot v_b)(Z_a\cdot Z_b)
        +(v_a\cdot Z_b)(v_b\cdot Z_a)\right]
        \\
        &-16(v_a\cdot v_b)\frac{\left(Z_a^\alpha v_b^\beta+Z_a^\beta v_b^\alpha\right)(v_a\cdot Z_b)}{\rho_a^3\rho_b^3
        }\\
        &+\left.\eta^{\alpha\beta}\frac{\left(-4(v_a\cdot v_b)^2+2\right)(Z_a\cdot Z_b)+8(v_a\cdot v_b)(v_b\cdot Z_a)(v_a\cdot Z_b)}{\rho_a^3\rho_b^3}\right]. \label{Eq:SourceLWPluggedIn}
    \end{aligned}
\end{equation}
In particular, this corresponds to the gravitational source terms at the corners of both null infinities. However, since determining the gravitational waveform requires integrating over the past lightcone, this neglects the effect of the scattering region on $\scri^+_+$. We address this missing contribution and how it relates to the graviton drag in Section \ref{Section:GravDrag}.

\subsection{Asymptotic Weyl tensor perturbations}
\label{subsection Lienard Wiechert fields}
The covariant metric perturbations
\begin{equation}
    g_{\mu \nu}= \eta_{\mu \nu} + G h^{(1)}_{\mu \nu} + G^2 h^{(2)}_{\mu \nu} + O(G^3) \label{Eq:PMExpansion}
\end{equation}
can be computed from the gothic metric $\mathfrak{h}^{\alpha \beta}$: 
\begin{equation}
    \begin{aligned}
        h^{(1)}_{\mu \nu} &= -\mathfrak{h}^{(1)}_{\mu \nu} + \dfrac12 \eta_{\mu \nu}\mathfrak{h}_{(1)}, \\
        h^{(2)}_{\mu \nu} &= -\mathfrak{h}^{(2)}_{\mu \nu} + \dfrac12 \eta_{\mu \nu}\mathfrak{h}_{(2)} + \mathfrak{h}^{(1)}_{\mu \lambda}\mathfrak{h}_{(1)\nu}^{\phantom{(1)\nu}\lambda} - \dfrac12 \mathfrak{h}_{(1)} \mathfrak{h}^{(1)}_{\mu \nu} + \eta_{\mu \nu} \left(\dfrac18 \mathfrak{h}_{(1)}^2 - \dfrac14 \mathfrak{h}^{(1)}_{\lambda \sigma} \mathfrak{h}_{(1)}^{\lambda \sigma} \right),
    \end{aligned}
\end{equation}
such that 
\begin{equation}
    \begin{aligned}
        &h^{(2), \text{matter}}_{\mu \nu} = -\mathfrak{h}^{(2), \text{matter}}_{\mu \nu} + \dfrac12 \eta_{\mu \nu}\mathfrak{h}_{(2), \text{matter}} \\
        &\hspace{113 pt}+ \mathfrak{h}^{(1)}_{\mu \lambda}\mathfrak{h}_{(1)\nu}^{\phantom{(1)\nu}\lambda} - \dfrac12 \mathfrak{h}_{(1)} \mathfrak{h}^{(1)}_{\mu \nu} + \eta_{\mu \nu} \left(\dfrac18 \mathfrak{h}_{(1)}^2 - \dfrac14 \mathfrak{h}^{(1)}_{\lambda \sigma} \mathfrak{h}_{(1)}^{\lambda \sigma} \right),\\
        &h^{(2), \text{grav.}}_{\mu \nu} = -\mathfrak{h}^{(2), \text{grav.}}_{\mu \nu} + \dfrac12 \eta_{\mu \nu}\mathfrak{h}^{(2), \text{grav.}}.\label{Eq:MetricFromGothicG2}
    \end{aligned}
\end{equation}
We now decompose the Weyl tensor into its various contributions
\begin{equation}
    \begin{aligned}
        C_{\mu \nu \rho \sigma}&= G \,C_{\mu \nu \rho \sigma}^{(1),\text{lin.}}[h_{(1)}]\\
        &+ G^2 \left(C_{\mu \nu \rho \sigma}^{(2),\text{non-lin.}}[h_{(1)}, h_{(1)}] + C_{\mu \nu \rho \sigma}^{(2),\text{lin.}}[h_{(2), \text{matter}}] + C_{\mu \nu \rho \sigma}^{(2),\text{lin.}}[h_{(2), \text{grav.}}]\right).
    \end{aligned}
\end{equation}
As already stated, we will be interested in the logarithmic pieces in the Weyl tensor at order $G^2$. This focus allows us to discard contributions that cannot produce such terms: first $C_{\mu \nu \rho \sigma}^{(2),\text{non-lin.}}[h_{(1)}, h_{(1)}]$, but also all terms in \eqref{Eq:MetricFromGothicG2} involving $\mathfrak{h}_{(1)}^{\mu \nu}$ and its contractions. The only object of interest is thus the linear vacuum  Weyl tensor at each order:
\begin{equation}
    C_{\mu \nu \rho \sigma}^{(i),\text{lin.}}[h^{(i)}]\equiv \dfrac12 \left( \partial_\rho \partial_\nu h^{(i)}_{\mu \sigma} + \partial_\sigma \partial_\mu h^{(i)}_{\nu \rho}  - \partial_\rho \partial_\mu h^{(i)}_{\nu \sigma} - \partial_\sigma \partial_\nu h^{(i)}_{\mu \rho}\right) 
\end{equation}
with $h^{(i)}$, $i=1,2$, truncated to the linearly-trace-reversed perturbation
\begin{equation}
    h^{(i)}_{\mu \nu} \equiv -\mathfrak{h}^{(i)}_{\mu \nu} + \dfrac12 \eta_{\mu \nu}\mathfrak{h}^{(i)}.
\end{equation}

\subsubsection{Matter Weyl perturbations from ingoing particles}

\begin{figure}[h!]
    \centering
    \includegraphics[width=1\linewidth]{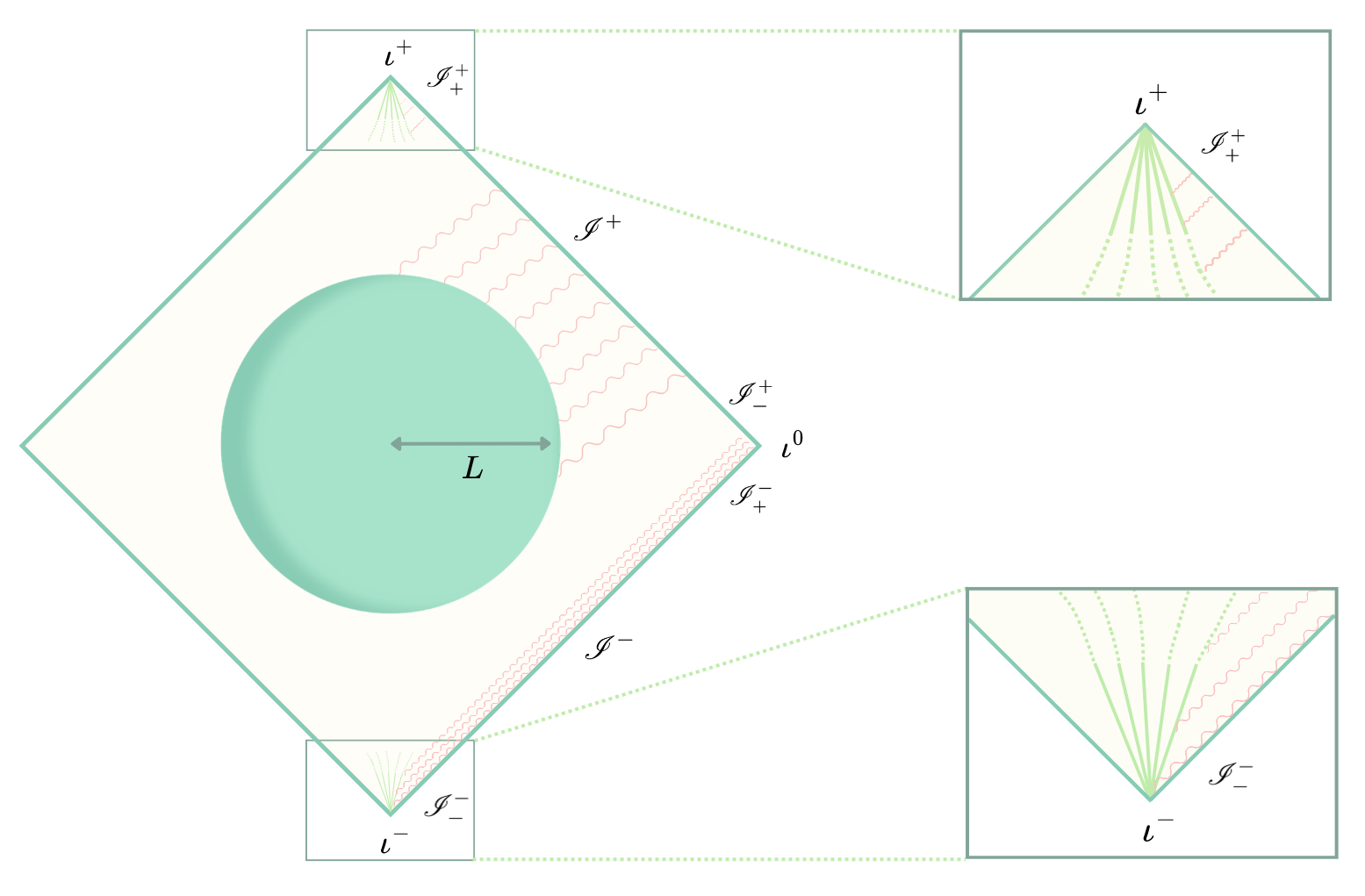}
    \caption{Ingoing particles at $\iota^-$ source perturbations at $\scri^-$ and $\scri^+_-$, while outgoing particles source perturbations at $\scri^+_+$.}
    \label{Fig:Scattering}
\end{figure}

We now consider the second-order perturbations created by ingoing particles at $\iota^-$ in the absence of incoming radiation at $\scri^-$. This is schematically represented in Figure \ref{Fig:Scattering}. This contribution to the Weyl tensor is generated by the accelerated trajectories of the massive particles in the far past $(s \to -\infty)$, as shown by Eq. \eqref{Eq:h2matter}. 

\paragraph{At past null infinity.}
To zoom onto $\scri^-_+$, we introduce the advanced coordinates 
\begin{equation}
    t= v -r, \quad x^i = r n^i
\end{equation}
and consider the large $r$ limit. In these coordinates, the retarded worldline parameter $s_{a,\text{ret}}$ is given by 
\begin{equation}
    s_{a,\text{ret}}= 2 r (\tilde{q} \cdot v_a) + O(r^0)
\end{equation}
and is negative given that $(\tilde{q} \cdot v_a)$ is negative for future-directed timelike particles. For the needs of the leading soft graviton theorem, we start by discussing the leading $O(G)$ contribution to $\Psi_2$. With 
\begin{equation}
    \Psi_2^{(1)}\equiv  -G\, C^{(1), \text{lin.}}_{l m\bar{m}n}[h_{(1)}],
\end{equation}
we find 
\begin{equation}
   \left. \Psi_2^{(1)}\right|_{\scri^-_+}= -G\, \sum_{a \in \text{in}} \dfrac{m_a}{r^3(\tilde{q}\cdot v_a)^3} + O(r^{-4}).
\end{equation}
In the notations of Section \ref{Section:AntipodalMatchings}, this is 
\begin{equation}
   \left. \overset{2,2}{\Psi}_{\hspace{-1 pt}2}\right|_{\scri^-_+}= -G\sum_{a \in \text{in}} \dfrac{m_a}{(\tilde{q}\cdot v_a)^3}.
\end{equation}
At next order, with 
\begin{equation}
    \Psi_3^{(2),\text{matter}}= -G^2\, C^{(2), \text{lin.}}_{l n \bar{m}n}[h_{(2), \text{matter}}],
\end{equation}
we find that $\Psi_3^{(2),\text{matter}}$, at $\scri^-_+$, is given by
\begin{equation}\label{eq:421}
    \left. \Psi_3^{(2),\text{matter}}\right|_{\scri^-_+}= G^2\sum_{a \in \text{in}}\dfrac{3m_a}{2 (\tilde{q} \cdot v_a)^4}\dfrac{\log r}{r^4}\bar{\theta}^A (\partial_A q^\mu)(c^a_{ \mu} v^a_{\nu}- v^a_{ \mu} c^a_{\nu} ) \tilde{q}^\nu + O\!\left(r^{-4}\right).
\end{equation}
This leading contribution and the $O(r^{-4})$ subleading contribution do not depend on $v$, so this result is valid everywhere at $\scri^-$. There are also no $\log v$ terms as these are switched on by the presence of incoming radiation. In the language of Eq. (2.13b) of \cite{Geiller:2024ryw}, this is part of $D_{AB}$ appearing in $\Psi_1$ at $\scri^+$ and thus in $\Psi_3$ at $\scri^-$. The $v$-independence is a consequence of the Einstein equations $\partial_v D_{AB}=0$. In the notations of Section \ref{Section:AntipodalMatchings}, this is  
\begin{equation}
    \left.(\hspace{-3 pt}\overset{3, \text{log.}}{\Psi}_{\hspace{-6 pt}3})_A\right|_{\scri^-_+}  = G^2 \sum_{a \in \text{in}}\dfrac{3m_a}{2 (\tilde{q} \cdot v_a)^4}(\partial_A q^\mu)(c^a_{ \mu} v^a_{\nu}- v^a_{ \mu} c^a_{\nu}) \tilde{q}^\nu\,. \label{Eq:Psi3insoft}
\end{equation}

\paragraph{At future null infinity.} We perform the same calculation in retarded coordinates
\begin{equation}
    t= u+r, \qquad x^i = r n^i.
\end{equation}
For large $r$, the retarded worldline parameter is 
\begin{equation}
    s_{a,\text{ret}} = -\left(\dfrac{u}{(q \cdot v_a)} + O(u^0) \right)+ O(r^{-1}),
\end{equation}
with $(q \cdot v_a)<0$, such that the accelerated trajectories \eqref{Eq:Trajectory} are only a valid approximation of the actual $O(G)$ massive particle trajectory deviations from straight lines for $u\to - \infty$. At order $G$, we have 
\begin{equation}
   \left. \Psi_2^{(1)}\right|_{\scri^+_-}= -G\sum_{a \in \text{in}} \dfrac{m_a}{r^3(q\cdot v_a)^3} + O(r^{-4}),
\end{equation}
or, in the notations of Section \ref{Section:AntipodalMatchings},
\begin{equation}
   \left. \overset{2,2}{\Psi}_{\!2}\right|_{\scri^+_-}= -G\sum_{a \in \text{in}} \dfrac{m_a}{(q\cdot v_a)^3}.
\end{equation}
At next order, with 
\begin{equation}
    \Psi_1^{(2),\text{matter}} = -G^2\, C^{(2), \text{lin.}}_{lnlm}[h_{(2), \text{matter}}],
\end{equation}
we get, at $\scri^+_-$ the logarithmic contribution
\begin{equation}
    \left. \Psi_1^{(2),\text{matter}}\right|_{\scri^+_-}=-G^2\sum_{a \in \text{in}}\dfrac{3m_a}{2 (q \cdot v_a)^4}\dfrac{\log \abs{u}}{r^4}\theta^A (\partial_A q^\mu)( c^a_{ \mu} v^a_{\nu}-v^a_{\mu} c^a_{\nu} ) q^\nu + O\!\left(r^{-4}\right),\label{Eq:Psi1insoft}
\end{equation}
where we checked that the $O(r^{-4})$ terms do not depend on $u$ and do not include $\log r$ contributions, switched on by advanced radiation. In the notations of Section \ref{Section:AntipodalMatchings}, this is 
\begin{equation}
    \left. (\hspace{-3 pt}\overset{2, \text{log.}}{\Psi}_{\hspace{-6 pt}1})_A\right|_{\scri^+_-}  = -G^2\sum_{a\in \text{in}}\dfrac{3m_a}{2 (q \cdot v_a)^4} (\partial_A q^\mu)( c^a_{ \mu} v^a_{\nu}-v^a_{\mu} c^a_{\nu} ) q^\nu.
\end{equation}
Again, in Eq. (2.13b) of \cite{Geiller:2024ryw}, this is picking the $\log u$ contribution of $\mathcal{P}_a$. These results agree with the antipodal matching conditions obtained in Section \ref{Section:AntipodalMatchings}. Namely, we have found 
\begin{equation}
\label{antipodal matching psi2 scattering}
     \left. \overset{2,2}{\Psi}_{\!2}\right|_{\scri^+_-}= \Upsilon^*\! \left. \overset{2,2}{\Psi}_{\!2}\right|_{\scri^-_+}
\end{equation}
and
\begin{equation}
    (\hspace{-6 pt}\left.\overset{2, \text{log.}}{\Psi}_{\hspace{-6 pt}1})_A\right|_{\scri^+_-} =\Upsilon^*\! \! \left.(\hspace{-4 pt}\overset{3, \text{log.}}{\Psi}_{\hspace{-6 pt}3})_A\right|_{\scri^-_+}. \label{Eq:AntipodalMatchingScattering}
\end{equation}

\subsubsection{Matter Weyl perturbations from outgoing particles}
We now consider outgoing particles in the far future $(s \to + \infty)$ as shown in the top right part of Figure \ref{Fig:Scattering}. At $\scri^+_+$, we find  
\begin{equation}
   \left. \Psi_2^{(1)}\right|_{\scri^+_+}= -G\, \sum_{a \in \text{out}} \dfrac{m_a}{r^3(q\cdot v_a)^3} + O(r^{-4}),
\end{equation}
that is, 
\begin{equation}
   \left. \overset{2,2}{\Psi}_{\!2}\right|_{\scri^+_+}= -G\sum_{a \in \text{out}} \dfrac{m_a}{(q\cdot v_a)^3}.
\end{equation}
We also have
\begin{equation}
    \left. \Psi_1^{(2),\text{matter}}\right|_{\scri^+_+}=G^2 \sum_{a \in \text{out}}\dfrac{3m_a}{2 (q \cdot v_a)^4}\dfrac{\log u}{r^4}\theta^A (\partial_A q^\mu)(v^a_{ \mu} c^a_{\nu} - c^a_{ \mu} v^a_{\nu}) q^\nu + O\!\left(r^{-4}\right),\label{Eq:Psi1outsoft}
\end{equation}
where, again, the $O(r^{-4})$ terms do not contain $u$-dependent terms or $\log r$ terms. We write this as 
\begin{equation}
    \left. (\hspace{-3 pt}\overset{2, \text{log.}}{\Psi}_{\hspace{-6 pt}1})_A^{\text{matter}}\right|_{\scri^+_+}  = G^2\sum_{a \in \text{out}}\dfrac{3m_a}{2 (q \cdot v_a)^4} (\partial_A q^\mu)(v^a_{ \mu} c^a_{\nu} - c^a_{ \mu} v^a_{\nu}) q^\nu.\label{Eq:Psi1outsoft1}
\end{equation}

\subsection{Gravitational Weyl perturbations}\label{Section:GravDrag}
\paragraph{Radiative tail at future null infinity.}
Let us now turn to the gravitational Weyl perturbations. In addition to that of the second-order trajectories, the classical soft factor is expected to receive a non-trivial contribution from the gravitational stress-tensor sourced by $\mathfrak{h}_1$. At $\scri^+_-$, the past lightcone on which the retarded Green function has support is essentially the asymptotic region corresponding to past null infinity. Therefore, at $\scri^+_-$, the gravitational source term is built upon the ingoing Liénard-Wiechert solution only. At $\scri^+_+$, however, the past lightcone covers $\scri^+$ and the resulting perturbation will thus depend on the memory contribution to the first-order waveform. Our scattering setup is the one represented in Figure \ref{Fig:Scattering}: $m$ ingoing massive hard particles at $\iota^-$ enter a scattering region of possibly large but finite characteristic size $L$, from which $n$ outgoing massive hard particles emerge and exit at $\iota^+$. With $\rho(x)= - (q \cdot v) r + O(r^0)$, we thus consider 
\begin{equation}
	\hspace{-5 pt}\mathfrak{h}_{(1)}^{\alpha \beta}(x) =  \dfrac{4}{r} \left(\sum_{i \in \text{in}}^{m} m_i\dfrac{ {v}_{i}^{\alpha} {v}_{i}^{\beta}}{ (q\cdot v_i)} + \mathcal{F}\left(\dfrac{u}{L}\right) \left[ \sum_{j \in \text{out}}^{n} m_j\dfrac{ {v}_j^{\alpha} {v}_j^{\beta}}{ (q\cdot v_j)}- \sum_{i \in \text{in}}^{m} m_i\dfrac{ {v}_{i}^{\alpha} {v}_{i}^{\beta}}{ (q\cdot v_i)}\right]\right)+ O(r^{-2}).
\end{equation}
Here, $\mathcal{F}$ is a profile interpolating smoothly between the ingoing and outgoing waveforms. It is defined such that 
\begin{align}
	\mathcal{F}(- \infty) &=0, \qquad \mathcal{F}(+\infty) = 1, \qquad 
\lim_{u \to \pm \infty} u\,  \dot{\mathcal{F}}(u)=0.\label{derF}
\end{align}
The presence of the transition function drastically affects the large $r$ expansion of the source $N^{\alpha \beta}$. Whereas the pure ingoing and outgoing contributions, i.e., Eq. \eqref{Eq:SourceLWPluggedIn}, give
\begin{equation}
	N^{\alpha \beta}|_{\scri^+_\pm} = \sum_{k=0}^\infty \sum_{p=0}^k \dfrac{u^{k-p}}{r^{k+4}} N^{+,\alpha\beta}_{k,p,\pm} (\vec{n})
\end{equation}
and 
\begin{equation}
	N^{\alpha \beta}|_{\scri^-_+} = \sum_{k=0}^\infty \sum_{p=0}^k \dfrac{v^{k-p}}{r^{k+4}} N^{-,\alpha\beta}_{k,p, -} (\vec{n}),
\end{equation}
the transition function $\mathcal{F}$ introduces new leading terms. Indeed, we have
\begin{equation}
	\partial_\mu \mathfrak{h}_{(1)}^{\alpha \beta}= -\dfrac{4 q_\mu}{r L}\dot{\mathcal{F}}\left(\dfrac{u}{L}\right) \left( \sum_{j \in \text{out}}^{n} m_j\dfrac{ {v}_j^{\alpha} {v}_j^{\beta}}{ (q\cdot v_j)}- \sum_{i \in \text{in}}^{m} m_i\dfrac{ {v}_{i}^{\alpha} {v}_{i}^{\beta}}{ (q\cdot v_i)}\right) + O(r^{-2})
\end{equation}
and 
\begin{equation}
	\partial_\mu \partial_\nu \mathfrak{h}_{(1)}^{\alpha \beta}= \dfrac{4 q_\mu q_\nu}{r L^2}\ddot{\mathcal{F}}\left(\dfrac{u}{L}\right) \left( \sum_{j \in \text{out}}^{n} m_j\dfrac{ {v}_j^{\alpha} {v}_j^{\beta}}{ (q\cdot v_j)}- \sum_{i \in \text{in}}^{m} m_i\dfrac{ {v}_{i}^{\alpha} {v}_{i}^{\beta}}{ (q\cdot v_i)}\right) + O(r^{-2}). \label{Eq:MemoryDoubleDiff}
\end{equation}
Let us, for now, only consider the first term of \eqref{Eq:SourceFormula}: 
\begin{equation}
	N^{\alpha \beta}[\mathfrak{h}_{(1)}, \mathfrak{h}_{(1)}] \supset -\mathfrak{h}_{(1)}^{\mu \nu} \partial_\mu \partial_\nu \mathfrak{h}_{(1)}^{\alpha \beta}.
\end{equation}
Given \eqref{Eq:MemoryDoubleDiff}, the leading contribution to the source is
\begin{equation}\label{Nab}
	\hspace{-8 pt} \begin{aligned}
		&N^{\alpha \beta}\supset -\dfrac{16}{r^2 L^2} \left(\sum_{i \in \text{in}}^{m} m_i (q \cdot{v}_{i})  + \mathcal{F}\left(\dfrac{u}{L}\right) \left[ \sum_{j \in \text{out}}^{n} m_j (q\cdot v_j)- \sum_{i \in \text{in}}^{m} m_i(q\cdot v_i)\right]\right)\\
		& \times \ddot{\mathcal{F}}\left(\dfrac{u}{L}\right) \left( \sum_{j \in \text{out}}^{n} m_j\dfrac{ {v}_j^{\alpha} {v}_j^{\beta}}{ (q\cdot v_j)}- \sum_{i \in \text{in}}^{m} m_i\dfrac{ {v}_{i}^{\alpha} {v}_{i}^{\beta}}{ (q\cdot v_i)}\right) +O\left(r^{-3}\right)\\
		&= -\dfrac{16}{r^2 L^2} \left(\sum_{i \in \text{in}}^{m} m_i (q \cdot{v}_{i})\right)\ddot{\mathcal{F}}\left(\dfrac{u}{L}\right) \left( \sum_{j \in \text{out}}^{n} m_j\dfrac{ {v}_j^{\alpha} {v}_j^{\beta}}{ (q\cdot v_j)}- \sum_{i \in \text{in}}^{m} m_i\dfrac{ {v}_{i}^{\alpha} {v}_{i}^{\beta}}{ (q\cdot v_i)}\right) + O\left(r^{-3}\right),
	\end{aligned}
\end{equation}
where we used momentum conservation to simplify the first line. With 
\begin{equation}
	\mathcal{A}^{\alpha \beta}(\vec n)\equiv \left(\sum_{i \in \text{in}}^{m} m_i (q \cdot{v}_{i})\right)\left( \sum_{j \in \text{out}}^{n} m_j\dfrac{ {v}_j^{\alpha} {v}_j^{\beta}}{ (q\cdot v_j)}- \sum_{i \in \text{in}}^{m} m_i\dfrac{ {v}_{i}^{\alpha} {v}_{i}^{\beta}}{ (q\cdot v_i)}\right),
\end{equation}
we thus consider the equation
\begin{equation}
	\Box \mathfrak{h}^{\alpha \beta}_{(2), \text{grav.}}(x) \supset -\dfrac{16}{r^2 L^2} \ddot{\mathcal{F}}\left(\dfrac{u}{L}\right) \mathcal{A}^{\alpha \beta}(\vec n). 
\end{equation}
The retarded Green function \eqref{Eq:GreenRet} then yields
\begin{equation}
	\mathfrak{h}^{\alpha \beta}_{(2), \text{grav.}}(x) \supset \dfrac{8}{\pi L^2} \int du' dr' d\Omega'\, \Theta(t-t') \delta\!\left((x-x')^2\right) \ddot{\mathcal{F}}\left(\dfrac{u'}{L}\right) \mathcal{A}^{\alpha \beta}(\vec n'),
\end{equation}
where we introduced the integration coordinates
\begin{equation}
	t'= u' + r', \qquad {x'}^i = r' {n'}^i
\end{equation}
together with the measure $d^4x'= {r'}^2 dr' du' d\Omega'$, cancelling the powers of $r'$ appearing in the source term. We can use the delta distribution to get rid of the integral over $r'$: 
\begin{equation}
	\delta\!\left((x-x')^2\right) = \dfrac{\delta(r'-r'_*)}{2\left|(u-u') + r(1-\vec{n} \cdot \vec{n}') \right|} \label{Eq:GravDelta}
\end{equation}
with
\begin{equation}
	r'_*\equiv \dfrac{(u-u')(2r + u-u')}{2((u-u')+ r(1-\vec{n} \cdot \vec{n}'))}. \label{Eq:GravRadiusLightcone}
\end{equation}
Then, we are left with 
\begin{equation}
	\mathfrak{h}^{\alpha \beta}_{(2), \text{grav.}}(x) \supset \dfrac{4}{\pi L^2} \int du'  d\Omega'\, \Theta(u-u') \ddot{\mathcal{F}}\left(\dfrac{u'}{L}\right) \dfrac{\mathcal{A}^{\alpha \beta}(\vec n')}{(u-u') + r(1-\vec{n} \cdot \vec{n}')}.
\end{equation}
First, we wish to determine the field at $\scri^+_+$. Before taking a large $r$ limit at fixed $u$ on the integrand, one should be mindful of the potential singular term arising from such a limit: 
\begin{equation}
	\dfrac{1}{(u-u') + r(1-\vec{n} \cdot \vec{n}')}= \dfrac{1}{ r(1-\vec{n} \cdot \vec{n}')} + O(r^{-2}),
\end{equation}
which is singular in the collinear regime $\vec{n}= \vec{n}'$. Hence, we delay such a limit and instead single out the collinear contribution to the integral by performing the following split:
\begin{equation}
	\mathfrak{h}^{\alpha \beta}_{(2), \text{grav.}}(x) \supset \dfrac{4}{\pi L^2} \int du'  d\Omega'\, \Theta(u-u') \ddot{\mathcal{F}}\left(\dfrac{u'}{L}\right) \dfrac{\mathcal{A}^{\alpha \beta}(\vec n) + (\mathcal{A}^{\alpha \beta}(\vec n')-\mathcal{A}^{\alpha \beta}(\vec n))}{(u-u') + r(1-\vec{n} \cdot \vec{n}')}.
\end{equation}
For the first term of the split, the angular integral can be performed before going to large $r$: with
\begin{equation}
	\hspace{-10 pt} \int d\Omega'\,  \dfrac{1}{(u-u') + r(1-\vec{n} \cdot \vec{n}')} = \dfrac{2 \pi}{r} \log\left( \dfrac{u-u'+ 2r}{u-u'} \right) = \dfrac{2 \pi}{r} \log\left( \dfrac{2r}{u-u'} \right) + O(r^{-2}),
\end{equation}
we find, for large $u$,
\begin{equation}
	\begin{aligned}
		\left.\mathfrak{h}^{\alpha \beta}_{(2), \text{grav.}}\right|_{\scri^+_+} &\supset  \dfrac{8}{r L^2} \int_{-\infty}^{+\infty} du'\,   \ddot{\mathcal{F}}\left(\dfrac{u'}{L}\right) \left(\log\left( \dfrac{2r}{u} \right)+ \dfrac{u'}{u} +O\left(u^{-2}\right)\right)\mathcal{A}^{\alpha \beta}(\vec n).
	\end{aligned}
\end{equation}
Only the transition function depends on $u'$ in the first term and thus its integral is proportional to 
\begin{equation}
	\dot{\mathcal{F}}(+\infty) -\dot{\mathcal{F}}(-\infty)=0
\end{equation} 
under our assumptions \eqref{derF}.  We can integrate the second term by parts: with $z=u'/L$, 
\begin{equation}
	\int_{-\infty}^{+\infty} du'\,   \ddot{\mathcal{F}}\left(\dfrac{u'}{L}\right) u' =L^2 \left.  \left(z \dot{\mathcal{F}}(z) -  \mathcal{F}(z)\right)\right|^{z\to + \infty}_{z\to - \infty}= -L^2.
\end{equation}
Then, we find a universal $(ru)^{-1}$ tail in the waveform at second post-Minkowskian order. This is exactly the contribution of the graviton drag derived in \cite{Saha:2019tub}, here fully derived in position space: 
\begin{equation}
	\left.\mathfrak{h}^{\alpha \beta}_{(2), \text{grav.}}\right|_{\scri^+_+} \supset -\dfrac{8}{r u}\left(\sum_{i \in \text{in}}^{m} m_i (q \cdot{v}_{i})\right)\left( \sum_{j \in \text{out}}^{n} m_j\dfrac{ {v}_j^{\alpha} {v}_j^{\beta}}{ (q\cdot v_j)}- \sum_{i \in \text{in}}^{m} m_i\dfrac{ {v}_{i}^{\alpha} {v}_{i}^{\beta}}{ (q\cdot v_i)}\right).
\end{equation}
When $\vec{n}= \vec{n}'$, the combination $(\mathcal{A}^{\alpha \beta}(\vec n')-\mathcal{A}^{\alpha \beta}(\vec n))$ vanishes. Thus, after performing the large $r$ limit, the collinear contribution to the integral is undetermined. In the neighbourhood of $\vec{n}'=\vec{n}$, that is $\theta=0$, we find
\begin{equation}
	1- \vec{n} \cdot \vec{n}'=1-\cos \theta =  \dfrac{\theta^2}{2} + O(\theta^4), \quad \sin \theta = \theta + O(\theta^3).
\end{equation}
On the sphere, two nearby points $n$ and $n'$ are joined by an arc of a great circle of length $\theta$. Then, if $e^\lambda_A= \lambda_i e_A^i$ is the tangent vector along this geodesic, we have 
\begin{equation}
	\mathcal{A}^{\alpha \beta}(\vec n')-\mathcal{A}^{\alpha \beta}(\vec n) = \theta e^\lambda_A D^A \mathcal{A}^{\alpha \beta}(\vec n) + O(\theta^2).
\end{equation}
Furthermore, the measure is $d\Omega'= -d(\cos \theta) d\phi= \sin \theta d\theta d\phi$. Then, all powers of $\theta$ cancel and the integrand is regular and finite in the neighbourhood of $\theta=0$. Then, the angular integral is finite and its contribution to the 2PM waveform vanishes due to the compact support of $\dot{\mathcal{F}}(u')$.\\

\noindent We now argue that the first term of the source \eqref{Eq:SourceFormula} is the only one contributing to a radiative $(ur)^{-1}$ tail leading to logarithmic contributions to the Weyl scalars discussed in the matter section. All other terms involve one derivative on each $\mathfrak{h}_{(1)}$, with various contractions. Furthermore, $q$ is null, and momentum conservation \eqref{Eq:MomentumConservation} implies that 
\begin{equation}
	q_\mu \partial_\nu\mathfrak{h}_{(1)}^{\mu \alpha}= O(r^{-2}).
\end{equation}
Then, the only terms that are not made automatically subleading are
\begin{equation}
	N^{\alpha \beta}(x) \supset  \frac{1}{2}
	\partial^\alpha \mathfrak h^{(1)}_{\mu\nu}
	\partial^\beta \mathfrak h^{\mu\nu}_{(1)}
	- \frac{1}{4}
	\partial^\alpha \mathfrak h_{(1)}
	\partial^\beta \mathfrak h_{(1)},
\end{equation}
for which the leading order in $r$ will carry $q^\alpha q^\beta$. Schematically, both of these terms will be of the form 
\begin{equation}
	\partial^\alpha \mathfrak h_{(1)} \partial^\beta \mathfrak h_{(1)} \sim \dfrac{1}{r^2 L^2} q^\alpha q^\beta \left(\dot{\mathcal{F}}(u/L)\right)^2 \mathcal{B}(\vec n).
\end{equation}
Performing the same procedure as before, one finds a non-universal contribution to the $1/(ru)$ tail. However, such terms cannot contribute to the Bondi shear as they vanish when projected on the angular directions and thus have no impact on the radiative Weyl scalars $\Psi_0, \Psi_1, \Psi_3$ and $\Psi_4$. \\

\noindent Let us now discuss the subleading order $O(r^{-3})$ of the gravitational source $N^{\alpha \beta}$ \eqref{Nab}. Consider a generic source of that order, such that the resulting gravitational perturbation takes the form
\begin{equation}
	\mathfrak{h}^{\alpha \beta}_{(2), \text{grav.}}(x) = \ \int du' dr' d\Omega'\, (r')^2 \Theta(t-t') \delta\!\left((x-x')^2\right) \dfrac{K^{\alpha \beta}(u', \vec n')}{(r')^3}.
\end{equation}
Since the radial power in the denominator does not cancel exactly with the Jacobian, Eqs. \eqref{Eq:GravDelta}-\eqref{Eq:GravRadiusLightcone} imply 
\begin{equation}
	\mathfrak{h}^{\alpha \beta}_{(2), \text{grav.}}(x) = \ \int du' d\Omega'\,  \Theta(u-u')  \dfrac{K^{\alpha \beta}(u', \vec n')}{(u-u')(2r + u-u')}.
\end{equation}
 This can now be expanded for large $r$ and large $u$ to find a $(ur)^{-1}$ tail. However, the result of this integral will not depend on the angle of observation $\vec n$, and the resulting tail will be annihilated by any derivative on the sphere. In what follows, we show that the radiative tails propagate as logarithms in the Weyl scalars through divergences on the sphere. Thus, the radiative tail originating from the subleading $O(r^{-3})$ source cannot offer any such contribution and we will also ignore it in the rest of our discussion.
 
\paragraph{Logarithmic tails and gravitational Weyl perturbations.} 
We now wish to determine whether the radiative tail we have just computed can offer a $r^{-4}\log u$ contribution to the Weyl scalar $\Psi_1$. Since Cartesian derivatives either decrease the power of $u$ or $r$, the only terms in the waveform that could produce such a contribution are proportional to an arbitrary function of the angles multiplied by 
\begin{equation}
	\dfrac{\log u}{r^2}, \quad  \dfrac{u \log u}{r^3} , \quad  \dfrac{u^2 \log u}{r^4}.
\end{equation}
Knowing the behaviour of the source term locally at $\scri^+_+$, we can derive these terms straight from the $(ru)^{-1}$ tail. Consider the ansatz 
\begin{equation}
	\left.\mathfrak{h}_{(2), \text{grav.}}^{\alpha \beta}\right|_{\scri^+_+} = \dfrac{A^{\alpha \beta}(u,\vec n)}{r} + \dfrac{B^{\alpha \beta}(u,\vec n)}{r^2} + \dfrac{C^{\alpha \beta}(u,\vec n)}{r^3} +\dfrac{D^{\alpha \beta}(u,\vec n)}{r^4} + O\left( r^{-5} \right).
\end{equation}
This ansatz ignores the potential existence of $\log r$ branches such as $\log r/r^2$ as they will not interfere with our computation of the $\log u/r^4$ contribution to the Weyl scalars. However, in the next paragraph, we show that such a branch does not appear at spatial infinity.\\

\noindent Since the derivatives of the transition function have no support at $\scri^+_+$, the source is given by the purely outgoing contribution
\begin{equation}
	N^{\alpha \beta}|_{\scri^+_+} = \sum_{k=0}^\infty \sum_{p=0}^k \dfrac{u^{k-p}}{r^{k+4}} N^{+,\alpha\beta}_{k,p,+} (\vec n).
\end{equation}
Using 
\begin{equation}
 \Box\left(\frac{F(u,\vec n)}{r^k}\right)
 =\frac{2(k-1)\partial_uF}{r^{k+1}}
 +\frac{[D^A D_A +k(k-1)]F}{r^{k+2}},
 \label{eq:wavehierarchy}
\end{equation}
for $k$ integer, the wave equation splits into a radial hierarchy:
\begin{align}
	&\partial_u B^{\alpha \beta} = -\dfrac12 D^A D_A A^{\alpha \beta}, \label{TailHierarchy1}\\
	&\partial_u C^{\alpha \beta} = \dfrac14\left( N^{+,\alpha\beta}_{0,0, +} -(D^A D_A+2) B^{\alpha \beta}\right),\label{TailHierarchy2}\\&\partial_u D^{\alpha \beta} = \dfrac{1}{6}\left(u N^{+,\alpha\beta}_{1,0, +} + N^{+,\alpha\beta}_{1,1,+} -(D^A D_A+6) C^{\alpha \beta} \right)\label{TailHierarchy3}.
\end{align}
In particular, we find the following homogeneous tower:
\begin{equation}
\begin{aligned}
			&\left. \mathfrak{h}_{(2), \text{grav.}}^{\alpha \beta}\right|_{\scri^+_+}\\
            &= -\dfrac{8\mathcal{A}^{\alpha \beta}(\vec n)}{ru} + \dfrac{4\log u}{r^2}D^A D_A \mathcal{A}^{\alpha \beta}(\vec n)   - \dfrac{u (\log u-1)}{r^3} (D^B D_B +2)D^AD_A \mathcal{A}^{\alpha \beta}(\vec n)\\
			&\hspace{50 pt}+ \dfrac{1}{12}\dfrac{u^2 \left(\log u- \frac32\right)}{r^4} (D^C D_C + 6)(D^B D_B +2) D^AD_A \mathcal{A}^{\alpha \beta}(\vec n) + O(r^{-5}),
	\end{aligned}
\end{equation}
since the Coulombic source does not contribute to logarithms after integration. From this expression, it is then possible to compute the $r^{-4}\log u$ contribution of the graviton drag to ${\Psi_1}|_{\scri^+_+}$, which, as we will see, is needed to prove the logarithmic soft graviton theorem. First, let us mention two following identities: 
\begin{equation}
    D^A D_A \left(\dfrac{1}{(q \cdot v)^k} \right) = -\dfrac{k(k-1)}{(q \cdot v)^k} - \dfrac{2 k^2 v^0}{(q \cdot v)^{k+1}} - \dfrac{k(k+1)}{(q \cdot v)^{k+2}}\,, 
\end{equation}
which in turn implies
\begin{equation}
\begin{aligned}
    D^A D_A \left(\dfrac{(q \cdot P)}{(q \cdot v)^k} \right) &= -\dfrac{2k (P\cdot v)}{(q \cdot v)^{k+1}} + \dfrac{2(k-1)P^0}{(q \cdot v)^{k}} - \dfrac{2 k(k-1)(P \cdot q)v^0}{(q \cdot v)^{k+1}} \\
    & - \dfrac{(k-1)(k-2)(P\cdot q)}{(q \cdot v)^k} - \dfrac{k(k+1)(P \cdot q)}{(q \cdot v)^{k+2}}\,,
\end{aligned}    
\end{equation}
for any 4-vectors $P$ and $v$ such that $v^2=-1$. With the conventions of Section \ref{sec:conventions}, we can also write
\begin{equation}
	\mathcal{A}^{\alpha \beta}(\vec n)= \sum_{a \in \pm} \eta_a m_a{v}_a^{\alpha} {v}_a^{\beta}\dfrac{ (P \cdot q)}{ (q\cdot v_a)}= \sum_{a \in \pm} {p}_a^{\alpha} {p}_a^{\beta}\dfrac{ (P \cdot q)}{ (q\cdot p_a)}\,.
\end{equation}
Then, in terms of the future-directed velocities, we find 
\begin{align}\hspace{-18 pt}\label{Eq:GothicTailTower}
			\mathfrak{h}_{(2), \text{grav.}}^{\alpha \beta}|_{\scri^+_+} &= -8\sum_{a \in \pm} \eta_a m_a{v}_a^{\alpha} {v}_a^{\beta}\left\{\dfrac{1}{ru}\dfrac{ (P \cdot q)}{ (q\cdot v_a)} +  \dfrac{\log u}{r^2} \left(\dfrac{(P \cdot v_a)}{(q \cdot v_a)^2} + \dfrac{(P \cdot q)}{(q \cdot v_a)^3}\right) \right.\nn\\
			+&\left. \dfrac{u \log u}{r^3} \left(3 \dfrac{(P \cdot q)}{(q \cdot v_a)^5} +3 \dfrac{(P \cdot v_a)}{(q \cdot v_a)^4} -  \dfrac{P^0}{(q \cdot v_a)^3} + 3 \dfrac{v_a^0(P \cdot q)}{(q \cdot v_a)^4}+ 2 \dfrac{v_a^0(P \cdot v_a)}{(q \cdot v_a)^3}\right)    \right.\\
           & \left. + \text{(non-log)}\right\}+O(r^{-4})\nn.
	\end{align} 
In the last expression we kept the $u^2 \log u/r^4$ term implicit as we explicitly show that its contribution to the leading logarithmic term vanishes in Appendix~\ref{Appendix:GravitationalTail}. In that same Appendix, we also compute the following drag contribution to $\Psi_1$ at $\scri^+_+$:
\begin{equation}
\begin{aligned}
     \left. \Psi_1^{(2),\text{grav.}}\right|_{\scri^+_+}&= G^2\, \sum_{a \in \pm}\dfrac{3 \eta_a m_a}{(q \cdot v_a)^4}\dfrac{\log u}{r^4}\theta^A (\partial_A q^\mu)( P_\mu v_{a\nu}-v_{a \mu} P_\nu ) q^\nu + O\!\left(r^{-4}\right)\\
     &= G^2\sum_{a \in \pm}\dfrac{3  m_a^4}{(q \cdot p_a)^4}\dfrac{\log u}{r^4}\theta^A (\partial_A q^\mu)( P_\mu p^a_{\nu}-p^a_{\mu} P_\nu ) q^\nu + O\!\left(r^{-4}\right).
\end{aligned}
    \label{Eq:Psi1outsoft2}
\end{equation}

\paragraph{Logarithmic branches at spatial infinity}
Finally, we prove that the source \eqref{Eq:SourceLWPluggedIn} does not produce a logarithmic branch in $r$ at spatial infinity. Consider the retarded integral sourced by Eq. \eqref{Eq:SourceLWPluggedIn} in which we first eliminate the $u'$ integral: 
\begin{equation}
    \mathfrak{h}^{\alpha \beta}_{(2),\text{grav.}}= -\dfrac{1}{4 \pi} \int_0^\infty dr' (r')^2 \int d \Omega' \; \dfrac{N^{\alpha \beta}\left(t- \abs{r \vec{n} - r' \vec{n}'}, r' \vec{n}'\right)}{\abs{r \vec{n} - r' \vec{n}'}}.
\end{equation}
The terms composing the source $N^{\alpha \beta}$ feature either, schematically, $\rho^{-4}$ or $Z^2 \rho^{-6}$. Neglecting the subleading offsets $y^\mu_a$, this implies that the source term can be written as 
\begin{equation}
    N^{\alpha \beta}(t', r' \vec{n}')= \dfrac{1}{r'^4} N^{\alpha \beta}\left(\dfrac{t'}{r'},\vec{n}'\right) \label{Eq:SourceScaling}.
\end{equation}
Let us split the radial integration in three separate regimes: 
\begin{equation}
    \int_0^\infty dr' =  \int_0^b dr' +  \int_b^{\sqrt{r\lambda}} dr' + \int_{\sqrt{r \lambda}}^\infty dr'.
\end{equation}
The lower cutoff $b$ is taken such that $b \gg \abs{u}$ and $b \sim r^0$ and such that the region $[0,b]$ is large enough to contain the bounded region in which particle singularities can occur. The intermediate cutoff $\sqrt{r \lambda}$, $\lambda >0$, is chosen such that for $r' \in [b,\sqrt{r \lambda}]$,  $r'\gg \abs{u}$ for fixed $u$ and $r'/r \to 0$ for $r$ large. As the bounds of the first integral $\int^b_0 dr'$ do not scale with $r$ and the integrand does not contain $\log r$, this part will not generate $\log r$ pieces. Let us now inspect the intermediate regime. Since $r' \ll r$, we have
\begin{equation}
    \begin{aligned}
        &\abs{r \vec{n} - r' \vec{n}'} = r - (\vec{n} \cdot \vec{n}') r' + \dfrac{1}{2} \dfrac{r'^2}{r}\left(1- (\vec{n} \cdot \vec{n}')^2\right) + O(r'^3/r^2), \\
     &\abs{r \vec{n} - r' \vec{n}'}^{-1}= \dfrac{1}{r} + \dfrac{(\vec{n} \cdot \vec{n}')r'}{r^2} + O(r'^2/r^3),\\
     &t'= u + (\vec{n}\cdot \vec{n}') r' -\dfrac{1}{2} \dfrac{r'^2}{r}\left(1- (\vec{n} \cdot \vec{n}')^2\right) + O(r'^3/r^2).
    \end{aligned}
\end{equation}
Through Taylor expanding the integrand around $\xi= u + r' (\vec{n} \cdot \vec{n}')$, we find
\begin{equation}\label{Eq:TaylorExpansion1}
    \begin{aligned}
         &I_{b, \lambda}\equiv \int_b^{\sqrt{r \lambda}} dr'\; r'^2 \,  \dfrac{N^{\alpha \beta}\left(t- \abs{r \vec{n} - r' \vec{n}'}, r' \vec{n}'\right)}{\abs{r \vec{n} - r' \vec{n}'}}=\int_b^{\sqrt{r \lambda}} dr' \; \dfrac{r'^2}{r}\left\{ N^{\alpha \beta}(\xi, r' \vec{n}')     \right.\\
           &\hspace{-5 pt} \left. +\dfrac{r'}{r} \left((\vec{n} \cdot \vec{n}')N^{\alpha \beta}(\xi, r' \vec{n}') +\dfrac{r'}{2}( (\vec{n}\cdot \vec{n}')^2-1) \partial_\xi N^{\alpha \beta}(\xi, r' \vec{n}') \right) +\dots  \right\}_{\xi= u+ r' (\vec{n}\cdot \vec{n}')} \! \! .
    \end{aligned}
\end{equation}
We are interested in the upper endpoint that scales with $r$. At large $r'$ and fixed $u$, we have 
\begin{equation}
    \dfrac{u + r'(\vec{n} \cdot \vec{n}')}{r'} = (\vec{n} \cdot \vec{n}')+ O(r'^{-1})
\end{equation}
such that, with Eq. \eqref{Eq:SourceScaling}, we have 
\begin{equation}
    \begin{aligned}
         I_{b,\lambda}&=  \int_{b}^{\sqrt{r \lambda}} dr'  \; \dfrac{1}{r r'^2}\left\{ N^{\alpha \beta}(\xi,\vec{n}')     \right.\\
           &\hspace{-5 pt} \left. + \dfrac{r'}{r} \left((\vec{n} \cdot \vec{n}')N^{\alpha \beta}(\xi, \vec{n}') +\dfrac{1}{2}( (\vec{n}\cdot \vec{n}')^2-1) \partial_\xi N^{\alpha \beta}(\xi, \vec{n}') \right)  +\dots \right\}_{\xi= (\vec{n}\cdot \vec{n}')} .
    \end{aligned}
\end{equation}
The first term converges and cannot contribute to a logarithm, whereas the second line constitutes a valid candidate for a $\log r/r^2$ branch. To confirm this, we now proceed to match the intermediate regime with the upper regime, in particular at its lowest endpoint, where $r'$ is comparable to $r$. To this effect, we introduce the dimensionless ratio
\begin{equation}
    \kappa \equiv \dfrac{r'}{r}.
\end{equation}
We have 
\begin{equation}
\begin{aligned}
    I_{\lambda, \infty}&\equiv \int_{\sqrt{r \lambda}}^{+ \infty} dr'\; r'^2 \,  \dfrac{N^{\alpha \beta}\left(t- \abs{r \vec{n} - r' \vec{n}'}, r' \vec{n}'\right)}{\abs{r \vec{n} - r' \vec{n}'}}\\
    &= \int_{\sqrt{\lambda/r}}^{+ \infty} d\kappa\; \dfrac{1}{r^2 \kappa^2} \,  \dfrac{N^{\alpha \beta}\left(\dfrac{1 + u/r - \abs{ \vec{n} - \kappa \vec{n}'}}{\kappa}, \vec{n}'\right)}{\abs{ \vec{n} - \kappa \vec{n}'}},
\end{aligned}
\end{equation}
where we used Eq. \eqref{Eq:SourceScaling}.  Around the lower endpoint of the integral, for $\sqrt{\lambda/r} \leq \kappa \leq \kappa_0$ with $\kappa_0$ fixed and small, we have 
\begin{equation}
    \begin{aligned}
    &\abs{ \vec{n} - \kappa \vec{n}'}= 1 - \kappa (\vec{n}\cdot \vec{n}') + \dfrac{\kappa^2}{2} (1- (\vec{n}\cdot \vec{n}')^2) + O(\kappa^3),\\
    & \dfrac{1 + u/r - \abs{ \vec{n} - \kappa \vec{n}'}}{\kappa }=  \vec{n} \cdot \vec{n}' + \dfrac{u}{r \kappa} - \dfrac{\kappa}{2} \left(1- (\vec{n}\cdot \vec{n}')^2 \right) + O(\kappa^2).
    \end{aligned}
\end{equation}
Focusing on the interval close to the lower endpoint of the integral, a Taylor expansion of the integrand yields 
\begin{equation}\label{Eq:TaylorExpansion2}
    \begin{aligned}
       I_{\lambda, \infty}^{(-)}&\equiv  \int_{\sqrt{\lambda/r}}^{\kappa_0} d \kappa\,  \dfrac{1}{r^2 \kappa^2} \left\{ \left(1 - \kappa (\vec{n}\cdot \vec{n}') + \dfrac{\kappa^2}{2}\left(1- (\vec{n}\cdot \vec{n}')^2 \right) + O(\kappa^3)\right)^{-1}   \right.\\
       &\hspace{35 pt}\left. \times\left( N^{\alpha \beta}(\xi, \vec{n}') + \left(\dfrac{u}{r\kappa} - \dfrac{\kappa}{2} (1-(\vec{n}\cdot n')^2) \right)\partial_\xi N^{\alpha \beta}(\xi, \vec{n}')  +...\right) \right\}_{\xi= (\vec{n}\cdot \vec{n}')}.
    \end{aligned}
\end{equation}
The $\log r/r^2$ branch of this integral is generated by the $\kappa^{-1}$ terms. Summed to the previous result obtained in the intermediate regime, we find 
\begin{equation}
  \hspace{-8 pt}  \left.\mathfrak{h}^{\alpha \beta}_{(2), \text{grav.}}\right|_{\scri^+_-, \frac{\log r}{r^2}} \propto \int d\Omega' \left\{ \xi N^{\alpha \beta}(\xi, \vec{n}') +\dfrac{1}{2}( \xi^2-1) \partial_\xi N^{\alpha \beta}(\xi, \vec{n}') \right\}_{\xi= (\vec{n}\cdot \vec{n}')}. \label{Eq:IntegrandLogR}
\end{equation}
Given Eq. \eqref{Eq:SourceLWPluggedIn} and given $\rho_a(x)$ and $Z^\mu_a(x)$ are respectively even and odd functions of $x$, we have 
\begin{equation}
    N^{\alpha \beta}(-(\vec{n}\cdot \vec{n}'), -\vec{n}')= N^{\alpha \beta}((\vec{n}\cdot \vec{n}'), \vec{n}'). 
\end{equation}
The whole integrand of Eq. \eqref{Eq:IntegrandLogR} is thus odd under $\vec{n}' \mapsto -\vec{n}'$ and the angular integral vanishes. The $\log r/r^2$ branch is not the only one potentially contributing to the $\log r/r^4$, as $u \log r/r^3$ or $u^2 \log r/r^4$ branches can also arise from subleading $u/r$ terms in the Taylor expansions \eqref{Eq:TaylorExpansion1} and \eqref{Eq:TaylorExpansion2}. However, the resulting angular integrals also vanish due to their integrand being odd under $\vec{n}' \mapsto -\vec{n}'$. We can thus conclude that
\begin{equation}\label{logrbranch}
    \Psi_1 \vert_{\scri^+_-, \frac{\log r}{r^4}}=0
\end{equation}
up to order $G^3$ contributions.

\subsection{From harmonic to radiative gauge}\label{sec2.5}
The proof of the logarithmic soft graviton theorem relies on the insertion of a flux operator at $\scri^+$ that is typically expressed in terms of the Bondi news tensor. However, our computations have, so far, all been done in the harmonic gauge, in which the retarded direction $u$ does not align with the physical light cone \cite{Blanchet:2013haa}. Ref. \cite{Blanchet:2020ngx} describes the change from the harmonic to the Newman-Unti gauge perturbatively in $G$. Focusing only on the logarithmic contributions at the leading order $O(G)$ and writing the retarded harmonic coordinates $(u,r,x^A)$ and the radiative ones $(U,R,X^A)$, we have 
\begin{equation}\label{mapcoord}
\begin{aligned}
    &U = u + 2 G (q \cdot P) \log\left(\dfrac{r}{b} \right)+ O(r^0),\\
    &R= r - 2 G n^i P_i\log\left(\dfrac{r}{b} \right) +O(r^0),\\
    &X^A = x^A - \dfrac{2G}{r} e^{A}_i P^i \log\left(\dfrac{r}{b} \right)+ O(r^{-1}),
\end{aligned}
\end{equation}
where $P^\mu=(M, P^i)$ is the ADM four-momentum and $b$ is an arbitrary scale. We can translate this to Cartesian coordinates: with $T= U+ R$, we find 
\begin{equation}
T = u + r - 2 G M \log\left(\dfrac{r}{b} \right) +\text{(non-log)} = t -2 GM \log\left(\dfrac{r}{b} \right)+O(r^0)\,.
\end{equation}
Likewise, with $X^i = R N^i[X^A]$, 
\begin{equation}
\begin{aligned}
    X^i &= \left(r - 2 G n^k P_k\log\left(\dfrac{r}{b} \right) \right)n^i\left[ x^A - \dfrac{2G}{r} e^A_j P^j \log\left(\dfrac{r}{b} \right)\right]+O(r^0)\\
    & =\left(r - 2 G n^k P_k\log\left(\dfrac{r}{b} \right) \right)\left(n^i -\dfrac{2G}{r} e_A^i e^A_j P^j \log\left(\dfrac{r}{b} \right) \right)+O(r^0)\\
    &= x^i - 2 G P^i \log\left(\dfrac{r}{b} \right) +O(r^0), 
\end{aligned}
\end{equation}
where we used $e_A^i e_j^A= \delta_{j}^i - n^i n_j$. This can be written covariantly as 
\begin{equation}
X^\mu= x^\mu - G \xi^\mu_{(1)} +O(G^2),
\end{equation}
with 
\begin{equation}
\xi_{(1)}^\mu = 2 P^\mu \log\left(\dfrac{r}{b} \right) + O(r^0).
\end{equation}
Under this transformation, the covariant metric perturbations transform as \cite{Sonego_1998,Bonetto:2021exn}
\begin{equation}
\begin{aligned}
    &h_{\mu \nu}^{(1), \text{rad.}}= h_{\mu \nu}^{(1), \text{harm.}} +\mathcal{L}_{\xi_{(1)}}\eta_\mn, \\
    &h_{\mu \nu}^{(2), \text{rad.}}= h_{\mu \nu}^{(2), \text{harm.}} + \mathcal{L}_{\xi_{(1)}}h_{\mu \nu}^{(1), \text{harm.}}  +\frac{1}{2}\cl_{\xi_{(1)}}^2\eta_\mn+\cl_{\xi_{(2)}}\eta_\mn. 
\end{aligned}
\end{equation}
The linearized Weyl tensor is invariant under linearized diffeomorphisms. We can therefore ignore the contributions of the form $\mathcal L_{\xi_{(1)}} \eta_{\mu\nu}$. Regarding the term $\mathcal L_{\xi_{(1)}}^2 \eta_{\mu\nu}$, there is a pure gauge term \(\mathcal L_\chi\eta_\mn\) with \(\chi^\nu=\xi_{(1)}^\rho\partial_\rho\xi_{(1)}^\n\). Only this term contains the relevant order $\log r/r^2$, but it will not contribute to the linearized Weyl tensor. Thus the quadratic Lie derivative will not be considered further. 
However, the variation of the first-order perturbation will generate a variation in the $O(G^2)$ Weyl tensor that we now set out to compute. In harmonic coordinates, in the far past/future, the first order metric perturbation is
\begin{equation}
h_{\mu \nu}^{(1), \text{harm.}}=\sum_{a \in \text{in/out}} \dfrac{4 m_a \left(v_{a\mu} v_{a\nu} + \frac12 \eta_{\mu \nu}\right)}{\rho_a}.
\end{equation} 
The variation of this perturbation is thus
\begin{equation}
\mathcal{L}_{\xi_{(1)}} h_{\mu \nu}^{(1), \text{harm.}} = \xi_{(1)}^\lambda \partial_\lambda h_{\mu \nu}^{(1), \text{harm.}} + (\text{non-log}) \equiv -\sum_{a \in \text{in/out}} A_{a \mu \nu}(r) P^\lambda \dfrac{Z_{a\lambda}}{\rho_a^3} + (\text{non-log})
\end{equation}
with 
\begin{equation}
A_{a \mu \nu}(r)=8 m_a \left(v_{a\mu} v_{a\nu} + \frac12 \eta_{\mu \nu}\right) \log\left(\dfrac{r}{b} \right).
\end{equation}
In outgoing radiative gauge, the linearised vacuum Weyl tensor then picks up an additional logarithmic contribution generated by this variation, which we now compute. Since no derivatives are allowed to hit $A_{a\mu \nu}(r)$ for our purpose, using Eq. \eqref{Eq:SourceObjectsDefinitions}, the only relevant derivatives are 
\begin{equation}
\partial_{\alpha } \partial_\beta \left(\dfrac{Z_{a\lambda}}{\rho_a^3}\right) = \dfrac{15}{\rho_a^7} Z_{a\lambda} Z_{a\alpha} Z_{a\beta} - \dfrac{3}{\rho_a^5}\left(\Pi_{a\lambda \beta} Z_{a \alpha} + \Pi_{a\alpha \beta} Z_{a\lambda} + \Pi_{a\lambda \alpha}Z_{a\beta}\right),
\end{equation}
such that the logarithmic part of the variation of the Weyl tensor is 
\begin{align}
&C^{(2), \text{lin.}}_{\mu \nu \gamma \sigma}\left[\mathcal{L}_{\xi_{(1)}}h_{(1)}\right]\\
    &= -\frac{1}{2}\sum_{a\in\text{in/out}} P^\lambda \left[ A_{a\mu \sigma} \left(\dfrac{15}{\rho_a^7} Z_{a\lambda} Z_{a\gamma} Z_{a\nu} - \dfrac{3}{\rho_a^5} \left(\Pi_{a \lambda \gamma }Z_{a\nu} +\Pi_{a\lambda \nu }Z_{a\gamma} +\Pi_{a\gamma \nu }Z_{a\lambda}  \right) \right)\right.\nonumber\\
    &\hspace{15 pt}+ (\gamma \nu \mu \sigma \to \sigma \mu \nu \gamma) -(\gamma \nu \mu \sigma \to \gamma \mu \nu \sigma) - (\gamma \nu \mu \sigma \to \sigma \nu \mu \gamma)\bigg].
\end{align}
Contracted with the appropriate fixed background frame elements, this is such that
\begin{equation}
\begin{split}
&\left. \Psi_1^{(2)}\left[\mathcal{L}_{\xi_{(1)}}h_{(1)}\right]\right|_{\scri^+_\mp}\\
&= -G^2\sum_{a \in \text{in/out}}\dfrac{3 m_a}{(q \cdot v_a)^4} \dfrac{\log r}{r^4} \theta^A e_A^\mu \left(v_{a\mu} P_\nu - v_{a\nu} P_\mu \right) q^\nu +O(r^{-4}),\label{eq2.101}
\end{split}
\end{equation}
which picks out (part of) the $D^B D_{AB}$ term of Eq.~(2.13b) of \cite{Geiller:2024ryw}. Thus, we have shown that going from the harmonic to the radiative gauge produces a $\log r/ r^4$ contribution in the Weyl scalar $\Psi_1$. This is the established leading order violation of peeling due to mass-quadrupolar interactions at order $G^2$  \cite{Damour:1985cm}.

\section{New proofs of classical soft graviton theorems}
\label{Section:SoftTheorems}

As a direct application of the antipodal matching relations obtained in Section~\ref{Section:AntipodalMatchings}, we revisit the derivation of the leading soft graviton theorem and the logarithmic soft graviton theorem in their classical form as given in \cite{Sahoo:2018lxl,Saha:2019tub}. This complements a recent asymptotic analysis performed in the metric formalism \cite{Boschetti:2026gfd}, as well as earlier derivations \cite{Agrawal:2023zea,Choi:2024ajz} where such antipodal relations were postulated rather than derived. For completeness, we start by revisiting Weinberg's soft graviton theorem \cite{Weinberg:1965nx} using the formalism developed in this work. This provides a version of the derivation found in \cite{Strominger:2013jfa,He:2014laa,Campiglia:2015kxa}, based here on Weyl scalar variables rather than metric variables, which is therefore naturally invariant under linearized diffeomorphisms.\\

\noindent In the following, we will restore the standard NP notations
\begin{equation}
\Psi^0_1\equiv \overset{2}{\Psi}_1\,, \qquad \Psi^0_2\equiv \overset{2}{\Psi}_2\,, \qquad  \Psi^0_3\equiv \overset{1}{\Psi}_3\,,
\end{equation}
and indicate the corner considered with the notation $\vert_{}$. The leading order NP evolution equations are given by 
\begin{align}
\label{evolution psi1}
\partial_u \Psi^0_1&=\frac{1}{2}\eth \Psi^0_2-\sigma_2 \Psi^0_3=\frac{1}{2}\theta^A \partial_A \Psi^0_2- \sigma_2 \Psi^0_3\,,\\
\label{evolution psi2}
\partial_u \Psi^0_2&=\frac{1}{2}\eth \Psi^0_3-\frac{1}{2} \sigma_2 \Psi^0_4=\frac{1}{2}(\theta^A \partial_A+\cot \theta) \Psi^0_3- \frac{1}{2}\sigma_2 \Psi^0_4\,,
\end{align}
where $\sigma_2 = -\frac{1}{4}C_{AB}\theta^A \theta^B$, in agreement with Eqs.~(2.16)-(2.17) and (A.21) of \cite{Geiller:2024ryw} (modulo normalization conventions).

\subsection{Leading soft graviton theorem}
The starting point of the derivation of the leading soft graviton theorem is a particular case of the antipodal matching relation \eqref{antipodal matching Psi2}, namely
\begin{equation}
\label{leading: starting point}
\overset{2,2}{\Psi}_{\hspace{-2 pt}2}\big|_{\scri^+_-} =\Upsilon^*\overset{2,2}{\Psi}_{\hspace{-2 pt}2}\big|_{\scri^-_+} \,.
\end{equation}
While Eq.~\eqref{leading: starting point} was derived at tree level, this antipodal relationship has been proven to hold non-perturbatively in General Relativity \cite{Ashtekar:1978zz,PhysRevLett.43.649,1979JMP....20.1362A,Herberthson:1992gcz,Friedrich:1998xmu,Friedrich:1999ax,Strominger:2013jfa,Troessaert:2017jcm,Prabhu:2021cgk,Capone:2022gme,Compere:2023qoa}. We therefore assume the validity of Eq.~\eqref{leading: starting point} at the non-perturbative level for a complete treatment.

We directly rewrite the left-hand side of \eqref{leading: starting point} as
\begin{equation}
\overset{2,2}{\Psi}_{\hspace{-2 pt}2} \big|_{\scri^+_-}=\lim_{u \to -\infty}\Psi^0_2=-\int_{-\infty}^\infty du\, \partial_u \Psi^0_2 +  \overset{2,2}{\Psi}_{\hspace{-2 pt}2}\big|_{\scri^+_+}\,,
\end{equation}
and we use the evolution equation \eqref{evolution psi2} such that 
\begin{equation}
\label{leading: step 2}
\overset{2,2}{\Psi}_{\hspace{-2 pt}2} \big|_{\scri^+_-}= -\frac{1}{2}\int_{-\infty}^\infty du\, (\eth \Psi^0_3-\sigma_2 \Psi^0_4) + \overset{2,2}{\Psi}_{\hspace{-2 pt}2}\big|_{\scri^+_+}\,.
\end{equation}
At this point, we make use of the relation between $\Psi^0_3, \Psi^0_4$ and the Bondi quantities using Eqs.~(2.13) of \cite{Geiller:2024ryw},
\begin{align}
 \Psi^0_3 &=\frac{1}{2}\bar\theta^A D^B N_{AB}\,,   \\
 \Psi^0_4 &=\frac{1}{2} \bar\theta^A \bar \theta^B \partial_u N_{AB}=- 2 \partial_u^2 \bar \sigma_2\,, 
\end{align}
where $N_{AB}=\partial_u C_{AB}$ is the news tensor. Thus, the first term on the right-hand side of Eq. \eqref{leading: step 2} may be written as
\begin{equation}
\label{eq 4.8}
\begin{split}
\int_{-\infty}^\infty du\, \eth \Psi^0_3&=\frac{1}{2}\int_{-\infty}^\infty du\, \eth(\bar \theta^A D^B N_{AB})\\
&=\frac{1}{2} \int_{-\infty}^\infty du\, (D^A D^B N_{AB}+ i \epsilon^{AB} D_A D^C N_{CB})\,,
\end{split}
\end{equation}
where we used the formula \eqref{eth identity}.
The term proportional to $\epsilon^{AB}$ may be shown to vanish on account of the falloff conditions of the shear tensor. Indeed, the latter is traceless and therefore admits the decomposition
\begin{equation}
\label{shear electric magnetic decomposition}
C_{AB}=(-2 D_A D_B +\gamma_{AB} D^2 )C+\epsilon_{C (A} D_{B)} D^C \Psi\,,
\end{equation}
where $C$ and $\Psi$ are called electric and magnetic potentials, respectively, due to their parity under the antipodal map $\Upsilon^*$. The term $\int du\, \epsilon^{AB} D_A D^C N_{CB}$ projects out the electric potential $C$, and is therefore only sensitive to the total difference $\Psi|_{\scri^+_+}-\Psi|_{\scri^+_-}$ in the magnetic potential, which is known to vanish in the context of particle scattering.  Introducing the `leading soft graviton operator' in the notation (6.6) of \cite{Choi:2024ajz},
\begin{equation}
\mathcal{J}_{AB}^{(-1)}\equiv \int_{-\infty}^\infty du\, N_{AB}\,,
\end{equation}
we can thus write
\begin{equation}
\int_{-\infty}^\infty du\, \eth \Psi^0_3=\frac{1}{2}D^A D^B \mathcal{J}_{AB}^{(-1)}\,.
\end{equation}
The second term on the right-hand side of \eqref{leading: step 2} may also be rewritten
\begin{equation}
\int_{-\infty}^\infty du\, \sigma_2 \Psi^0_4=-2\int_{-\infty}^\infty du\, \sigma_2\, \partial_u^2 \bar \sigma_2=-2\left[\sigma_2\, \partial_u \bar \sigma_2\right]^{+\infty}_{-\infty}+2\int_{-\infty}^\infty du\, \partial_u \sigma_2\, \partial_u \bar \sigma_2\,.
\end{equation}
The first term vanishes under the hypothesis that the news decays as $\partial_u \bar \sigma_2 \to 0$ as $u \to \pm \infty$.  In order to rewrite the second term we introduce \(N=N_{AB}\theta^A\theta^B\) and its complex conjugate so that we can decompose the news as
\begin{align}
N_{AB}=\frac{1}{4}\left(N\bar{\theta}_A\bar{\theta}_B+\bar N\theta_A\theta_B\right)
\end{align}
in the basis of STF tensors spanned by $\theta_A\theta_B$ and its conjugate. Substituting $\partial_u\sigma_2=-N/4$, the second term is then recognized to be the integral of the energy flux density of the graviton field, 
\begin{equation}
T_{uu}^{(\text{grav})} \equiv  \frac{N \bar N}{16} = \frac{1}{8} N_{AB} N^{AB}\,,
\end{equation}
in the conventions of \cite{Compere:2018aar}.
Putting everything together, equation \eqref{leading: step 2} becomes 
\begin{equation}
\overset{2,2}{\Psi}_{\hspace{-2 pt}2} \big|_{\scri^+_-}=-\frac{1}{4}D^A D^B \mathcal{J}_{AB}^{(-1)}+\int_{-\infty}^\infty du\, T_{uu}^{(\text{grav})} + \overset{2,2}{\Psi}_{\hspace{-2 pt}2}\big|_{\scri^+_+}\,,
\end{equation}
and the antipodal matching relation \eqref{leading: starting point} can thus be brought to the form
\begin{equation}
\label{leading: step 3}
\Upsilon^*\overset{2,2}{\Psi}_{\hspace{-2 pt}2}(\vec n)\big|_{\scri^-_+}=-\frac{1}{4}D^A D^B \mathcal{J}_{AB}^{(-1)}(\vec n)+\int_{-\infty}^\infty du\, T_{uu}^{(\text{grav})}(u,\vec n) + \overset{2,2}{\Psi}_{\hspace{-2 pt}2}(\vec n)\big|_{\scri^+_+}\,.
\end{equation}
We now show that \eqref{leading: step 3} is directly related to a classical and gauge-invariant version of Weinberg's soft theorem, restricting to the case where all finite-energy scattering states are massive particles. In that restricted framework, the graviton energy flux density $T_{uu}^{(\text{grav})}$ is set to zero such that Eq. \eqref{leading: step 3} can be rewritten
\begin{equation}
\frac{1}{4} D^A D^B \mathcal{J}_{AB}^{(-1)}(\vec n)=\overset{2,2}{\Psi}_{\hspace{-2 pt}2}(\vec n)\big|_{\scri^+_+}-\Upsilon^*\overset{2,2}{\Psi}_{\hspace{-2 pt}2}(\vec n)\big|_{\scri^-_+}\,.
\end{equation}
We can use the results of Section~\ref{subsection Lienard Wiechert fields} to directly evaluate the right-hand side of this equation for the scattering of massive particles, where we found
\begin{equation}
\begin{split}
\overset{2,2}{\Psi}_{\hspace{-2 pt}2}(\vec n)\big|_{\scri^+_+}&= -G\sum_{a \in \text{out}}  \frac{m_a}{(q \cdot v_a)^3}\,,\\
\overset{2,2}{\Psi}_{\hspace{-2 pt}2}(\vec n)\big|_{\scri^-_+}&= - G\sum_{a \in \text{in}}  \frac{m_a}{(\tilde{q} \cdot v_a)^3}\,,
\end{split}
\end{equation}
where we recall $q^\mu=(1,\vec n)$ and $\tilde{q}^\mu=(1,-\vec n)$.
The antipodal map $\Upsilon^*$ effectively implements $\tilde{q}^\mu \mapsto q^\mu$ such that 
\begin{equation}
\boxed{D^A D^B \mathcal{J}_{AB}^{(-1)}(\vec n)=-4G \sum_{a \in \text{out}}  \frac{m_a}{(q \cdot v_a)^3}+4G\sum_{a \in \text{in}}  \frac{m_a}{(q \cdot v_a)^3}=-4G\sum_{a \in \pm}  \frac{m_a^4}{(q \cdot p_a)^3}\,.}
\end{equation}
In Appendix~\ref{app: soft factor identities}, we show that the right-hand side may be replaced by\footnote{A version of formula \eqref{soft identity 1} adapted to complex stereographic coordinates was derived in \cite{Campiglia:2015kxa,Campiglia:2015lxa,Boschetti:2026gfd}. In flat complex stereographic coordinates $z,\bar z$, the null vector $q$ takes the form
\begin{equation*}
q^\mu=(1+ z\bar z, z+\bar z, -i(z-\bar z), 1- z \bar z)\,, 
\end{equation*} 
the frame field
\begin{equation*}
e^\mu_z=\partial_z q^\mu=(\bar z,1,-i,-\bar z)\,, \qquad e^\mu_{\bar z}=\partial_{\bar z} q^\mu=(z,1,i,-z)\,,
\end{equation*}
and the components of the polarization tensor
\begin{equation*}
\varepsilon^{\mu\nu}_{zz}=e^\mu_z e^\nu_z\,, \qquad \varepsilon^{\mu\nu}_{\bar z \bar z}=e^\mu_{\bar z} e^\nu_{\bar z}\,, \qquad \varepsilon^{\mu\nu}_{z\bar z}=0\,.
\end{equation*}}
\begin{equation}
\label{soft identity 1}
\sum_{a \in \pm} \frac{m_a^4}{(q \cdot p_a)^3}=\sum_{a \in \pm}  D^A D^B \left(\frac{\varepsilon_{AB}^{\mu\nu}(q)\, p^a_\mu\, p^a_\nu}{q \cdot p_a}\right)\,,
\end{equation}
where we introduce the polarization tensor
\begin{equation}
\label{polarization}
\varepsilon_{AB}^{\mu\nu}(q)\equiv e^\m_{\langle A}e^\n_{B\rangle}=e^{(\mu}_A e^{\nu)}_B-\frac{1}{2}\gamma_{AB} \gamma^{CD} e^{\mu}_C e^{\nu}_D\,,
\end{equation}
symmetrical in both sets of indices and satisfying
\begin{equation}
q_\mu\, \varepsilon_{AB}^{\mu\nu}(q)=0\,, \qquad \gamma^{AB}\varepsilon_{AB}^{\mu\nu}(q)=0\,.
\end{equation} 
In summary, we have obtained 
\begin{equation}
\label{GI leading soft theorem}
\boxed{
D^A D^B \mathcal{J}_{AB}^{(-1)}(\vec n)=-4G \sum_{a \in \pm}  D^A D^B \left(\frac{\varepsilon_{AB}^{\mu\nu}(q)\, p^a_\mu\, p^a_\nu}{q \cdot p_a}\right)\,.}
\end{equation}
Weinberg's soft factor \cite{Weinberg:1965nx} may be written in the form
\begin{equation}
\label{leading soft theorem}
\mathcal{J}_{AB}^{(-1)}(\vec n)=-4G \sum_{a \in \pm} \frac{\varepsilon_{AB}^{\mu\nu}(q)\, p^a_\mu\, p^a_\nu}{q \cdot p_a}\,,
\end{equation}
where $q^\mu(\vec n)$ is interpreted as the momentum direction of a soft graviton and $\varepsilon^{\mu\nu}_{AB}(q)$ as the corresponding polarization tensor. Our Eq. \eqref{GI leading soft theorem} is therefore a direct consequence of the leading soft graviton theorem. The operator $D^AD^B$ admits as kernel the magnetic STF tensors. Now, the right-hand side of Eq. \eq{leading soft theorem} is purely electric
\begin{align}
\frac{\ve^\mn_{AB}p_\m p_\n}{q\cdot p}=D_{\langle A}D_{B\rangle}[(q\cdot p)\log|q\cdot p|],
\end{align}
consistently with our working assumption that there is no magnetic memory. Under that assumption, the equations \eqref{GI leading soft theorem} and  Eq. \eqref{leading soft theorem} are therefore equivalent.

\subsection{Logarithmic soft graviton theorem}
The starting point for this new derivation is a particular instance of the antipodal matching \eqref{Eq:PsiAMatchingLog}, namely 
\begin{equation}
\label{log: starting point}
\left.     (\hspace{-6 pt}\overset{2, \text{log.}}\Psi_{\hspace{-5 pt}1})_A\right|_{\mathscr{I}^+_-} = \Upsilon^* \left.(\hspace{-5 pt}\overset{3, \text{log.}}\Psi_{\hspace{-5 pt}3})_A\right|_{\mathscr{I}^-_+}\,.
\end{equation}
We directly rewrite the left-hand side of this equation as
\begin{equation}
\left. \hspace{-6 pt}\overset{2, \text{log.}}\Psi_{\hspace{-5 pt}1}\right|_{\mathscr{I}^+_-}=-\lim_{u \to -\infty} u^2\partial_u^2 \Psi^0_1=\int_{-\infty}^\infty du\, \partial_u\big[u^2\partial_u^2 \Psi^0_1 \big] + \left.\hspace{-6 pt}\overset{2, \text{log.}}\Psi_{\hspace{-5 pt}1}\right|_{\mathscr{I}^+_+}\,,
\end{equation}
and we use the evolution equations \eqref{evolution psi1}-\eqref{evolution psi2} to write
\begin{align}
\partial_u^2 \Psi^0_1 =\frac{1}{2} \partial_u (\eth \Psi^0_2-2 \sigma_2 \Psi^0_3)=\frac{1}{4}\eth (\eth \Psi^0_3-\sigma_2 \Psi^0_4)- \partial_u (\sigma_2 \Psi^0_3)\,,
\end{align}
such that
\begin{equation}
\label{log: step 2}
\left. \hspace{-6 pt}\overset{2, \text{log.}}\Psi_{\hspace{-5 pt}1}\right|_{\mathscr{I}^+_-}=\frac{1}{4}\eth^2 \int_{-\infty}^\infty du\, \partial_u (u^2 \Psi^0_3)-\frac{1}{4}\left[ u^2\left(\eth (\sigma_2 \Psi^0_4)+4\partial_u(\sigma_2 \Psi^0_3)\right)\right]_{-\infty}^{+\infty} + \left. \hspace{-6 pt}\overset{2, \text{log.}}\Psi_{\hspace{-5 pt}1}\right|_{\mathscr{I}^+_+}\,.
\end{equation}
The second term on the right-hand side of this equation vanishes since $u^2 \sigma_2 \Psi^0_4 \to 0$ and $u^2\partial_u(\sigma_2\Psi_3^0)\to 0$ as $u \to \pm \infty$. Using \eqref{eq 4.8} and the definition \eqref{eth definition}, the first term may be written
\begin{equation}
\eth^2 \int_{-\infty}^\infty du\, \partial_u (u^2 \Psi^0_3) =\frac{1}{2}\theta^A \partial_A \int_{-\infty}^\infty du\, \partial_u \big[u^2 (D^B D^C N_{BC}+i \epsilon^{BC} D_B D^D N_{CD})\big]\,.
\end{equation}
Introducing the `logarithmic soft graviton operator' following the notation (6.23) of \cite{Choi:2024ajz},
\begin{equation}\label{defJ}
\mathcal{J}_{AB}^{(\log)}\equiv \int_{-\infty}^\infty du\, \partial_u (u^2 N_{AB})\,,
\end{equation}
equation \eqref{log: step 2} then becomes 
\begin{equation}
\begin{split}
\left. \hspace{-6 pt}\overset{2, \text{log.}}\Psi_{\hspace{-5 pt}1}\right|_{\mathscr{I}^+_-}&=\frac{1}{8}\theta^A \partial_A \left[\left(D^B D^C - i \epsilon^{CD}D_D D^B \right) \mathcal{J}_{BC}^{(\log)}\right] + \left. \hspace{-6 pt}\overset{2, \text{log.}}\Psi_{\hspace{-5 pt}1}\right|_{\mathscr{I}^+_+}\\
&=\frac{1}{8}\theta^A \partial_A \left(D^B D^C \mathcal{J}_{BC}^{(\log)} \right) -\frac{1}{8}\theta^A \epsilon\indices{_A^B}\partial_B \left( \epsilon^{CD}D_D D^E \mathcal{J}_{EC}^{(\log)} \right) +\left. \hspace{-6 pt}\overset{2, \text{log.}}\Psi_{\hspace{-5 pt}1}\right|_{\mathscr{I}^+_+}\,.
\end{split}
\end{equation}
In the second equality, we have used the property $\theta^A=- i \theta^B \epsilon\indices{_B^A}$. In components, this reads
\begin{equation}
\begin{split}
\left. (\hspace{-6 pt}\overset{2, \text{log.}}\Psi_{\hspace{-5 pt}1})_A\right|_{\mathscr{I}^+_-}&= \frac{1}{8} \mathcal{D}\indices{_A^{BC}} \mathcal{J}_{BC}^{(\log)}+\left. (\hspace{-6 pt}\overset{2, \text{log.}}\Psi_{\hspace{-5 pt}1})_A\right|_{\mathscr{I}^+_+}\,,
\end{split}
\end{equation}
where the operator $\mathcal{D}\indices{_A^{BC}}$ is defined as 
\begin{align}
\mathcal{D}\indices{_A^{BC}} \equiv D_A D^B D^C - \epsilon\indices{_A^E}     \epsilon^{CD}D_E D_D D^B. 
\end{align}
Using the property (43) of \cite{Compere:2026mdj} we can substitute $\mathcal{D}\indices{_A^{BC}}T_{BC}=2D_C D_{\langle A}D_{B \rangle}T^{BC}$ when acting on a tracefree symmetric tensor $T_{BC}$. 

Thus, the antipodal matching relation \eqref{log: starting point} may be written as
\begin{equation}
\label{log: step 3}
\mathcal{D}\indices{_A^{BC}} \mathcal{J}_{BC}^{(\log)}(\vec n)  =8 \Upsilon^* \left.(\hspace{-6 pt}\overset{3, \text{log.}}\Psi_{\hspace{-5 pt}3})_A\right|_{\mathscr{I}^-_+}-8\left. (\hspace{-6 pt}\overset{2, \text{log.}}\Psi_{\hspace{-5 pt}1})_A\right|_{\mathscr{I}^+_+}\,.
\end{equation}
We now show that Eq. \eqref{log: step 3} is directly related to a classical version of the logarithmic soft graviton theorem, restricting to the case where all finite-energy scattering states are massive particles. We can use the results of Section~\ref{Section:GravitationalScattering} to evaluate the right-hand side of this equation, that is summing up the contributions from Eqs. \eq{Eq:Psi1outsoft} and \eq{Eq:Psi1outsoft2} to obtain
\begin{align}
\left.(\hspace{-6 pt}\overset{2, \text{log.}}\Psi_{\hspace{-5 pt}1})_A\right|_{\mathscr{I}^+_+}=& -\frac{3}{2} G^2 \sum_{a \in \text{out}} m_a\, \frac{\partial_A q^\mu (c^a_\mu v^a_\nu - v^a_\mu c^a_\nu) q^\nu}{(q\cdot v_a)^4}\\
\nonumber
&+3G^2\sum_{a\in \pm} m_a^4 \dfrac{\partial_A q^\mu(P_\mu p^a_{\nu} -P_\nu  p^a_{\mu} )q^\nu}{(q\cdot p_a)^4}\,,
\end{align}
where $P_\mu$ is the total momentum, and Eq. \eq{Eq:Psi3insoft}, namely
\begin{equation}
\label{Psi3log}
\left.(\hspace{-5 pt}\overset{3, \text{log.}}\Psi_{\hspace{-5 pt}3})_A\right|_{\scri^-_+}=\dfrac{3}{2} G^2 \sum_{a \in \text{in}} m_a\, \frac{\partial_A q^\mu (c^a_\mu v^a_\nu - v^a_\mu c^a_\nu) \tilde{q}^\nu}{(\tilde{q}\cdot v_a)^4}\,.
\end{equation}
The antipodal map $\Upsilon^*$ on Eq. \eqref{Psi3log} effectively implements $\tilde{q}^\mu \mapsto q^\mu$ and $\partial_A q^\mu \mapsto - \partial_A q^\mu$ and thus yields
\begin{equation}
 \Upsilon^* \left.(\hspace{-5 pt}\overset{3, \text{log.}}\Psi_{\hspace{-5 pt}3})_A\right|_{\scri^-_+}=- \dfrac{3}{2} G^2\sum_{a \in \text{in}} m_a\, \frac{ \partial_A q^\mu (c^a_\mu v^a_\nu - v^a_\mu c^a_\nu) q^\nu}{(q\cdot v_a)^4}\,.
\end{equation}
Thus, \eqref{log: step 3} becomes 
\begin{equation}
\label{log: step 4}
\boxed{
\mathcal{D}\indices{_A^{BC}} \mathcal{J}_{BC}^{(\log)}(\vec n)  = 12 G^2 \sum_{a \in \pm} m_a^4\, \frac{e_A^\mu J_{\mu\nu}^a q^\nu}{(q\cdot p_a)^4}-24G^2\sum_{a\in \pm} m_a^4 \dfrac{e_A^\mu (P_\mu p^a_{\nu} -P_\nu  p^a_{\mu})q^\nu}{(q\cdot p_a)^4}\,,}
\end{equation}
where we introduced the `logarithmically divergent' angular momentum 
\begin{equation}
J_{\mu\nu}^a \equiv c^a_\mu p^a_\nu-c^a_\nu p^a_\mu\,.
\end{equation}
Importantly, it may be shown to automatically satisfy a conservation law
\begin{equation}
\label{angular momentum conservation}
\sum_{a \in \pm} J_{\mu\nu}^a=0\,,
\end{equation}
using the explicit expression of the deviation vectors \eqref{deviation vector}.
In Appendix~\ref{app: soft factor identities}, we show that the terms on the right-hand side of \eqref{log: step 4} may be rewritten using
\begin{subequations}
\begin{align}
\mathcal{D}\indices{_A^{BC}} \left(\sum_{a\in \pm} \frac{\varepsilon_{BC}^{\mu\nu}(q)\, p^a_\mu J_{\nu \rho}^a q^\rho}{q \cdot p_a}\right)&=-3 \sum_{a \in \pm} m_a^4 \frac{e_A^\mu J_{\mu\nu}^a q^\nu}{(q\cdot p_a)^4}\,,\label{log id 1}\\
\mathcal{D}\indices{_A^{BC}} \left( (q \cdot P) \sum_{a \in \pm} \frac{\varepsilon_{BC}^{\mu\nu}(q)\, p^a_\mu\, p^a_\nu}{q \cdot p_a} \right)&=3\sum_{a\in \pm} m_a^4\, \frac{e_A^\mu (P_\mu p^a_\nu-P_\nu p^a_\mu) q^\nu}{(q \cdot p_a)^4}\,,\label{log id 2}
\end{align}
\end{subequations}
such that Eq. \eqref{log: step 4} becomes 
\begin{equation}
\hspace{-7 pt}\boxed{\mathcal{D}\indices{_A^{BC}} \mathcal{J}_{BC}^{(\log)}(\vec n)  \!= \!-\mathcal{D}\indices{_A^{BC}} \!\left(\! 4G^2  \sum_{a\in \pm} \frac{\varepsilon_{BC}^{\mu\nu}(q)\, p^a_\mu J_{\nu \rho}^a q^\rho}{q \cdot p_a}\!+\!8G^2(q \cdot P) \sum_{a\in \pm} \frac{\varepsilon_{BC}^{\mu\nu}(q)\, p^a_\mu\, p^a_\nu}{q \cdot p_a}\right)\!.\!}
\end{equation}
Now we show that the operator $\mathcal{D}\indices{_A^{BC}}$ acting on a symmetric tracefree tensor has a vanishing kernel. Defining $V_A \equiv D^B T_{AB}$, $E\equiv D^AV_A$, $B\equiv\epsilon^{AB}D_AV_B$, we have $\mathcal{D}\indices{_{A}^{BC}} T_{BC}=D_A E+\epsilon_A^{\;\;B}D_B B$. If this quantity vanishes, its curl and divergence both vanish which imply $D^2 E=0=D^2B$. Therefore, $E,B$ are constants. Furthermore, since they are defined as a divergence and curl, they contain no $\ell=0$ harmonic and therefore vanish. This implies that $V_A=0$ and therefore $T_{AB}$ has vanishing divergence. It remains to show that a symmetric tracefree tensor on \(S^2\) with vanishing divergence is identically zero. Any smooth symmetric tracefree tensor admits the tensor-harmonic decomposition
\begin{align}
T_{AB}=\sum_{\ell=2}^\infty\sum_{m=-\ell}^\ell
\left[a_{\ell m}D_{\langle A}D_{B\rangle}Y_{\ell m}+b_{\ell m}\epsilon_{C(A}D_{B)}D^CY_{\ell m}
\right].
\end{align}
Using \(D^2Y_{\ell m}=-\ell(\ell+1)Y_{\ell m}\) and \(R_{AB}=\gamma_{AB}\), one finds 
\begin{subequations}
\begin{align}
D^BD_{\langle A}D_{B\rangle}Y_{\ell m}&=-\frac12(\ell-1)(\ell+2)D_AY_{\ell m},\\
D^B\!\left(\epsilon_{C(A}D_{B)}D^CY_{\ell m}\right)&=\frac12(\ell-1)(\ell+2)\epsilon_A{}^CD_CY_{\ell m}.
\end{align}
\end{subequations}
Since \((\ell-1)(\ell+2)\neq0\) for  \(\ell\geq2\), \(D^BT_{AB}=0\) implies \(a_{\ell m}=b_{\ell m}=0\) for all \(\ell,m\), and hence \(T_{AB}=0\). We therefore proved that \(\mathcal D_A{}^{BC}\) has a trivial kernel on smooth symmetric tracefree tensors. We have therefore proven the classical logarithmic soft graviton theorem, written in the form 
\begin{equation}
\label{log soft theorem}
\mathcal{J}_{AB}^{(\log)}(\vec n)=-4G^2 \sum_{a\in \pm} \frac{\varepsilon_{AB}^{\mu\nu}(q)\, p^a_\mu J_{\nu \rho}^a q^\rho}{q \cdot p_a}-8G^2 (q \cdot P) \sum_{a\in \pm} \frac{\varepsilon_{AB}^{\mu\nu}(q)\, p^a_\mu\, p^a_\nu}{q \cdot p_a}\,,
\end{equation}
This expression matches with Eq. (5.14) and Eqs. (3.3)-(3.6) of \cite{Boschetti:2026gfd}. From the definition \eqref{defJ} one also has 
\begin{align}
\mathcal{J}_{AB}^{(\log)}(\vec n)= C^{(1)}_{AB} \vert_{\scri^+_+} -C^{(1)}_{AB} \vert_{\scri^+_-}\,,     
\end{align}
where $C^{(1)}_{AB}$ is the leading tail defined from the expansion of the news $N_{AB}(u,x^A)=u^{-2}\, C_{AB}^{(1)}\vert_{\scri^+_\pm}+o(u^{-2})$ as $u\to \pm \infty$\footnote{We define expansions at spatial infinity using the positive expansion parameter $-u \to +\infty$. Our convention for $C^{(1)}_{AB}$ differs from \cite{Boschetti:2026gfd} by a minus sign, see their Eq. (4.61).}.

\section{Summary and discussion}

We have developed a position-space description of antipodal matching relationships for massive gravitational scattering with no incoming radiation, using the multipolar formalism in harmonic gauge and the asymptotic Weyl tensor. At linear order in $G$, this yields an infinite hierarchy of matching relations for all five Newman--Penrose scalars at each order in the radial expansion. We expect that these maps extend in the non-linear theory to a hierarchy of antipodal relationships of logarithmic branches of the gravitational field at spatial infinity. In this paper, we established the leading order such antipodal map through order $G^2$ which governs the logarithmic soft theorem. Moreover, we computed from first principles in position space the nonlinear gravitational contributions at order $G^2$ that appear in the logarithmic soft theorem, namely the graviton drag, in terms of scattering data, thereby confirming the logarithmic soft factor computed using Fourier-domain methods \cite{Laddha:2018myi,Sahoo:2018lxl,Saha:2019tub}. The fate of the full hierarchy of antipodal maps in the non-linear theory is open as it relies upon the validity of the polyhomogeneous expansion of the gravitational field at spatial infinity, which needs confirmation from an all-order gravitational perturbation analysis (see recent work \cite{Alessio:2024onn,Fucito:2024wlg}). The progress on all-order soft theorems for electromagnetism \cite{Karan:2025ndk} demonstrates that the leading radial electromagnetic waveform at future null infinity admits an infinite polyhomogeneous tower organised in powers of the coupling constant. This suggests that this property might also hold for the gravitational waveform. While a large literature already exists on higher-spin charges, also called celestial multipoles,    \cite{Strominger:2021mtt,Freidel:2021ytz,Compere:2022zdz,Geiller:2024ryw,Geiller:2024bgf} in gravity, such charges have been defined without including the effects of tails and violation of peeling. The inclusion of these effects in a consistent nonlinear theory of scattering that admits polyhomogeneous expansions remains open. It was shown recently \cite{Akhtar:2026pkm} that the backreaction of the radiation emitted by compact binaries at (2.5+1.5$n$)PN provides a $\omega^{n-1} (\log \omega)^n$ term to the radiation amplitude, consistent with the structure presented in \cite{Sahoo:2020ryf, Sahoo:2021ctw}. However, in such a setting the asymptotic trajectories of hard particles do not acquire logarithmic corrections due to their gravitational binding and the resulting soft theorems only encompass the graviton drag contribution. An all-order treatment of the gravitational corrections to the asymptotic trajectories of hard particles such as the one presented in \cite{Karan:2025ndk} is required to extend this result to an asymptotically-free scattering scenario.\\

\noindent Our analysis reveals the simplicity of the antipodal relationships in harmonic gauge as compared with ingoing and outgoing radiative gauges, compare Eq. \eqref{log: starting point0} with Eq. \eqref{DAmatch}. This suggests an alternative to the gravitational scattering framework set in \cite{Compere:2023qoa,Compere:2026jmk} that maps incoming Bondi-Sachs gauge to outgoing Bondi-Sachs gauge through the intermediate Beig-Schmidt gauge: there might exist a nonlinear expansion of the gravitational field in harmonic gauge that exhibits a more natural map of fields at spatial infinity. Such a direction remains to be explored. \\

\noindent The considerations of the present work were purely classical, and indeed we have only retrieved the classical version of the logarithmic soft graviton theorem. For quantum gravitational scattering based on evaluation of Feynman diagrams, Sahoo and Sen obtained additional quantum contributions compared to the classical result \cite{Sahoo:2018lxl}. This requires using Feynman propagators instead of retarded ones which leads to corrections to the antipodal map relationships \cite{AtulBhatkar:2021txo}. A more systematic understanding of these quantum terms and their inclusion within the multipolar formalism remain to be explored.

\acknowledgments

We thank Tim Adamo, Marc Geiller, and S\'ebastien Robert for useful discussions.  The work of DF is supported by F.R.S.-FNRS FRIA grant. The work of KN is supported by a postdoctoral fellowship of the
F.R.S.-FNRS (Belgium). G.C. is F.R.S.-FNRS Research Director.

\appendix

\section{Multipole expansion of the Weyl scalars}
\label{App:Multipole}

\subsection{Mass multipole sector}
In this appendix, we provide the details of the computation of the Weyl scalars used in the bulk of this paper. We start by rewriting the metric perturbations using the identity
\begin{equation}\label{Eq:akl}
	\p_L\left(\frac{M_{L}(u)}{r} \right)=(-1)^\ell\,n_L\sum_{k=0}^{\ell}a_{k\ell} \,\frac{M_L^{(\ell-k)}(u)}{r^{k+1}}\,,
\end{equation}
valid for any STF tensor $M_L$. The coefficients $a_{k\ell}$ are defined as
\begin{align}
   a_{k\ell} \equiv \frac{(\ell+k)!}{2^k k!(\ell-k)!},\qquad 0 \leq k \leq \ell.
\end{align}
By definition, $a_{k \ell}= 0$ for $k>\ell$ or $k<0$. All the Weyl scalars can be written as projections of the electric and magnetic Weyl tensors $E_{ij}$ and $B_{ij}$ on tetrad elements. We have 
\begin{equation}
    E_{ij}= \dfrac{1}{2} \left(\partial_0 \partial_j h_{0i} + \partial_0 \partial_i h_{0j} - \partial_i \partial_j h_{00} - \partial_0^2 h_{ij} \right)
\end{equation}
and, exploiting the antisymmetry of the Levi-Civita tensor,
\begin{equation}
    B_{ij}= \dfrac{1}{2}\epsilon_{ipq}(\partial_0 \partial_q h_{jp} + \partial_j \partial_p h_{0q}).
\end{equation}
We work with the mass multipole moments exclusively. Then, with Eq. \eqref{Eq:akl}, we have 
\begin{align}
    h_{00}&= 2 \sum_{\ell=0}^\infty \sum_{k=0}^\ell \dfrac{a_{k\ell}}{\ell!} n_L \dfrac{M_L^{(\ell-k)}}{r^{k+1}},\\
    h_{0p}&= -4 \sum_{\ell=0}^\infty \sum_{k=0}^\ell \dfrac{a_{k \ell}}{(\ell+1)!} n_L \dfrac{M_{pL}^{(\ell-k+1)}}{r^{k+1}},\\
    h_{jk}&= 2 \sum_{\ell=0}^\infty \sum_{k=0}^\ell \dfrac{a_{k \ell}}{\ell!} n_L \left(\delta_{jk} \dfrac{M_L^{(\ell-k)}}{r^{k+1}} + \dfrac{2}{(\ell+1)(\ell+2)} \dfrac{M_{jkL}^{(\ell-k+2)}}{r^{k+1}} \right).
\end{align}
To compute the derivatives of the perturbations, we will use the following results: 
\begin{equation}
    \partial_i M_L(u)= -n_i \dfrac{d}{du} M_L(u), \quad \partial_i r= n_i, \quad \partial_i n_L = \dfrac{\ell}{r}(\delta_{i(i_1}n_{i_2}...n_{i_\ell)}-n_i n_L).
\end{equation}
Then, 
\begin{equation}
    \partial_i h_{00}= 2\sum_{\ell=0}^\infty \sum_{k=0}^\ell \dfrac{a_{k \ell}}{\ell!}\left(\ell n_{L-1} \dfrac{M_{iL-1}^{(\ell-k)}}{r^{k+2}} - (\ell+k+1) n_i n_L \dfrac{M_L^{(\ell-k)}}{r^{k+2}} - n_i n_L \dfrac{M_L^{(\ell-k+1)}}{r^{k+1}}\right),
\end{equation}
and 
\begin{align}\label{Eq:didjh00}
     &\partial_j \partial_i h_{00}=2\sum_{\ell=0}^\infty \sum_{k=0}^\ell \dfrac{a_{k \ell}}{\ell!}\left( \ell(\ell-1) n_{L-2} \dfrac{M_{ijL-2}^{(\ell-k)}}{r^{k+3}} - \delta_{ij} n_L \left[(\ell+k+1) \dfrac{M_L^{(\ell-k)}}{r^{k+3}}  + \dfrac{M_L^{(\ell-k+1)}}{r^{k+2}}\right]\right. \nonumber\\
     &+ n_in_j n_L \left[ (\ell+k+1)(\ell+k+3) \dfrac{M_L^{(\ell-k)}}{r^{k+3}} + (2\ell+2k +3) \dfrac{M_L^{(\ell-k+1)}}{r^{k+2}} + \dfrac{M_L^{(\ell-k+2)}}{r^{k+1}}\right]
     \\
     &-\left. 2 \ell n_{L-1}n_{(i} \left[(\ell+k+1) \dfrac{M_{j)L-1}^{(\ell-k)}}{r^{k+3}} + \dfrac{M_{j)L-1}^{(\ell-k+1)}}{r^{k+2}} \right]    \right)\nonumber.
\end{align}
Since $h_{0p}$ is essentially $h_{00}$ with $2 \frac{M_L^{\ell-k}}{\ell!}$ replaced by $-4 \frac{M_{pL}^{(\ell-k+1)}}{(\ell+1)!}$, one can read out $\partial_i h_{0p}$ and $\partial_j \partial_i h_{0p}$ from these last two equations. This is enough to compute the mass part of the electric Weyl tensor: 
   \begin{align}
        &E_{ij}= \sum_{\ell=0}^\infty \sum_{k=0}^\ell \dfrac{a_{k \ell}}{\ell! } \left(2\ell (\ell+k+1) n_{L-1} \dfrac{n_{(i}M_{j)L-1}^{(\ell-k)}}{r^{k+3}} + 2\ell  n_{L-1} \dfrac{n_{(i}M_{j)L-1}^{(\ell-k+1)}}{r^{k+2}} -\ell (\ell-1) n_{L-2} \dfrac{M_{ijL-2}^{(\ell-k)}}{r^{k+3}}\right.\nn \\
      &+\delta_{ij} n_L \left[(\ell+k+1) \dfrac{M_{L}^{(\ell-k)}}{r^{k+3}} + \dfrac{M_{L}^{(\ell-k+1)}}{r^{k+2}}- \dfrac{M_{L}^{(\ell-k+2)}}{r^{k+1}}\right] - \dfrac{2 }{(\ell+1)(\ell+2)}n_L \dfrac{M_{ijL}^{(\ell-k+4)}}{r^{k+1}} \nn\\
       &- \dfrac{4 \ell}{(\ell+1)} n_{L-1} \dfrac{M_{ijL-1}^{(\ell-k+2)}}{r^{k+2}} + \dfrac{4}{(\ell+1)} n_L \left[(\ell+k+1)\dfrac{n_{(i}M_{j)L}^{(\ell-k+2)}}{r^{k+2}} + \dfrac{n_{(i}M_{j)L}^{(\ell-k+3)}}{r^{k+1}} \right]\\
       &- \left.n_i n_j n_L \left[(\ell+k+1)(\ell+k+3) \dfrac{M_{L}^{(\ell-k)}}{r^{k+3}} + (2 \ell + 2k +3) \dfrac{M_{L}^{(\ell-k+1)}}{r^{k+2}} + \dfrac{M_{L}^{(\ell-k+2)}}{r^{k+1}}\right]\right).\nn
    \end{align}
We also have 
\begin{equation}
 \begin{aligned}
        \partial_0 \partial_q h_{jp}&= 2 \sum_{\ell=0}^\infty \sum_{k=0}^\ell\dfrac{a_{k\ell}}{\ell!} \left( \ell \delta_{j{p}} n_{L-1}\dfrac{M_{qL-1}^{(\ell-k+{1})}}{r^{k+2}}  + \dfrac{2 \ell}{(\ell+1)(\ell+2)}n_{L-1} \dfrac{M_{jpqL-1}^{(\ell-k+3)}}{r^{k+2}} \right.\\
        &\hspace{50 pt}-n_q n_L \delta_{jp} \left[(\ell+k+1) \dfrac{M_L^{(\ell-k+1)}}{r^{k+2}} + \dfrac{M_L^{(\ell-k+2)}}{r^{k+1}} \right]\\ 
        &\hspace{60 pt} \left. - \dfrac{2}{(\ell+1)(\ell+2)} n_q n_L \left[(\ell+k+1)\dfrac{M_{jpL}^{(\ell-k+3)}}{r^{k+2}} +\dfrac{M_{jpL}^{(\ell-k+4)}}{r^{k+1}}  \right]\right),
    \end{aligned}
\end{equation}
such that the magnetic Weyl tensor $B_{ij}$ is
\begin{align}\label{MagneticWeyl}
&B_{ij}= \epsilon_{ipq} \sum_{\ell=0}^{+ \infty} \sum_{k=0}^{\ell}\dfrac{a_{k \ell}}{\ell!} \left( -\dfrac{2}{(\ell+1)(\ell+2)} n_q n_L \left[ \dfrac{M_{jpL}^{(\ell-k+4)}}{r^{k+1}} + (\ell +k +1) \dfrac{M_{jpL}^{(\ell-k+3)}}{r^{k+2}} \right]\right. \nonumber\\
& +\dfrac{2}{(\ell+1)} \left[ n_j n_q n_L \left( \dfrac{M_{pL}^{(\ell-k+3)}}{r^{k+1}} +(2 \ell + 2k +3 )\dfrac{M_{pL}^{(\ell-k+2)}}{r^{k+2}} +(\ell+k+1)(\ell +k +3)\dfrac{M_{pL}^{(\ell-k+1)}}{r^{k+3}}\right)     \right. \nn\\
&\left.- \delta_{qj} n_L \left(\dfrac{M_{pL}^{(\ell-k+2)}}{r^{k+2}} + (\ell+k+1) \dfrac{M_{pL}^{(\ell-k+1)}}{r^{k+3}}\right) - \ell n_q n_{L-1}\left(\dfrac{M_{jpL-1}^{(\ell-k+2)}}{r^{k+2}} + (\ell+k+1)\dfrac{M_{jpL-1}^{(\ell-k+1)}}{r^{k+3}}  \right) \right] \nn\\
&\left. - \delta_{jp} \left[n_q n_L \left(\dfrac{M_{L}^{(\ell-k+2)}}{r^{k+1}} +(\ell+k+1) \dfrac{M_{L}^{(\ell-k+1)}}{r^{k+2}}  \right)- \ell n_{L-1} \dfrac{M_{qL-1}^{(\ell-k+1)}}{r^{k+2}} \right] \right). 
\end{align}

\paragraph{Projections of Weyl scalars on harmonics.}

Let us define the scalar, vector and tensor projections 
\begin{align}
M_L \oint d\Omega\,n_L \wh n_{L'} &= C^{(0)}_{\ell'} \delta_{\ell \ell'} M_{L'}, \label{ScalarProjection} \\ 
M_{iL}\oint d\Omega\,e^i_{ A} n_{L} D^A \widehat{n}_{L'} &= C^{(1)}_{\ell'} \delta_{\ell',\ell+1}M_{L'}, \label{VectorProjection}\\
M_{ijL}\oint d\Omega\,e^i_{\langle A}e^j_{B\rangle} n_{L} D^AD^B {\widehat{n}}_{L'} &= C^{(2)}_{\ell'} \delta_{\ell',\ell+2}M_{L'},\label{TensorProjection}
\end{align}
where the normalizations are
\begin{align}
C^{(0)}_\ell\equiv \frac{4\pi \ell !}{(2\ell +1)!!}, \quad C^{(1)}_\ell\equiv(\ell+1)C^{(0)}_\ell, \quad C^{(2)}_\ell\equiv\frac{\ell+2}{2}C^{(1)}_\ell.
\end{align}
The second and third normalizations can be found from the first by identifying
\begin{equation}
\ell e_A^i n_{L-1}M_{iL-1}=D_A (M_L n_L)\,, \quad \ell(\ell-1) e_{\langle A}^i e^j_{B \rangle} n_{L-2}M_{ijL-2}=D_{\langle A}D_{B \rangle} (M_L n_L)\,,
\end{equation}
and integrating by parts. Following Eq. \eqref{Eq:Psi0EM}, $\Psi_0$ can be obtained as a projection of a combination of $E_{ij}$ and $B_{ij}$: 
\begin{equation}
    \Psi_0 =- \dfrac{1}{2} \theta^A \theta^B e_A^i e_B^j (E_{ij}+i B_{ij}) \equiv \Psi_0^{(E)} + \Psi_0^{(B)},
\end{equation}
with 
\begin{equation}
    \Psi_0^{(E)}= -\dfrac{1}{2} \theta^A \theta^B e_A^i e_B^j E_{ij}, \qquad  \Psi_0^{(B)}= -\dfrac{i}{2} \theta^A \theta^B e_A^i e_B^j B_{ij}.
\end{equation}
Since $e_A^i n_i=0$, we find 
\begin{equation}
   \hspace{-5 pt} \begin{aligned}
        \Psi_0^{(E)}= & \,\theta^A \theta^B e_A^i e_B^j \sum_{\ell=0}^\infty \sum_{k=0}^\ell \dfrac{a_{k\ell}}{\ell!} \left(\dfrac{\ell(\ell-1)}{2}n_{L-2} \dfrac{M_{ijL-2}^{(\ell-k)}}{r^{k+3}} + \dfrac{2 \ell}{(\ell+1)}n_{L-1} \dfrac{M_{ijL-1}^{(\ell-k+2)}}{r^{k+2}} \right. \\
        &\hspace{205 pt} \left.+\dfrac{1}{(\ell+1)(\ell+2)}n_L  \dfrac{M_{ijL}^{(\ell-k+4)}}{r^{k+1}}\right) 
    \end{aligned}
\end{equation}
and 
\begin{align}
        \Psi_0^{(B
        )}= &- \, \theta^A \theta^B e_A^i e_B^j \sum_{\ell=0}^\infty \sum_{k=0}^\ell \dfrac{a_{k\ell}}{\ell!} \left(  \dfrac{\ell}{(\ell+1)}n_{L-1} \left[(\ell+k+1) \dfrac{M_{ijL-1}^{(\ell-k+1)}}{r^{k+3}} + \dfrac{M_{ijL-1}^{(\ell-k+2)}}{r^{k+2}} \right] \right. \nn\\
        &\left. \hspace{70 pt}+ \dfrac{1}{(\ell+1)(\ell+2)} n_L\left[ (\ell+k+1)\dfrac{M_{ijL}^{(\ell-k+3)}}{r^{k+2}} +  \dfrac{M_{ijL}^{(\ell-k+4)}}{r^{k+1}}\right] \right). 
    \end{align}
Then, $\Psi_0$ can be written as
\begin{equation}
    \begin{aligned}
   \hspace{-16 pt} \Psi_0&= -\theta^A \theta^B e_A^i e_B^j \sum_{\ell=0}^\infty \sum_{k=0}^\ell \dfrac{a_{k \ell}}{\ell!} \left( \dfrac{\ell(\ell+k+1)}{(\ell+1)}n_{L-1} \dfrac{M_{ijL-
   1}^{(\ell-k+1)}}{r^{k+3}} - \dfrac{\ell(\ell-1)}{2} n_{L-2} \dfrac{M_{ijL-2}^{(\ell-k)}}{r^{k+3}} \right.\\
    & \hspace{115 pt}+\left. \dfrac{(\ell+k+1)}{(\ell+1)(\ell+2)}n_{L} \dfrac{M_{ijL}^{(\ell-k+3)}}{r^{k+2}} - \dfrac{\ell}{(\ell+1)} n_{L-1}\dfrac{M_{ijL-1}^{(\ell-k+2)}}{r^{k+2}} \right).
    \end{aligned}
\end{equation}
Introducing the notation
\begin{equation}
    \Psi_0= \theta^A \theta^B \sum_{n=0}^\infty \dfrac{1}{r^{n+2}} (\overset{n}{\Psi}_0)_{AB},
\end{equation}
we find that 
\begin{equation}
\begin{aligned}
 \hspace{-10 pt}  (\overset{n}{\Psi}_0)_{AB}&= -e_{\langle A}^i e_{B\rangle}^j \left(\sum_{\ell=n}^\infty \dfrac{a_{n \ell}}{\ell!} \left[\dfrac{(\ell+n+1)}{(\ell+1)(\ell+2)} n_L M_{ijL}^{(\ell-n+3)} - \dfrac{\ell}{(\ell+1)} n_{L-1} M_{ij L-1}^{(\ell-n+2)}  \right] \right. \\
  & \hspace{24 pt}+\sum_{\ell=n-1}^\infty \left. \dfrac{a_{n-1 \ell}}{\ell!} \left[\dfrac{\ell(\ell+n)}{(\ell+1)} n_{L-1} M_{ijL-1}^{(\ell-n+2)} - \dfrac{\ell(\ell-1)}{2} n_{L-2} M_{ij L-2}^{(\ell-n+1)}  \right]\right).
  \end{aligned}
\end{equation}
Although this expansion appears to violate peeling, one can show that 
\begin{equation}
    (\overset{0}{\Psi}_0)_{AB}=(\overset{1}{\Psi}_0)_{AB}= (\overset{2}{\Psi}_0)_{AB}=0
\end{equation}
for any mass multipole moment $M_L(u)$. A detailed discussion of the non-fictitious fall-offs for all Weyl scalars is given in Appendix \ref{App:FallOffs}. We now project this on the STF tensor harmonics $D^A D^B {\widehat{n}}_L'$ (the other, odd, tensor harmonics yielding a vanishing projection) by using Eq. \eqref{TensorProjection}. The full expression for $C_{\ell}^{(2)}$ will be kept implicit,  as it does not affect the antipodal matching relations. We find 
\begin{equation}\label{Eq:ProjectedPsi0}
 \begin{aligned}
      & \left\langle (\overset{n}{\Psi}_0)_{AB}, D^A D^B {\widehat{n}}_L \right\rangle = -\dfrac{C^{(2)}_\ell}{\ell!} M_L^{(\ell-n+1)}\bigg(a_{n, \ell-2}(\ell+n-1)  \\
       & \left.\hspace{85 pt}- a_{n, \ell-1} (\ell-1)+ a_{n-1, \ell-1} (\ell-1)(\ell+n-1) - a_{n-1, \ell}\dfrac{\ell(\ell-1)}{2} \right).
    \end{aligned}
\end{equation}
We now turn to Eq. \eqref{Eq:Psi1EM} to compute $\Psi_1$: with 
\begin{equation}
    \begin{aligned}
        &\Psi_1^{(E)}= \sum_{\ell=0}^\infty \sum_{k=0}^\ell \theta^A e_A^i \dfrac{a_{k \ell}}{\ell!} \left(\dfrac{\ell}{2} n_{L-1} \left[(k+2) \dfrac{M_{iL-1}^{(\ell-k)}}{r^{k+3}} + \dfrac{M_{iL-1}^{(\ell-k+1)}}{r^{k+2}} \right]  \right.\\
        &\hspace{-3pt}\left.+ \dfrac{1}{(\ell+1)}n_L \left[(k+1-\ell) \dfrac{M_{iL}^{(\ell-k+2)}}{r^{k+2}} + \dfrac{M_{iL}^{(\ell-k+3)}}{r^{k+1}} \right] - \dfrac{1}{(\ell+1)(\ell+2)} n_{L+1} \dfrac{M_{iL+1}^{(\ell-k+{4})}}{r^{k+{1}}}\right)
    \end{aligned}
\end{equation}
and
\begin{equation}
    \begin{aligned}
       &\Psi_1^{(B)}= \sum_{\ell=0}^\infty \sum_{k=0}^\ell \theta^A e_A^i \dfrac{a_{k \ell}}{\ell!} \left(\dfrac{n_{L+1}}{(\ell+1)(\ell+2)}  \left[ \dfrac{M_{iL+1}^{(\ell-k+4)}}{r^{k+1}} +(\ell+k+1) \dfrac{M_{iL+1}^{(\ell-k+3)}}{r^{k+2}} \right]  \right.\\
        &\hspace{-5pt}- \dfrac{n_L}{(\ell+1)} \left[ \dfrac{M_{iL}^{(\ell-k+3)}}{r^{k+1}} + (\ell+2k+2)\dfrac{M_{iL}^{(\ell-k+2)}}{r^{k+2}} + (k+2)(\ell+k+1) \dfrac{M_{iL}^{(\ell-k+1)}}{r^{k+3}}  \right]\\
        &\left. +\dfrac{\ell}{2} n_{L-1} \dfrac{M_{iL-1}^{(\ell-k+1)}}{r^{k+2}}\right).
    \end{aligned}
\end{equation}
Expanding $\Psi_1$ as 
\begin{equation}
    \Psi_1= \theta^A\sum_{n=0}^\infty \dfrac{1}{r^{n+2}} (\overset{n}{\Psi}_1)_A,
\end{equation}
we have 
   \begin{align}
        &(\overset{n}{\Psi}_1)_A= e_A^i \sum_{\ell=n}^\infty \dfrac{1}{\ell!} a_{n \ell} \left[ \ell n_{L-1} M_{iL-1}^{(\ell-n+1)} -  \dfrac{(2\ell+n+1)}{(\ell+1)} n_L M_{iL}^{(\ell-n+2)}  \right. \nn\\ 
      &\hspace{-5 pt}+\left.  \dfrac{(\ell+n+1)}{(\ell+1)(\ell+2)} n_{L+1} M_{iL+1}^{(\ell-n+3)} \right] \nonumber\\
      &+ {e_A^i\sum_{\ell=n-1}^\infty}{ \dfrac{1}{\ell!}}a_{n-1, \ell} \left[\dfrac{\ell}{2} (n+1) n_{L-1} M_{i L-1}^{(\ell-n+1)} - \dfrac{(n+1)(\ell+n)}{(\ell+1)} n_LM_{iL}^{(\ell-n+2)} \right] . 
    \end{align}
Using the projection \eqref{VectorProjection}, we can project $(\Psi_1)_A$ on electric vector harmonics: 
\begin{equation}\label{Eq:ProjectedPsi1}
    \begin{aligned}
       &\left\langle \! (\overset{n}{\Psi}_1)_{A}, D^A {\widehat{n}}_L \right\rangle =- \dfrac{C^{(1)}_\ell}{\ell!}M_L^{(\ell-n+1)} \Bigg( a_{n, \ell-1} (2 \ell+n -1) - a_{n\ell} \ell  \\
       &\hspace{60 pt} \left.- (\ell+n-1) a_{n, \ell-2} + a_{n-1, \ell-1} (n+1)(\ell+n-1) - a_{n-1, \ell} \dfrac{\ell (n+1)}{2} \right).
    \end{aligned}
\end{equation}
In the case of $\Psi_2$, one can check that the magnetic part of \eqref{Eq:Psi2EM} vanishes. Then, $\Psi_2$ is entirely determined by $\Psi_2^{(E)}$. Expanding it as
\begin{equation}
    \Psi_2 = \sum_{n=0}^\infty \dfrac{1}{r^{n+1}} \overset{n}{\Psi}_2,
\end{equation}
we have 
\begin{equation}
\begin{aligned}
  \overset{n}{\Psi}_2&= \sum_{\ell=0}^\infty \dfrac{1}{\ell!}\left( n_L M_L^{(\ell-n+2)}\left[a_{n-2,\ell} \dfrac{n(n-1)}{2} + n a_{n-1, \ell} + a_{n\ell}\right] \right.\\
  &\left.- \dfrac{2}{(\ell+1)} n_{L+1} M_{L+1}^{(\ell-n+3)} \left( n a_{n-1, \ell} + a_{n \ell}  \right) + \dfrac{{1}}{(\ell+1)(\ell+2)}{a_{n \ell}} n_{L+2} M_{L+2}^{(\ell-n+4)}\right)
\end{aligned}
\end{equation}
and after using Eq. \eqref{ScalarProjection}  
its projection on harmonics is given by
\begin{equation}\label{Eq:ProjectedPsi2}
\begin{aligned}
  &\langle \overset{n}{\Psi}_2, \widehat n_{L} \rangle= -\dfrac{C^{(0)}_\ell}{\ell!}M_L^{(\ell-n+2)} \bigg(2( a_{n, \ell-1} + {n} a_{n-1, \ell-1})  \\
  &\left. \hspace{160 pt} - a_{n, \ell-2} - a_{n \ell}- na_{n-1, \ell} - \dfrac{n (n-1)}{2} a_{n-2, \ell} \right). 
  \end{aligned}
\end{equation}
We have actually computed all the projections that determine the remaining Weyl scalars: given Eqs. \eqref{Eq:Psi3EM} and \eqref{Eq:Psi1EM}, it is clear that $\Psi_3^{(E)}=-(\Psi_1^{(E)})_{\theta^A \to \bar{\theta}^A}$. For the magnetic part, since every projected term in $\Psi_1^{(B)}$ involves the identity \eqref{LeviCivitaId}, we have \linebreak $\Psi_3^{(B)}=(\Psi_1^{(B)})_{\theta^A \to \bar{\theta}^A}$. Conversely, Eqs. \eqref{Eq:Psi4EM} and \eqref{Eq:Psi0EM} imply $\Psi_4^{(E)}=(\Psi_0^{(E)})_{\theta^{A,B} \to \bar{\theta}^{A,B}}$ and $\Psi_4^{(B)}=-(\Psi_0^{(B)})_{\theta^{A,B} \to \bar{\theta}^{A,B}}$. With the expansions
\begin{equation}
    \Psi_3= \bar{\theta}^A \sum_{n=0}^\infty \dfrac{1}{r^{n+1}} (\overset{n}{\Psi}_3)_A, \qquad \Psi_4 = \bar{\theta}^A \bar{\theta}^B \sum_{n=0}^\infty \dfrac{1}{r^{n+1}} (\overset{n}{\Psi}_4)_{AB},
\end{equation}
we find the following results: 
\begin{equation}
    \begin{aligned}
    &(\overset{n}{\Psi}_3)_A=- e_A^i \sum_{\ell={n-2}}^\infty \dfrac{1}{\ell!}  a_{n-2, \ell} \left[\dfrac{n \ell}{2} n_{L-1}M_{i L-1}^{(\ell-n+2)} + \dfrac{n(\ell+n-1)}{(\ell+1)} n_L M_{iL}^{(\ell-n+3)} \right]  \\ 
  & -e_A^i \sum_{\ell={n-1}}^\infty {\frac{1}{\ell!}}a_{n-1, \ell} \left[ \dfrac{3 n}{(\ell+1)}n_L M_{iL}^{(\ell-n+3)} - \dfrac{(\ell+n)}{(\ell+1)(\ell+2)} n_{L+1} M_{iL+1}^{(\ell-n+4)}\right]\\ 
  & - e_A^i \sum_{\ell=n}^\infty{\frac{1}{\ell!}}\dfrac{2 a_{n\ell}}{(\ell+1)}\left[n_L M_{iL}^{(\ell-n+3)} - \dfrac{n_{L+1}}{(\ell+2)}M_{iL+1}^{(\ell-n+4)} \right]      ,
    \end{aligned}
    \end{equation}
    which, once projected on vector harmonics, become
\begin{equation}\label{Eq:ProjectedPsi3}
\begin{aligned}
    &\left\langle \! (\overset{n}{\Psi}_3)_{A}, D^A \widehat{n}_L \right\rangle =- \dfrac{C^{(1)}_\ell}{\ell!}M_L^{(\ell-n+2)} \left( \dfrac{n \ell}{2} a_{n-2, \ell} + 3n a_{n-1, \ell-1}   \right. \\
    &  \hspace{60 pt}+ 2 a_{n, \ell-1} - 2 a_{n,\ell-2} - (\ell+n-2) a_{n-1, \ell-2} + n (\ell+n-2) a_{n-2, \ell-1} \bigg)
\end{aligned}
\end{equation}
and 
\begin{align}
 &  (\overset{n}{\Psi}_4)_{AB}= e_{\langle A}^i e_{B\rangle}^j \sum_{\ell={n-2}}^\infty \frac{1}{\ell !}\left(\dfrac{\ell(\ell-1)}{2}a_{n-2,\ell}n_{L-2} M_{ijL-2}^{(\ell-n+2)} + \dfrac{3 \ell}{(\ell+1)} a_{n-1, \ell} n_{L-1}M_{ijL-1}^{(\ell-n+3)} \right.\nn \\
  &\hspace{15 pt} \left.+\dfrac{ n_L M_{ijL}^{(\ell-n+4)}}{(\ell+1)(\ell+2)}\left(2a_{n \ell} + (\ell+n) a_{n-1, \ell} \right)  + \dfrac{ \ell(\ell+n-1)}{(\ell+1)} a_{n-2, \ell} n_{L-1} M_{ijL-1}^{(\ell-n+3)}\right), 
  \end{align}
which can be projected on tensor harmonics:
\begin{equation}\label{Eq:ProjectedPsi4}
\begin{aligned}
    &\left\langle \! (\overset{n}{\Psi}_4)_{AB}, D^A D^B \widehat{n}_L \right\rangle = \dfrac{C^{(2)}_\ell}{\ell!}M_L^{(\ell-n+2)} \left( \dfrac{\ell(\ell-1)}{2}a_{n-2,\ell} + 2 a_{n, \ell-2}   \right. \\
    &  \hspace{30 pt}+ 3 (\ell-1)a_{n-1, \ell-1} + (\ell+n-2) a_{n-1, \ell-2} +  (\ell-1)(\ell+n-2) a_{n-2, \ell-1} \bigg).
\end{aligned}\end{equation}
Using the explicit definition of the $a_{n\ell}$ coefficients, these expressions can be substantially shortened. We have 
\begin{equation}\label{Eq:MultipoleShortExpansion}
   \left\langle (\overset{n}{\Psi}_I)_{(s)}, \widehat{n}_L^{(s)}\right \rangle= \dfrac{C_\ell^{(s)}}{\ell!} M_{L}^{(\ell-n+1+\Theta_{I-2})} R_I(n,\ell)
\end{equation}
where the index $s=\vert I-2\vert = (2,1,0,1,2)$ denotes scalar, vector or tensor indices and 
\begin{align}
    R_0(n,\ell)&= \dfrac{(\ell+n-1)!}{2^n (n-3)!(\ell-n+1)!}  \qquad &\ell\ge2,\ 3\le n\le\ell+1,\label{Eq:Kernel0}\\
    R_1(n,\ell)&= \dfrac{(\ell+2)(\ell+n-1)!}{2^n (n-2)!(\ell-n+1)!}  \qquad &\ell\ge1,\ 2\le n\le\ell+1, \label{Eq:Kernel1}\\
    R_2(n,\ell)&= \dfrac{(\ell+1)(\ell+2)(\ell+n-2)!}{2^{n-1} (n-2)!(\ell-n+2)!} \qquad &\ell\ge0,\ 2\le n\le\ell+2,\label{Eq:Kernel2}\\
    R_3(n,\ell)&= -\dfrac{\ell(\ell+1)(\ell+2)(\ell+n-2)!}{2^{n-1} (n-1)!(\ell-n+2)!}  \qquad &\ell\ge1,\ 1\le n\le\ell+2,
    \label{Eq:Kernel3}\\
    R_4(n,\ell)&= \dfrac{\ell(\ell-1)(\ell+1)(\ell+2)(\ell+n-2)!}{2^{n-1} n!(\ell-n+2)!}  \qquad &\ell\ge2,\ 0\le n\le\ell+2.
    \label{Eq:Kernel4}
\end{align}
Outside the above range, $R_I$ vanishes. From these expressions, it is then straightforward to see that 
\begin{equation}
     \overset{0}{\Psi}_0= \overset{1}{\Psi}_0= \overset{2}{\Psi}_0=\overset{0}{\Psi}_1= \overset{1}{\Psi}_1 =  \overset{0}{\Psi}_2=\overset{1}{\Psi}_2= \overset{0}{\Psi}_3=0.
\end{equation}

\subsection{Current multipole sector}
\label{App:Current}
For completeness, we derive the Weyl scalars for non-vanishing current multipole moments. This section also serves as a non-trivial consistency check of the gravitational electromagnetic duality. Setting the mass multipole moments to zero, we have 
\begin{align}
    h_{00}&=0,\\
    h_{0j}&=4\sum_{\ell=0}^{\infty} \frac{(-1)^{\ell+1}}{\ell! (\ell+2)}\p_{p L}\left(\frac{\varepsilon_{jpq}S_{q L}}{r} \right) =-4\sum_{\ell=0}^{\infty}\sum_{k=0}^{\ell} \dfrac{a_{k\ell}}{\ell! (\ell+2)} \epsilon_{jpq} \partial_p \left(\dfrac{S_{qL}^{(\ell-k)}n_L}{r^{k+1}} \right) ,\\
    h_{jk}&=8\sum_{\ell=0}^{\infty} \frac{(-1)^{\ell}}{(\ell+1)!}\frac{1}{\ell+3}\p_{p L}\left(\frac{\varepsilon_{pq(j}S^{(1)}_{k)q L}}{r} \right)\nn\\
    &= 8\sum_{\ell=0}^{\infty}\sum_{k=0}^{\ell} \dfrac{a_{k\ell}}{(\ell+1)!(\ell+3)} \partial_p \left(\epsilon_{pq(j}\dfrac{S_{k)qL}^{(\ell-k+1)}n_L}{r^{k+1}} \right).
\end{align}
Again, we compute the electric and magnetic Weyl tensors. Since $h_{00}=0,$ we now have
\begin{equation}
E_{ij}= \dfrac{1}{2} \left(\partial_0 \partial_j h_{0i} + \partial_0 \partial_i h_{0j} - \partial_0^2 h_{ij} \right)
\end{equation}
while $B_{ij}$ remains the same. The result of $\partial_i h_{0j}$ can be read straight from \eqref{Eq:didjh00} by replacing $M_L^{(\ell-k)}$ with $-\frac{2}{(\ell+2)} \epsilon_{jpq} S_{qL}(u)$:
\begin{align}
 \hspace{-10 pt}   &\partial_i h_{0j}= -4 \sum_{\ell=0}^\infty \sum_{k=0}^\ell \dfrac{a_{k\ell}}{\ell!(\ell+2)} \left( \epsilon_{ijq} n_L \left[(\ell+k+1) \dfrac{S_{qL}^{(\ell-k)}}{r^{k+3}} + \dfrac{S_{qL}^{(\ell-k+1)}}{r^{k+2}} \right]     \right.\nn \\
    &- \ell \epsilon_{jpq} n_p  n_{L-1}\left[ (\ell+k+1) \dfrac{S_{iqL-1}^{(\ell-k)}}{r^{k+3}} + \dfrac{S_{iqL-1}^{(\ell-k+1)}}{r^{k+2}} \right]\\
    & \left. +\epsilon_{jpq}n_i n_p n_L \left[(\ell+k+1)(\ell+k+3) \dfrac{S_{qL}^{(\ell-k)}}{r^{k+3}} + (2\ell +2k +3) \dfrac{S_{qL}^{(\ell-k+1)}}{r^{k+2}} + \dfrac{S_{qL}^{(\ell-k+2)}}{r^{k+1}}\right]         \right).\nn
\end{align}
The electric tensor can be read straightforwardly from this result and the expression of the metric perturbations:
      \begin{align} \hspace{-10 pt}
     &E_{ij}= 4 \sum_{\ell=0}^\infty \sum_{k=0}^\ell \dfrac{a_{k\ell}}{\ell! (\ell+2)} \left(\ell n_{L-1} n_p \epsilon_{pq(j} \left[(\ell+k+1)\dfrac{S_{i)qL-1}^{(\ell-k+1)}}{r^{k+3}} + \dfrac{S_{i)qL-1}^{(\ell-k+2)}}{r^{k+2}}  \right] \right.\nn \\
        &\hspace{-25 pt}-\left. n_p n_L n_{(i} \epsilon_{j)pq} \left[(\ell+k+1)(\ell+k+3) \dfrac{S_{qL}^{(\ell-k+1)}}{r^{k+3}} + (2 \ell +2k +3) \dfrac{S_{qL}^{(\ell-k+2)}}{r^{k+2}} + \dfrac{S_{qL}^{(\ell-k+3)}}{r^{k+1}} \right]    \right)\nn\\
        +& 4 \sum_{\ell=0}^\infty \sum_{k=0}^\ell \dfrac{a_{k\ell}}{(\ell+1)! (\ell+3)}n_p n_L \epsilon_{pq(i} \left(\dfrac{S_{j)qL}^{(\ell-k+4)}}{r^{k+1}} + (\ell+k+1) \dfrac{S_{j)qL}^{(\ell-k+3)}}{r^{k+2}} \right).
    \end{align}
The computation of the magnetic tensor $B_{ij}$ is more tedious, but the two relevant contributions are 
\begin{equation}
    \hspace{-20 pt}\begin{aligned}
        &\dfrac{1}{2}\epsilon_{ipq}{\p_0} \partial_q h_{jp} = 4 \sum_{\ell=0}^\infty \sum_{k=0}^\ell \dfrac{a_{k\ell}}{(\ell+1)! (\ell+3)} \times \\
        &\left( n_L \left[\dfrac{S_{ijL}^{(\ell-k+4)}}{r^{k+1}} + \dfrac{(4k + 3 \ell +3)}{2}\dfrac{S_{ijL}^{(\ell-k+3)}}{r^{k+2}} + \dfrac{(\ell+k+1)(\ell+2k+3) }{2}\dfrac{S_{ijL}^{(\ell-k+2)}}{r^{k+3}}\right] \right. \\
        & +\dfrac12 \delta_{ij} n_{L+2}\left[\dfrac{S_{L+2}^{(\ell-k+4)}}{r^{k+1}} + (2\ell+2k+3)\dfrac{S_{L+2}^{(\ell-k+3)}}{r^{k+2}} +(\ell+k+1)(\ell+k+3) \dfrac{S_{L+2}^{(\ell-k+2)}}{r^{k+3}}\right] \\
        & - n_i n_{L+1} \left[\dfrac{S_{jL+1}^{(\ell-k+4)}}{r^{k+1}} + (2\ell+2k+3)\dfrac{S_{jL+1}^{(\ell-k+3)}}{r^{k+2}} +(\ell+k+1)(\ell+k+3) \dfrac{S_{jL+1}^{(\ell-k+2)}}{r^{k+3}}\right] \\
        &\left. -\dfrac12 n_j n_{L+1} \left[\dfrac{S_{iL+1}^{(\ell-k+4)}}{r^{k+1}} + (2\ell+2k+3)\dfrac{S_{iL+1}^{(\ell-k+3)}}{r^{k+2}} +(\ell+k+1)(\ell+k+3) \dfrac{S_{iL+1}^{(\ell-k+2)}}{r^{k+3}}\right]  \right)
    \end{aligned}
\end{equation}
and 
\begin{equation}
    \hspace{-20 pt}\begin{aligned}
        &\dfrac{1}{2}\epsilon_{ipq} \partial_j \partial_p h_{0q} = 2\sum_{\ell=0}^\infty \sum_{k=0}^\ell \dfrac{a_{k\ell}}{\ell! (\ell+2)} \times \\
        &\left( \ell \, n_{L-1} \left[ \dfrac{S_{ijL-1}^{(\ell-k+2)}}{r^{k+2}} + (\ell +2k +1) \dfrac{S_{ijL-1}^{(\ell-k+1)}}{r^{k+3}} + (\ell+k+1)(k+1)\dfrac{S_{ijL-1}^{(\ell-k)}}{r^{k+4}} \right] \right. \\
        & -(\ell+1) n_i n_{L} \left[\dfrac{S_{jL}^{(\ell-k+2)}}{r^{k+2}} + (2\ell+2k+3)\dfrac{S_{jL}^{(\ell-k+1)}}{r^{k+3}} +(\ell+k+1)(\ell+k+3) \dfrac{S_{jL}^{(\ell-k)}}{r^{k+4}}\right]\\
        &-\delta_{ij}n_{L+1} \left[\dfrac{S_{L+1}^{(\ell-k+2)}}{r^{k+2}} + (2\ell+2k+3)\dfrac{S_{L+1}^{(\ell-k+1)}}{r^{k+3}} +(\ell+k+1)(\ell+k+3) \dfrac{S_{L+1}^{(\ell-k)}}{r^{k+4}}\right]\\
        &- n_j n_L \left[ \dfrac{S_{iL}^{(\ell-k+3)}}{r^{k+1}} + (2 \ell +3k +2) \dfrac{S_{iL}^{(\ell-k+2)}}{r^{k+2}} +(k+1)(\ell+k+1)(\ell+k+3) \dfrac{S_{iL}^{(\ell-k)}}{r^{k+4}}\right.\\
        &\left. + (\ell^2 +4 (k+1)\ell + k(3k +7) +3) \dfrac{S_{iL}^{(\ell-k+1)}}{r^{k+3}} \right]\\ 
        & + n_i n_j n_{L+1} \left[\dfrac{S_{L+1}^{(\ell-k+3)}}{r^{k+1}}+ 3(\ell+k+2) \dfrac{S_{L+1}^{(\ell-k+2)}}{r^{k+2}} + 3((\ell+k)^2 + 5(\ell+k) +5)\dfrac{S_{L+1}^{(\ell-k+1)}}{r^{k+3}} \right.\\
        & \hspace{180 pt }\left. \left.+ (\ell+k+1)(\ell+k+3)(\ell+k+5)\dfrac{S_{L+1}^{(\ell-k)}}{r^{k+4}} \right]\right).
    \end{aligned}
\end{equation}
One can then compute the current part of the Weyl scalars: 
\begin{equation}
\begin{aligned}
    (\Psi_0)_{AB}&={ -i e_{\langle A}^ie^j_{B\rangle}}\sum_{\ell=0}^{\infty} \sum_{k=0}^\ell \dfrac{a_{k\ell}}{\ell!}\left( \dfrac{n_L}{(\ell+1)(\ell+3)}  \left[ (2k + \ell+1)\dfrac{S_{ijL}^{(\ell-k+3)}}{r^{k+2}} \right. \right.  \\
    &\left. \hspace{160 pt} +(k + \ell+1)(2k + \ell+3) \dfrac{S_{ijL}^{(\ell-k+2)}}{r^{k+3}} \right]  \\
    &\left. -\dfrac{\ell n_{L-1}}{(\ell+2)}  \left[ \dfrac{S_{ijL-1}^{(\ell-k+2)}}{r^{k+2}} + (\ell+1) \dfrac{S_{ijL-1}^{(\ell-k+1)}}{r^{k+3}} -(k+1)(\ell+k+1) \dfrac{S_{ijL-1}^{(\ell-k)}}{r^{k+4}}  \right] \right)
\end{aligned}
\end{equation}
and 
\begin{equation}
\begin{aligned}
    (\Psi_4)_{AB}&={ -i e_{\langle A}^ie^j_{B\rangle}}\sum_{\ell=0}^{\infty} \sum_{k=0}^\ell \dfrac{a_{k\ell}}{\ell!}\left(\dfrac{n_L}{(\ell+1)(\ell+3)}  \left[ (6k + 5\ell+5)\dfrac{S_{ijL}^{(\ell-k+3)}}{r^{k+2}} \right. \right.  \\
    &\left. \hspace{90 pt} + 4\dfrac{S_{ijL}^{(\ell-k+4)}}{r^{k+1}} +(k + \ell+1)(2k + \ell+3) \dfrac{S_{ijL}^{(\ell-k+2)}}{r^{k+3}} \right]  \\
    &\hspace{-30 pt}\left. +\dfrac{\ell n_{L-1}}{(\ell+2)}  \left[ 3\dfrac{S_{ijL-1}^{(\ell-k+2)}}{r^{k+2}} + (3\ell+4k +3) \dfrac{S_{ijL-1}^{(\ell-k+1)}}{r^{k+3}} +(k+1)(\ell+k+1) \dfrac{S_{ijL-1}^{(\ell-k)}}{r^{k+4}}  \right] \right).
\end{aligned}\label{eqA57}
\end{equation}
Projecting on tensor harmonics, we find
\begin{equation}
 \begin{aligned}
      & \left\langle (\overset{n}{\Psi}_0)_{AB}, D^A D^B\wh n_L \right\rangle = i \dfrac{C^{(2)}_\ell}{(\ell-1)!(\ell+1)} S_L^{(\ell-n+1)} \bigg( (\ell-1)(a_{n, \ell-1}  + \ell  a_{n-1, \ell-1})  \\
     &\hspace{-10 pt} - (\ell-1)(n-1)(\ell+n-2) a_{n-2, \ell-1} - (2n + \ell-1) (a_{n, \ell-2} +(n+\ell-2) a_{n-1, \ell-2})\bigg) 
    \end{aligned}
\end{equation}
and 
\begin{equation}
 \begin{aligned}
      & \left\langle (\overset{n}{\Psi}_4)_{AB}, D^A D^B\wh n_L \right\rangle =-i \dfrac{C^{(2)}_\ell}{(\ell-1)!(\ell+1)} S_L^{(\ell-n+2)} \bigg((6n + 5\ell -11) a_{n-1,\ell-2}   \\
     &  +(\ell-1) ((\ell+n-3)(n-2) a_{n-3, \ell-1} + (3 \ell + 4n -8) a_{n-2, \ell-1} + 3 a_{n-1, \ell-1}   )\\
      &+ (n + \ell-3)( 2 n +\ell-3) a_{n-2, \ell-2} + 4 a_{n, \ell-2}\bigg). 
    \end{aligned}
\end{equation}
 We also have 
 \begin{equation}
\hspace{-20 pt}\begin{aligned}
    &(\Psi_1)_{A}= i e_A^i \sum_{\ell=0}^{\infty} \sum_{k=0}^\ell \dfrac{a_{k\ell}}{\ell!}\left(\dfrac{n_{L+1}}{(\ell+1)(\ell+3)}  \left[ (k-1)\dfrac{S_{iL+1}^{(\ell-k+3)}}{r^{k+2}} + k(k+ \ell+1)\dfrac{S_{iL+1}^{(\ell-k+2)}}{r^{k+3}} \right] \right.  \\
    &\hspace{-25 pt}\left. -\dfrac{ n_{L}}{(\ell+2)}  \left[ (k-1)\dfrac{S_{iL}^{(\ell-k+2)}}{r^{k+2}} + k(2k+ \ell+3)\dfrac{S_{iL}^{(\ell-k+1)}}{r^{k+3}} + (k^2+4k+3)(k+ \ell+1)\dfrac{S_{iL}^{(\ell-k)}}{r^{k+4}} \right] \right)
\end{aligned}
\end{equation}
and 
\begin{equation}
\begin{aligned}
    (\Psi_3)_{A}&=i e_A^i \sum_{\ell=0}^{\infty} \sum_{k=0}^\ell \dfrac{a_{k\ell}}{\ell!}\left(\dfrac{n_L}{(\ell+2)}  \left[ 2 \dfrac{S_{iL}^{(\ell-k+3)}}{r^{k+1}} + (2 \ell+ 5k +5)\dfrac{S_{iL}^{(\ell-k+2)}}{r^{k+2}}  \right. \right.  \\
   & \left. + (k+2)(3 \ell + 4 k +3) \dfrac{S_{iL}^{(\ell-k+1)}}{r^{k+3}} + (k+1)(k+3)(\ell+k+1) \dfrac{S_{iL}^{(\ell-k)}}{r^{k+4}} \right] \\
 & \hspace{-30 pt}\left.  - \dfrac{n_{L+1}}{(\ell+1)(\ell+3)} \left[ 2 \dfrac{S_{iL+1}^{(\ell-k+4)}}{r^{k+1}} + (3k +2\ell +1) \dfrac{S_{iL+1}^{(\ell-k+3)}}{r^{k+2}} + k(k+\ell+1) \dfrac{S_{iL+1}^{(\ell-k+2)}}{r^{k+3}}\right] \right).
\end{aligned}
\end{equation}
 This can be projected on vector harmonics: 
 \begin{equation}
    \hspace{-10 pt} \begin{aligned}
        \left\langle (\overset{n}{\Psi}_1)_{A}, D^A\wh n_L \right\rangle &=i \dfrac{C^{(1)}_\ell(n-1)}{(\ell-1)!(\ell+1)} S_L^{(\ell-n+1)} \left( a_{n, \ell-2} + (n+ \ell-2) a_{n-1, \ell-2} - a_{n, \ell-1} \right. \\
        & \hspace{70 pt}\left. - (2n + \ell) a_{n-1, \ell-1} - (n+1)(\ell+n-2) a_{n-2, \ell-1}\right)
     \end{aligned}
 \end{equation}
 and 
 \begin{equation}
   \hspace{-30 pt}  \begin{aligned}
        \left\langle (\overset{n}{\Psi}_3)_{A}, D^A\wh n_L \right\rangle &=i \dfrac{C^{(1)}_\ell}{(\ell-1)!(\ell+1)} S_L^{(\ell-n+2)} \left( 2 (a_{n, \ell-1}- a_{n, \ell-2}) - (3 n +2 \ell-6) a_{n-1, \ell-2}    \right. \\
        &\hspace{30 pt} \left.  - (n-2)(n+ \ell-3) a_{n-2, \ell-2} + (2\ell +5n-2) a_{n-1, \ell-1} \right.\\
        &\hspace{60 pt}\left. + n(4n+ 3 \ell-8) a_{n-2, \ell-1} + n(n-2)(n+\ell-3) a_{n-3, \ell-1}\right). 
     \end{aligned}
 \end{equation}
Finally, 
 \begin{equation}
    \begin{aligned} \Psi_2&= i \sum_{\ell=0}^\infty \sum_{k=0}^\ell \dfrac{a_{k \ell}}{(\ell+1)!}  \left(n_{L+2} \left[ \dfrac{S_{L+2}^{(\ell-k+3)}}{r^{k+2}} + (k+\ell+1)\dfrac{S_{L+2}^{(\ell-k+2)}}{r^{k+3}}  \right]  \right.\\
     & \left.  \hspace{-10pt}- { (\ell+1)}n_{L+1} \left[\dfrac{S_{L+1}^{(\ell-k+2)}}{r^{k+2}} + (\ell + 2k +3) \dfrac{S_{L+1}^{(\ell-k+1)}}{r^{k+3}} + { (k+3)}(\ell+k+1) \dfrac{S_{L+1}^{(\ell-k)}}{r^{k+4}}  \right] \right),
     \end{aligned}
 \end{equation}
 which, projected on scalar harmonics, yields
 \begin{equation}\label{Eq:ProjectedPsi2S}
\begin{aligned}
  &\langle \overset{n}{\Psi}_2,\wh n_{L} \rangle=i\dfrac{C^{(0)}_\ell}{(\ell-1)!}S_L^{(\ell-n+2)} \left( a_{n-1, \ell-2} + (n+ \ell-3) a_{n-2, \ell-2} - a_{n-1, \ell-1} \right. \\
 & \left. \hspace{130 pt} - (\ell+2n-2) a_{n-2, \ell-1} -n(\ell+n-3) a_{n-3, \ell{ -1}}\right).
  \end{aligned}
\end{equation}
From all these projections, one can explicitly check that the expressions obtained in this Appendix obey
\begin{align}
\Psi_I^S(S_L)
=i\Psi_I^M
\left(M_L\longrightarrow-\frac{2\ell}{\ell+1}S_L\right),
\end{align}
as expected from the electric-magnetic duality. The mass multipoles only appear in the real part of the Weyl scalar components, while the current multipoles only appear in the imaginary part of the Weyl scalar components, consistently with gravitational electric-magnetic duality. As done in the main text, we will exclusively focus on the mass multipole sector.\\

\noindent The above five projections can be unified as
\begin{align}
\langle (\overset{n}{\Psi}_I)_{(s)}, \widehat{n}_L^{(s)}\rangle= -\frac{2i\ell}{\ell+1}\dfrac{C_\ell^{(s)}}{\ell!} S_{L}^{(\ell-n+1+\Theta_{I-2})} R_I(n,\ell).
\end{align}

\subsection{Fall-offs of the matter contribution through \(O(G^2)\) at future null infinity}
\label{App:FallOffs}
As mentioned in Appendix \ref{App:Current}, we exclusively focus on the mass sector. Let us now specify the multipole moments:
\begin{equation}
    M_L(u) = G\sum_{k=0}^{\ell}M_{L,k} u^k + G^2\left( M_{L,\ell-1}^{\text{log.}} \log \abs{u} \, u^{\ell -1} \Theta_{\ell-1}+ O(u^{\ell-1})\right).\label{expU}
\end{equation}
In this appendix, we spell out the factor $G$ to explicitly display the dependency in $G$, which differs from the main text. Then, at first non-vanishing radial order for $\Psi_0$, i.e. $  \overset{3}{\Psi}_{\hspace{-1 pt}0}$, the relevant derivative of the multipole moment is 
\begin{equation}
\hspace{-7 pt}\begin{aligned}
   M_{ijL}^{(\ell)} &= G\sum_{k=\ell}^{\ell+2} M_{ijL,k} \dfrac{k!}{(k-\ell)!}u^{k-\ell}  + G^2 \left( M_{ijL,\ell+1}^{\text{log.}} (\ell+1)! u 
\log \abs{u} +O(u)\right).
    \end{aligned}
\end{equation}
Hence, we can write 
\begin{equation}
    \overset{3}{\Psi}_{0} = G\left(u^2 \overset{3,1}{\Psi}_{\hspace{-1 pt}0} + u \hspace{-2 pt} \overset{3,2}{\Psi}_{\hspace{-1 pt}0} +  \hspace{-2 pt} \overset{3,3}{\Psi}_{\hspace{-1 pt}0} \right) + G^2 \left( u \log \abs{u} \hspace{-6 pt} \overset{3, \text{log.}}{\Psi}_{\hspace{-6 pt}0} + O(u)\right).
\end{equation}
At next order $O(r^{-6})$, the relevant multipole derivative is $M_{ijL}^{(\ell-1)}$. Then, 
\begin{equation}
    \overset{4}{\Psi}_{0} = G \left( u^3   \hspace{-1 pt} \overset{4,1}{\Psi}_{\hspace{-1 pt}0} + u^2  \hspace{-2 pt} \overset{4,2}{\Psi}_{\hspace{-1 pt}0} + u  \hspace{-2 pt} \overset{4,3}{\Psi}_{\hspace{-1 pt}0} +\hspace{-2 pt} \overset{4,4}{\Psi}_{\hspace{-1 pt}0}\right) + G^2 \left( u^2 \log \abs{u} \hspace{-6 pt} \overset{4, \text{log.}}{\Psi}_{\hspace{-6 pt}0} +O(u^2) \right)
\end{equation}
and thus 
\begin{equation}\label{Eq:Psi0FallOffs}
  \hspace{-15 pt}  \begin{aligned}
        &\Psi_{0} = \dfrac{1}{r^5} \underbrace{\left( G\left( u^2   \hspace{-1 pt} \overset{3,1}{\Psi}_{\hspace{-1 pt}0} + u   \hspace{-2 pt} \overset{3,2}{\Psi}_{\hspace{-1 pt}0} +  \hspace{-2 pt} \overset{3,3}{\Psi}_{\hspace{-1 pt}0} \right) + G^2 \left( u \log \abs{u} \hspace{-4 pt} \overset{3, \text{log.}}{\Psi}_{\hspace{-6 pt}0} + O(u)\right)\right)}_{ \overset{3}{\Psi}_{\hspace{-1 pt}0}}  \\
        &\hspace{-6 pt}+ \dfrac{1}{r^6}\underbrace{\left( G \left(u^3  \hspace{-1 pt} \overset{4,1}{\Psi}_{\hspace{-1 pt}0} + u^2   \hspace{-2 pt} \overset{4,2}{\Psi}_{\hspace{-1 pt}0} + u  \hspace{-2 pt} \overset{4,3}{\Psi}_{\hspace{-1 pt}0} + \hspace{-2 pt} \overset{4,4}{\Psi}_{\hspace{-1 pt}0} \right) + G^2 \left( u^2 \log \abs{u}\hspace{-4 pt} \overset{4, \text{log.}}{\Psi}_{\hspace{-6 pt}0} +O(u^2)\right) \right)}_{  \overset{4}{\Psi}_{\hspace{-1 pt}0}} + O \left(r^{-7}\right).
    \end{aligned}
\end{equation}
Similarly, our multipole ansatz \eqref{expU} leads to the following fall-offs for the other Weyl scalars: 
\begin{equation}\label{Eq:Psi1FallOffs}
    \begin{aligned}
        \Psi_{1} &= \dfrac{1}{r^4} \underbrace{\left( G\left( u  \hspace{-1 pt} \overset{2,1}{\Psi}_{\hspace{-1 pt}1} + \hspace{-2 pt} \overset{2,2}{\Psi}_{\hspace{-1 pt}1} \right) + G^2 \left( \log \abs{u}\hspace{-6 pt} \overset{2, \text{log.}}{\Psi}_{\hspace{-6 pt}1} +O(u^0)\right)\right)}_{  \overset{2}{\Psi}_{\hspace{-1 pt}1}}  \\
        &\hspace{10 pt}+ \dfrac{1}{r^5}\underbrace{\left( G \left(u^2  \hspace{-1 pt} \overset{3,1}{\Psi}_{\hspace{-1 pt}1} + u   \hspace{-2 pt} \overset{3,2}{\Psi}_{\hspace{-1 pt}1} + \hspace{-2 pt} \overset{3,3}{\Psi}_{\hspace{-1 pt}1} \right) + G^2 \left( u \log \abs{u} \hspace{-6 pt} \overset{3, \text{log.}}{\Psi}_{\hspace{-6 pt}1} +O(u)\right) \right)}_{  \overset{3}{\Psi}_{\hspace{-1 pt}1}} + O \left(r^{-6}\right),
    \end{aligned}
\end{equation} 
\begin{equation}\label{Eq:Psi2FallOffs}
     \begin{aligned}
         \Psi_{2} &= \dfrac{1}{r^3} \underbrace{\left( G  \hspace{-1 pt} \overset{2,2}{\Psi}_{\hspace{-1 pt}2} + G^2 \left(u^{-1}  \hspace{-0 pt} \overset{2,3}{\Psi}_{\hspace{-1 pt}2} + O\left(u^{-2}\right)\right)\right)}_{  \overset{2}{\Psi}_{\hspace{-1 pt}2}}  \\
        &\hspace{50 pt}+ \dfrac{1}{r^4}\underbrace{\left(G \left( u  \hspace{-1 pt} \overset{3,2}{\Psi}_{\hspace{-1 pt}2} +   \hspace{-2 pt} \overset{3,3}{\Psi}_{\hspace{-1 pt}2} \right) + G^2 \left( \log \abs{u}\hspace{-6 pt} \overset{3, \text{log.}}{\Psi}_{\hspace{-6 pt}2} +O(u^0)\right) \right)}_{  \overset{3}{\Psi}_{\hspace{-1 pt}2}} + O \left(r^{-5}\right),
    \end{aligned}
\end{equation}
\begin{equation}\label{Eq:Psi3FallOffs}
     \begin{aligned}
         \Psi_{3} &= \dfrac{1}{r^2} \underbrace{ \left( G^2 u^{-2}  \hspace{-0 pt} \overset{1, 3}{\Psi}_{\hspace{-1 pt}3} + O\left(u^{-3}\right) \right)}_{ \overset{1}{\Psi}_{\hspace{-1 pt}_3}} + \dfrac{1}{r^3}\underbrace{\left( G \hspace{-1 pt} \overset{2,2}{\Psi}_{\hspace{-1 pt}3}  + G^2 \left( u^{-1}  \overset{2, 3}{\Psi}_{\hspace{-2 pt}3} + O\left( u^{-2}\right)\right) \right)}_{  \overset{2}{\Psi}_{\hspace{-1 pt}3}}\\ & \hspace{80 pt}+ \dfrac{1}{r^4} \underbrace{\left( G \left( \overset{3,3}{\Psi}_{\hspace{-1 pt}3} + u \overset{3,2}{\Psi}_{\hspace{-1 pt}3}\right) + G^2 \left( \log \abs{u}\hspace{-3 pt} \overset{3, \text{log.}}{\Psi}_{\hspace{-6 pt}3} +O(u^0) \right) \right)}_{  \overset{3}{\Psi}_{\hspace{-1 pt}3}}+O(r^{-5})
    \end{aligned}
\end{equation}
and
\begin{equation}\label{Eq:Psi4FallOffs}
     \begin{aligned}
        \Psi_{4} &= \dfrac{1}{r} \underbrace{G^2  \left( u^{-3}  \overset{0, 3}{\Psi}_{\hspace{-2 pt}4} + O\left( u^{-4}\right) \right)}_{  \overset{0}{\Psi}_{\hspace{-1 pt}4}} + \dfrac{1}{r^2}\underbrace{ G^2 \left(  u^{-2}  \overset{1, 3}{\Psi}_{\hspace{-2 pt}4} +O\left( u^{-3}\right)\right)}_{ \overset{1}{\Psi}_{\hspace{-1 pt}4}} + O \left(r^{-3}\right).
    \end{aligned}
\end{equation}

\section{Soft factor identities}
\label{app: soft factor identities}
For the subsequent derivations, we introduce the following shorthand notation for the angular projection of constant 4-vector $p^\mu$ and $c^\mu$,
\begin{equation}
p_A\equiv e_A^\mu\, p_\mu=e_A \cdot p\,, \qquad c_A\equiv e_A^\mu\, c_\mu=e_A \cdot c\,,
\end{equation}
where we recall $e^\mu_A=(0, e^i_A)$.
Using $D_A\, e_B^i=-\gamma_{AB}\, n^i$, we have the useful identity
\begin{equation}
D_A\, p_B=-\gamma_{AB}\, (\vec n \cdot \vec p)\,.
\end{equation}
Since $\partial_A n^i=e^i_A$, we also have
\begin{equation}
\partial_A (\vec n \cdot \vec p)=p_A\,.
\end{equation}
These constitute the basic derivative identities used in this appendix. In addition, manifest Lorentz invariance may be restored by writing
\begin{equation}
\label{n dot p}
\vec n \cdot \vec p=\frac{q \cdot p-\tilde{q} \cdot p}{2}\,.
\end{equation}
Finally, the square $p_A p^A =\gamma_{AB} p^A p^B$ may be computed using the completeness relation
\begin{equation}
\label{completeness relation}
e^\mu_A \gamma^{AB} e^\nu_B=\eta^{\mu\nu} +\frac{1}{2}(\tilde{q}^\mu q^\nu +\tilde{q}^\nu q^\mu)\,, \qquad (\tilde{q} \cdot q=-2)\,,
\end{equation}
which implies
\begin{equation}
\label{pA pA}
p_A p^A=p^\mu p_\mu+(\tilde{q} \cdot p) (q\cdot p)\,.
\end{equation}

\subsection{Leading soft factor}
\label{app D1}
We want to establish the identity \eq{soft identity 1}, namely
\begin{equation}
\label{leading soft factor identity}
\sum_{a\in\pm}  D^A D^B \left(\frac{\varepsilon_{AB}^{\mu\nu}(q)\, p^a_\mu\, p^a_\nu }{q \cdot p_a}\right)=\sum_{a\in\pm} \frac{m_a^4}{(q \cdot p_a)^3}\,.
\end{equation}
Plugging in Eq. \eqref{polarization}, we can write the left-hand side as 
\begin{equation}
\sum_{a\in\pm}  D^A D^B \left(\frac{\varepsilon_{AB}^{\mu\nu}(q)\, p^a_\mu\, p^a_\nu}{q \cdot p_a}\right)=\sum_{a\in\pm} [D^A D^B]^{\rm TF} \left(\frac{(e_A \cdot p_a) (e_B \cdot p_a)}{q \cdot p_a} \right)\,,
\end{equation}
where we define the traceless derivative operator
\begin{equation}
[D^A D^B]^{\rm TF}=\left(D^A D^B- \frac{1}{2}\gamma^{AB} D^2 \right)\,.
\end{equation}
For a single term in the sum, we have 
\begin{equation}
\begin{split}
D^A D^B \left(\frac{p_A\, p_B}{q \cdot p} \right)&=D^A \left(-\frac{3\, p_A (\vec n \cdot \vec p)}{q \cdot p}-\frac{p_A\, p_B p^B}{(q \cdot p)^2} \right)\\
&=\frac{6(\vec n \cdot \vec p)^2-3 p_A p^A}{q \cdot p}+\frac{7(\vec n \cdot p)p_A p^A}{(q \cdot p)^2}+\frac{2(p_A p^A)^2}{(q \cdot p)^3} \,. 
\end{split}
\end{equation}
Using \eqref{n dot p} and \eqref{pA pA}, this becomes
\begin{equation}
\label{D10}
\begin{split}
&D^A D^B \left(\frac{p_A\, p_B}{q \cdot p} \right)\\
&=\frac{2(p_A p^A)^2}{(q \cdot p)^3}-\frac{7}{2}\frac{(\tilde{q} \cdot p)\, p_A p^A}{(q \cdot p)^2} +\frac{1}{2}\frac{p_A p^A}{q \cdot p}+\frac{3}{2} \frac{(\tilde{q} \cdot p)^2}{q \cdot p}+\frac{3}{2} (q \cdot p)-3 (\tilde{q} \cdot p)\\
&=\frac{2m^4}{(q \cdot p)^3}-\frac{m^2}{2} \frac{(\tilde{q} \cdot p)}{(q \cdot p)^2}-\frac{m^2}{2}\frac{1}{(q \cdot p)}+ \frac{3}{2} (q \cdot p)-\frac{5}{2} (\tilde{q} \cdot p)\,.
\end{split}
\end{equation}
Similarly, we compute
\begin{equation}
\label{D11}
\begin{split}
D^2 \left(\frac{p_A p^A}{q \cdot p} \right)&=D^2 \left(-\frac{m^2}{q \cdot p}+\tilde{q} \cdot p \right)=D^A \left(m^2\,\frac{p_A}{(q \cdot p)^2} -p_A\right)\\
&=-2m^2 \frac{p_A p^A}{(q \cdot p)^3}-2m^2 \frac{ (\vec n \cdot \vec p)}{(q \cdot p)^2}+2 (\vec n \cdot \vec p)\\
&=\frac{2m^4}{(q \cdot p)^3}-m^2 \frac{(\tilde{q} \cdot p)}{(q \cdot p)^2}-\frac{m^2}{(q \cdot p)}+(q \cdot p)-(\tilde{q} \cdot p)\,.    
\end{split}
\end{equation}
The relevant combination of \eqref{D10} and \eqref{D11} therefore yields
\begin{equation}
[D^A D^B]^{\rm TF}\left(\frac{p_A\, p_B}{q \cdot p} \right)=\frac{m^4}{(q \cdot p)^3}+(q \cdot p)-2(\tilde{q} \cdot p)\,.
\end{equation}
When summing over all particles, we may use momentum conservation $\sum_{a\in\pm} p_a=0$ to obtain
\begin{equation}
[D^A D^B]^{\rm TF}\sum_{a\in\pm} \left(\frac{(e_A \cdot p_a) (e_B \cdot p_a)}{q \cdot p_a} \right)=\sum_{a\in\pm} \frac{m_a^4}{(q \cdot p_a)^3}\,.
\end{equation}
Note that the same formula applies if we replace $q^\mu \mapsto \tilde{q}^\mu$ in the denominator.

\subsection{Logarithmic soft factor: matter contribution}
Here we will prove the identity \eq{log id 1}, i.e.,
\begin{equation}
\label{log soft facor identity}
\sum_{a\in\pm} \mathcal{D}\indices{_A^{BC}} \left( \frac{\varepsilon_{BC}^{\mu\nu}(q)\, p^a_\mu J_{\nu \rho}^a q^\rho}{q \cdot p_a} \right)=-3 \sum_{a\in\pm} m_a^4 \frac{e_A^\mu J_{\mu\nu}^a q^\nu}{(q\cdot p_a)^4}\,,
\end{equation}
with
\begin{equation}
\label{DABC}
\mathcal{D}\indices{_{A}^{BC}} \equiv D_A D^B D^C-\epsilon\indices{_A^E}\partial_E (\epsilon^{CD} D_D D^B)\,,
\end{equation}
and $J_{\mu\nu}=c_\mu p_\nu-c_\nu p_\mu$. It will be useful to unpack the numerator appearing on the left-hand side of this equation, 
\begin{equation}
\label{unpacking numerator}
\begin{split}
&\varepsilon_{AB}^{\mu\nu}(q)\, p_\mu J_{\nu \rho} q^\rho\\
&=\frac{1}{2}[(p_A\, c_B+c_A\, p_B-\gamma_{AB}\, p_D c^D) (q \cdot p)+(\gamma_{AB}\, p_C p^C-2p_A\, p_B) (q \cdot c)]\,.
\end{split}
\end{equation}

\paragraph{Electric contribution.}
Let us look at the electric contribution
\begin{equation}
\sum_{a\in\pm} D_A D^B D^C \left( \frac{\varepsilon_{BC}^{\mu\nu}(q)\, p^a_\mu J_{\nu \rho}^a q^\rho}{q \cdot p_a} \right)\,.
\end{equation}
For a single term in the sum, using \eqref{unpacking numerator} we get
\begin{align}
D^A D^B \left( \frac{\varepsilon_{AB}^{\mu\nu}(q)\, p_\mu J_{\nu \rho} q^\rho}{q \cdot p} \right)=[D^A D^B]^{\rm TF}  \left(\frac{p_A\, c_B+c_A\, p_B}{2}-\frac{p_A\, p_B\, (q \cdot c)}{q \cdot p} \right)\,.    
\end{align}
The first term yields
\begin{equation}
\begin{split}
[D^A D^B]^{\rm TF}  \left(\frac{p_A\, c_B+c_A\, p_B}{2} \right)&=-D^A\left[c_A\, (\vec n \cdot p)+p_A\, (\vec n \cdot c) \right]\\
&=4(\vec n \cdot \vec p)(\vec n \cdot c)-2 p_A c^A\,,
\end{split}
\end{equation}
while the second term yields
\begin{equation}
\begin{split}
& [D^A D^B]^{\rm TF}\left(\frac{p_A\, p_B\, (q \cdot c)}{q \cdot p} \right)\\
&=(q \cdot c) [D^A D^B]^{\rm TF} \left(\frac{p_A\, p_B}{q \cdot p} \right)+2c^A D^B \left(\frac{p_A\, p_B}{q \cdot p} \right)-c_B D^B \left(\frac{p_A p^A}{q \cdot p} \right)\\
&=(q \cdot c) \left(\frac{m^4}{(q \cdot p)^3}+(q \cdot p)-2(\tilde{q} \cdot p)\right)-p_A c^A \left(\frac{6 (\vec n \cdot \vec p)}{q \cdot p}+\frac{2p_B p^B}{(q \cdot p)^2} +\frac{m^2}{(q \cdot p)^2} -1\right)\\
&=(q \cdot c) \left(\frac{m^4}{(q \cdot p)^3}+(q \cdot p)-2(\tilde{q} \cdot p)\right)+p_A c^A \left(\frac{m^2}{(q \cdot p)^2}+\frac{\tilde q \cdot p}{q \cdot p}-2 \right)\,.
\end{split}
\end{equation}
Taken together, we thus get
\begin{equation}
\begin{split}
&[D^A D^B]^{\rm TF}  \left(\frac{p_A\, c_B+c_A\, p_B}{2}-\frac{p_A\, p_B\, (q \cdot c)}{q \cdot p} \right)\\
&=4(\vec n \cdot \vec p)(\vec n \cdot c)-(q \cdot c) \left(\frac{m^4}{(q \cdot p)^3}+(q \cdot p)-2(\tilde{q} \cdot p)\right)-p_A c^A \left(\frac{m^2}{(q \cdot p)^2}+\frac{\tilde q \cdot p}{q \cdot p}\right)\,.
\end{split}
\end{equation}
The gradient of this expression yields
\begin{equation}
\label{gradient}
\begin{split}
&D_A [D^B D^C]^{\rm TF} \left(\frac{p_B\, c_C+c_B\, p_C}{2}-\frac{p_B\, p_C\, (q \cdot c)}{q \cdot p} \right)\\
&=c_A\left[(\vec n \cdot \vec p) \left(\frac{m^2}{(q \cdot p)^2}+\frac{\tilde q \cdot p}{q \cdot p} +4\right)-\frac{m^4}{(q \cdot p)^3}-(q \cdot p)+2(\tilde{q} \cdot p)\right]\\
&+p_A \left[3(q \cdot c)\left(\frac{m^4}{(q \cdot p)^4}-1 \right)+(\vec n \cdot \vec c)\left(\frac{m^2}{(q \cdot p)^2}+\frac{\tilde q \cdot p}{q \cdot p}+4 \right)+p_B c^B \left(\frac{2m^2}{(q \cdot p)^3}+\frac{\tilde q \cdot p}{(q \cdot p)^2}+\frac{1}{q \cdot p} \right) \right]\\
&=c_A\left[-\frac{m^4}{(q \cdot p)^3}+\frac{m^2}{2}\left(\frac{1}{q \cdot p}-\frac{\tilde q \cdot p}{(q \cdot p)^2}\right)-\frac{1}{2}\frac{(\tilde q \cdot p)^2}{q \cdot p}+(q \cdot p)+\frac{1}{2} (\tilde q \cdot p)\right]\\
&+p_A \left[3(q \cdot c)\left(\frac{m^4}{(q \cdot p)^4}-1 \right)+(\vec n \cdot \vec c)\left(\frac{m^2}{(q \cdot p)^2}+\frac{\tilde q \cdot p}{q \cdot p}+4 \right)+p_B c^B \left(\frac{2m^2}{(q \cdot p)^3}+\frac{\tilde q \cdot p}{(q \cdot p)^2}+\frac{1}{q \cdot p} \right) \right]\,.
\end{split}
\end{equation}

\paragraph{Magnetic contribution.}
Let us now look at the magnetic contribution
\begin{equation}
\sum_{a\in\pm} \epsilon\indices{_A^E}\partial_E \left[\epsilon^{CD} D_D D^B \left( \frac{\varepsilon_{BC}^{\mu\nu}(q)\, p^a_\mu J_{\nu \rho}^a q^\rho}{q \cdot p_a} \right)\right]\,.
\end{equation}
For a single term in the sum, using again \eqref{unpacking numerator}, we have
\begin{align}
\epsilon^{BC} D_C D^A \left( \frac{\varepsilon_{AB}^{\mu\nu}(q)\, p_\mu J_{\nu \rho} q^\rho}{q \cdot p} \right)=\epsilon^{BC} D_C D^A \left(\frac{p_A\, c_B + c_A\, p_B}{2}-\frac{p_A\, p_B\, (q \cdot c)}{q \cdot p} \right)\,.    
\end{align}
It is easily shown that the first term vanishes identically. The second term yields
\begin{equation}
\label{D29}
\begin{split}
&\epsilon^{BC} D_C D^A \left( \frac{p_A\, p_B\, (q \cdot c)}{q \cdot p} \right)\\
&=\epsilon^{BC} D_C \left(-\frac{3 p_B\, (\vec n \cdot \vec p) (q \cdot c)}{q \cdot p}+\frac{p_Ac^A\, p_B}{q \cdot p}-\frac{p_Ap^A\, p_B\, (q \cdot c)}{(q \cdot p)^2} \right)\\
&=\epsilon^{CD} p_C c_D \left(\frac{m^2}{(q \cdot p)^2}-\frac{\tilde q \cdot p}{q \cdot p}-\frac{4(\vec n \cdot \vec p)}{q \cdot p} \right)=\epsilon^{CD} p_C c_D \left(\frac{m^2}{(q \cdot p)^2}+\frac{\tilde q \cdot p}{q \cdot p} -2\right)\,.
\end{split}
\end{equation}
We note 
\begin{equation}
\epsilon^{CD} p_C c_D=\epsilon_{ijk} n^i p^j c^k\,,
\end{equation}
such that we can derive the useful identity
\begin{equation}
\label{epsilon identity}
\begin{split}
\epsilon\indices{_A^B}\partial_B (\epsilon^{CD} p_C c_D)=\epsilon\indices{_A^B} e^i_B \epsilon_{ijk}  p^j c^k=e_A^p e^{Bq} n^r \epsilon_{rpq} e^i_B \epsilon_{ijk}  p^j c^k   \\
=(\delta^{iq}-n^i n^q)\, \epsilon_{rpq} \epsilon_{ijk} e_A^p n^r p^j c^k=c_A\, (\vec n \cdot \vec p)-p_A\, (\vec n\cdot \vec c)\,.
\end{split}
\end{equation}
This allows to evaluate the rotational of \eqref{D29}, 
\begin{equation}
\label{rotational}
\begin{split}
&\epsilon\indices{_A^E}D_E\left[\epsilon^{CD} D_D D^B \left( \frac{p_B\, p_C\, (q \cdot c)}{q \cdot p} \right)\right]\\
&=\left( c_A\, (\vec n \cdot \vec p)-p_A\, (\vec n\cdot \vec c) \right) \left(\frac{m^2}{(q \cdot p)^2}+\frac{\tilde q \cdot p}{q \cdot p} -2\right)\\
&-\epsilon\indices{_A^B} p_B\, \epsilon^{CD} p_C c_D \left(\frac{2m^2}{(q \cdot p)^3}+\frac{(\tilde q \cdot p)}{(q \cdot p)^2}+\frac{1}{q \cdot p} \right)\\
&=\left( c_A\, (\vec n \cdot \vec p)-p_A\, (\vec n\cdot \vec c) \right) \left(\frac{m^2}{(q \cdot p)^2}+\frac{\tilde q \cdot p}{q \cdot p} -2\right)\\
&+(c_A\, p_Bp^B-p_A\, p_Bc^B)\left(\frac{2m^2}{(q \cdot p)^3}+\frac{(\tilde q \cdot p)}{(q \cdot p)^2}+\frac{1}{q \cdot p} \right)\\
&=c_A \left[-\frac{2m^4}{(q \cdot p)^3}+\frac{m^2}{2}\left(\frac{\tilde q \cdot p}{(q \cdot p)^2}-\frac{1}{q \cdot p} \right)+\frac{1}{2}\frac{(\tilde q \cdot p)^2}{q \cdot p}+\frac{5}{2}(\tilde q \cdot p)-(q \cdot p) \right]\\
&-p_A \left[(\vec n \cdot \vec c) \left(\frac{m^2}{(q \cdot p)^2}+\frac{\tilde q \cdot p}{q \cdot p} -2\right)+p_Bc^B\left(\frac{2m^2}{(q \cdot p)^3}+\frac{(\tilde q \cdot p)}{(q \cdot p)^2}+\frac{1}{q \cdot p} \right)\right]\,,
\end{split}
\end{equation}
where we used $\epsilon\indices{_A^B} \epsilon^{CD}=\delta_A^C \gamma^{BD}-\delta_A^D \gamma^{BC}$.

\paragraph{Total result.}
Summing the electric contribution \eqref{gradient} and the magnetic contribution \eqref{rotational}, many terms cancel out, such that we obtain 
\begin{equation}
\begin{split}
&D_A D^B D^C \left( \frac{\varepsilon_{BC}^{\mu\nu}(q)\, p_\mu J_{\nu \rho} q^\rho}{q \cdot p} \right)-\epsilon\indices{_A^E}\partial_E \left[\epsilon^{CD} D_D D^B \left( \frac{\varepsilon_{BC}^{\mu\nu}(q)\, p_\mu J_{\nu \rho} q^\rho}{q \cdot p} \right)\right]\\
&=3\, p_A \left(\frac{m^4 (q \cdot c)}{(q \cdot p)^4}-\tilde q \cdot c \right)-3c_A \left( \frac{m^4}{(q \cdot p)^3}-\tilde q \cdot p\right)\\
&=-3m^4\, \frac{e_A^\mu J_{\mu\nu} q^\nu}{(q\cdot p)^4}+ 3 e_A^\mu J_{\mu\nu} \tilde q^\nu\,,
\end{split}
\end{equation}
where we used
\begin{equation}
\label{eqD18}
\frac{e_A^\mu J_{\mu\nu} q^\nu}{(q\cdot p)^4}=\frac{c_A}{(q \cdot p)^3}-\frac{p_A\, (q\cdot c)}{(q \cdot p)^4}\,.
\end{equation}
Summing over particles and using angular momentum conservation \eqref{angular momentum conservation}  therefore yields the identity \eqref{log soft facor identity} which we aimed to derive.

\subsection{Logarithmic soft factor: graviton drag contribution}
In this subsection, we establish the identity \eq{log id 2}, namely
\begin{equation}
\label{log soft facor identity 2}
\mathcal{D}\indices{_A^{BC}} \left( (q \cdot P) \sum_{a\in\pm} \frac{\varepsilon_{BC}^{\mu\nu}(q)\, p^a_\mu\, p^a_\nu}{q \cdot p_a} \right)=3\sum_{a\in\pm} m_a^4\, \frac{e_A^\mu (P_\mu p^a_\nu-P_\nu p^a_\mu) q^\nu}{(q \cdot p_a)^4}\,,
\end{equation}
with $\mathcal{D}\indices{_A^{BC}}$ given in \eqref{DABC}.

\paragraph{Electric contribution.}
We start by computing
\begin{equation}
D_A D^B D ^C \left( (q \cdot P) \sum_{a\in\pm} \frac{\varepsilon_{BC}^{\mu\nu}(q)\, p^a_\mu\, p^a_\nu}{q \cdot p_a} \right)\,.
\end{equation}
We have
\begin{equation}
\label{D27}
\begin{split}
&D^B D ^C \left( (q \cdot P) \sum_{a\in\pm} \frac{\varepsilon_{BC}^{\mu\nu}(q)\, p^a_\mu\, p^a_\nu}{q \cdot p_a} \right)\\
&=(q \cdot P) \sum_{a\in\pm} \frac{m_a^4}{(q \cdot p_a)^3}+P^A \sum_{a\in\pm} p^a_A \left(\frac{m_a^2}{(q \cdot p_a)^2}+ \frac{\tilde q \cdot p_a}{q \cdot p_a}\right)\,,
\end{split}
\end{equation}
where we used \eqref{leading soft factor identity} and
\begin{equation}
\begin{split}
D^A (q \cdot P) D^B \left(\frac{\varepsilon_{AB}^{\mu\nu}(q)\, p_\mu\, p_\nu}{q \cdot p} \right)&= P^A D^B \left( \frac{p_A p_B-\frac{1}{2}\gamma_{AB}\, p_C p^C}{q \cdot p} \right)\\
&=P^A\, p_A \left(\frac{1}{2} \frac{m^2}{(q \cdot p)^2}+\frac{1}{2} \frac{\tilde q \cdot p}{q \cdot p}-1 \right)\,.
\end{split}
\end{equation}
Taking the gradient of \eqref{D27} then yields
\begin{align}
&D_A D^B D ^C \left( (q \cdot P) \sum_{a\in\pm} \frac{\varepsilon_{BC}^{\mu\nu}(q)\, p^a_\mu\, p^a_\nu}{q \cdot p_a} \right)\\
\nonumber
&=P_A \sum_{a\in\pm} \frac{m_a^4}{(q \cdot p_a)^3}-3 (q \cdot P) \sum_{a\in\pm} p_A^a \frac{m_a^4}{(q \cdot p_a)^4}-(\vec n \cdot \vec P) \sum_{a\in\pm} p^a_A \left(\frac{m_a^2}{(q \cdot p_a)^2}+ \frac{\tilde q \cdot p_a}{q \cdot p_a}\right)\\
\nonumber
&-P_A \sum_{a\in\pm} (\vec n \cdot \vec p_a) \left(\frac{m_a^2}{(q \cdot p_a)^2}+ \frac{\tilde q \cdot p_a}{q \cdot p_a}\right)-P^B \sum_{a\in\pm} p^a_A p^a_B \left(\frac{2m_a^2}{(q \cdot p_a)^3}+ \frac{\tilde q \cdot p_a}{(q \cdot p_a)^2}+\frac{1}{q \cdot p_a}\right)\,.
\end{align}

\paragraph{Magnetic contribution.}
We then compute
\begin{equation}
\epsilon\indices{_A^E}D_E \left[\epsilon^{CD} D_D D^B \left( (q \cdot P) \sum_{a\in\pm} \frac{\varepsilon_{BC}^{\mu\nu}(q)\, p^a_\mu\, p^a_\nu}{q \cdot p_a} \right)\right]\,.
\end{equation}
For a single term in the sum, we have
\begin{equation}
\begin{split}
&\epsilon^{CD} D_D D^B \left[ (q \cdot P)  \frac{\varepsilon_{BC}^{\mu\nu}(q)\, p_\mu\, p_\nu}{q \cdot p} \right]\\
&=\epsilon^{CD} D_D \left[ \frac{P^B \varepsilon_{BC}^{\mu\nu}(q)\, p_\mu\, p_\nu}{q \cdot p}+(q \cdot P)\, p_C \left(\frac{1}{2} \frac{m^2}{(q \cdot p)^2}+\frac{1}{2} \frac{\tilde q \cdot p}{q \cdot p}-1 \right) \right]\\
&=\epsilon^{CD}\left[ P^B D_D \left(\frac{\varepsilon_{BC}^{\mu\nu}(q)\, p_\mu\, p_\nu}{q \cdot p}\right)+P_D\, p_C \left(\frac{1}{2} \frac{m^2}{(q \cdot p)^2}+\frac{1}{2} \frac{\tilde q \cdot p}{q \cdot p}-1 \right)\right]\\
&=\epsilon^{CD}p_C P_D \left(\frac{m^2}{(q\cdot p)^2}+\frac{\tilde q \cdot p}{q \cdot p}-2  \right)\,,
\end{split}
\end{equation}
where we have used 
\begin{equation}
\begin{split}
&D_D \left(\frac{\varepsilon_{BC}^{\mu\nu}(q)\, p_\mu\, p_\nu}{q \cdot p} \right)= D_D \left( \frac{p_B p_C-\frac{1}{2}\gamma_{BC}\, p_A p^A}{q \cdot p} \right)\\
&=\frac{\gamma_{BD}p_C+\gamma_{CD}p_B}{2}\left(\frac{\tilde q \cdot p}{q \cdot p}-1 \right)-\frac{p_B p_C p_D}{(q \cdot p)^2}-\frac{\gamma_{BC} p_D}{2}\left(\frac{m^2}{(q \cdot p)^2}-1 \right)\,.
\end{split}
\end{equation}
Differentiating yields
\begin{equation}
\begin{split}
&\epsilon\indices{_A^E}D_E \left[\epsilon^{CD} D_D D^B \left( (q \cdot P) \frac{\varepsilon_{BC}^{\mu\nu}(q)\, p_\mu\, p_\nu}{q \cdot p} \right)\right]\\
&=\left(P_A (\vec n \cdot \vec p)-p_A (\vec n \cdot \vec P) \right)\left(\frac{m^2}{(q\cdot p)^2}+\frac{\tilde q \cdot p}{q \cdot p}-2  \right)\\
&+\left(P_A\, p_Bp^B -p_A\, p_B P^B\right)\left(\frac{2m^2}{(q \cdot p)^3}+\frac{\tilde q \cdot p}{(q \cdot p)^2}+\frac{1}{q \cdot p} \right)\,,
\end{split}
\end{equation}  
where we used again the identity \eqref{epsilon identity} and $\epsilon\indices{_A^B} \epsilon^{CD}=\delta_A^C \gamma^{BD}-\delta_A^D \gamma^{BC}$.

\paragraph{Total result.}
Summing the electric and magnetic contributions yields
\begin{align}
\nonumber
&\mathcal{D}\indices{_A^{BC}} \left( (q \cdot P) \sum_{a\in\pm} \frac{\varepsilon_{BC}^{\mu\nu}(q)\, p^a_\mu\, p^a_\nu}{q \cdot p_a} \right)=-3 (q \cdot P) \sum_{a\in\pm} p_A^a \frac{m_a^4}{(q \cdot p_a)^4}\\
\nonumber
&+P_A \sum_{a\in\pm} \Bigg[ \frac{m_a^4}{(q \cdot p_a)^3}-2 (\vec n \cdot \vec p_a) \left(\frac{m_a^2}{(q \cdot p_a)^2}+ \frac{\tilde q \cdot p_a}{q \cdot p_a}\right)- p^a_B p^{aB} \left(\frac{2m_a^2}{(q \cdot p_a)^3}+ \frac{\tilde q \cdot p_a}{(q \cdot p_a)^2}+\frac{1}{q \cdot p_a}\right)\Bigg]\\
&=P_A \sum_{a\in\pm} \left[  \frac{3m_a^4}{(q \cdot p_a)^3} -2 (\tilde q \cdot p_a)\right]-3 (q \cdot P) \sum_{a\in\pm} p_A^a \frac{m_a^4}{(q \cdot p_a)^4}\\
\nonumber
&=3\sum_{a\in\pm} m_a^4\, \frac{e_A^\mu (P_\mu p^a_\nu-P_\nu p^a_\mu) q^\nu}{(q \cdot p_a)^4}\,, 
\end{align}
where we again used momentum conservation $\Sigma_a p_a=0$ in the last equality. This completes the proof of the identity \eqref{log soft facor identity 2}.

\section{Chasing the tails: graviton drag contribution to the Weyl scalars} \label{Appendix:GravitationalTail}
\subsection{From tail to tail}
Let us consider a homogeneous tower of metric perturbations
\begin{equation}
\! \! h^{\alpha \beta}= \dfrac{1}{ru} A^{\alpha \beta}(\vec n) \!+ \dfrac{\log u}{r^2} B^{\alpha \beta}(\vec n) \!+ \dfrac{u (\log u-1)}{r^3} C^{\alpha \beta}(\vec n)\! + \dfrac{u^2(\log u-\frac32)}{r^4} D^{\alpha \beta}(\vec n)\!+ O(r^{-5}).
\end{equation}
In this Appendix, we determine the leading logarithmic contribution of such a tower to the Weyl scalar $\Psi_1$
\begin{equation}
    \Psi_1= \dfrac12 \theta^A n^i e_A^j (C_{0i0j}-n^p C_{0ijp}),
\end{equation}
with 
\begin{equation}
    C_{0i0j}= \dfrac12(\partial_0 \partial_i h_{0j} + \partial_0 \partial_j h_{0i} - \partial_0^2 h_{ij} - \partial_i \partial_j h_{00})
\end{equation}
and 
\begin{equation}
    C_{0ijp}= \dfrac12(\partial_j \partial_i h_{0p} + \partial_0 \partial_p h_{ij} - \partial_j\partial_0 h_{ip} - \partial_i \partial_p h_{0j}).
\end{equation}
Furthermore,
\begin{equation}
    \partial_t= \partial_u, \quad \partial_i = n_i (\partial_r- \partial_u) + \dfrac{1}{r} e^A_i D_A .
\end{equation}

\paragraph{$B^{\alpha \beta}$ tail.}
Given that any $u$-derivatives will annihilate logarithms, there can be no such derivatives acting on the $B^{\alpha \beta}$ tail. Thus, 
\begin{equation}
\begin{aligned}
    C_{0i0j,B}&\equiv -\dfrac12 \partial_i \partial_j h_{00}= -\dfrac12\left(n^i \partial_r + \dfrac{1}{r}e^A_iD_A\right)\left(n^j \partial_r + \dfrac{1}{r}e^B_jD_B\right)\left(\dfrac{\log u}{r^2}B_{00}\right)\\
    &= \dfrac{\log u}{2 r^4} \left((2\delta_{ij} - 8 n_i n_j) B_{00} + 3(n_i e_j^A + n_j e_i^A) D_A B_{00} - e_i^A e_j^B D_A D_B B_{00} \right).
    \end{aligned}
\end{equation}
Likewise, 
\begin{equation}
\begin{aligned}
    C_{0ijp,B}&\equiv \dfrac12 \left(\partial_i \partial_j h_{0p} -(j \leftrightarrow p)\right)\\
    &=\dfrac12\left[\left(n^i \partial_r + \dfrac{1}{r}e^A_iD_A\right)\left(n^j \partial_r + \dfrac{1}{r}e^B_jD_B\right) \left(\dfrac{\log u}{r^2}B_{0p}\right) -(j \leftrightarrow p)\right]\\
    &= -\dfrac{\log u}{2 r^4} \left((2\delta_{ij} - 8 n_i n_j) B_{0p} + 3(n_i e_j^A + n_j e_i^A) D_A B_{0p} - e_i^A e_j^B D_A D_B B_{0p} \right)\\
    &+ \dfrac{\log u}{2 r^4} \left((2\delta_{ip} - 8 n_i n_p) B_{0j} + 3(n_i e_p^A + n_p e_i^A) D_A B_{0j} - e_i^A e_p^B D_A D_B B_{0j} \right).
    \end{aligned}
\end{equation}
Then, the leading logarithmic contribution of the $B^{\alpha \beta}$ tail to $\Psi_1$ is 
\begin{equation}
     \overset{2, \log}{\Psi}_{\! \!1,B}\equiv \dfrac{3}{4}\theta^A (D_A B_{00} + n^p D_A B_{0p} + 2 e_A^j B_{0j}).
\end{equation}

\paragraph{$C^{\alpha \beta}$ tail.}
At most one $u$-derivative can act on this tail. First, we have 
\begin{equation}
\begin{aligned}
    C_{0i0j,C}&\equiv \dfrac12 \left(\partial_0 \partial_i h_{0j} + \partial_0 \partial_j h_{0i} - \partial_i \partial_j h_{00} \right)\\
    &=\dfrac12\left[\left(n_i \partial_r+ \frac{1}{r}e_i^A D_A \right)\left(\dfrac{\log u}{r^3}C_{0j}\right) +\left(n_j \partial_r+ \frac{1}{r}e_j^A D_A \right)\left(\dfrac{\log u}{r^3}C_{0i} \right) \right. \\
    & \left.- \left(n_i(\partial_r-\partial_u)+ \frac{1}{r}e_i^A D_A \right)\left(n_j(\partial_r-\partial_u)+ \frac{1}{r}e_j^B D_B \right)\left(\dfrac{u \log u}{r^3}C_{00}\right)\right]\\
    &= \dfrac{\log u}{2 r^4} \Bigg( -3 n_i C_{0j} - 3 n_j C_{0i} + e^A_i D_A C_{0j} + e_j^A D_A C_{0i} \\
   & \hspace{140 pt}-(7 n_i n_j - \delta_{ij}) C_{00} + (n_j e_i^A + n_i e^{A}_j) D_A C_{00}\Bigg).
\end{aligned}
\end{equation}
Likewise, 
\begin{equation}
    \begin{aligned}
        C_{0ijp,C}&\equiv \dfrac{\log u}{2 r^4} \Bigg( (7 n_i n_j - \delta_{ij}) C_{0p} - (n_j e_i^A + n_i e_j^A) D_A C_{0p} - (7n_i n_p - \delta_{ip}) C_{0j} \\
   & + (n_p e_i^A + n_i e_p^A) D_A C_{0j} - 3 n_p C_{ij} + e_p^A D_A C_{ij} + 3 n_j C_{ip} - e_j^A D_A C_{ip}\Bigg).
    \end{aligned}
\end{equation}
Then, 
\begin{equation}
   \overset{2, \log}{\Psi}_{\! \!1,C}\equiv \dfrac{1}{4}\theta^A \left(3 e_A^i C_{0i} + 2n^i D_A C_{0i} + D_A C_{00} + 3 n^i e^j_A C_{ij} + n^i n^j D_A C_{ij} \right).
\end{equation}

\paragraph{$D^{\alpha \beta}$ tail.}
The leading logarithmic contribution from the $D^{\alpha \beta}$ tail requires two $u$-derivatives. We have 
\begin{equation}
        C_{0i0j,D}\equiv - \dfrac{\log u}{r^4} \Bigg( n_i D_{0j} + n_j D_{0i} + D_{ij} + n_i n_j D_{00} \Bigg)
\end{equation}
and
\begin{equation}
        C_{0ijp,D}\equiv  \dfrac{\log u}{r^4} \Bigg( n_i n_j D_{0p} - n_p D_{ij} + n_j D_{ip} - n_i n_p D_{0j} \Bigg).
\end{equation}
When projected and summed, these two contributions result in 
\begin{equation}
    \overset{2, \log}{\Psi}_{\! \!1,D}\equiv 0.
\end{equation}
All in all, the homogeneous tower sums to 
\begin{equation}
\begin{aligned}
     \overset{2, \text{log.}}{\Psi}_{\hspace{-6 pt}1}&= \theta^A \left(\dfrac34 \left(D_A B_{00} + n^i D_A B_{0i} + 2 e_A^i B_{0i} \right) \right.\\
     &\left. \hspace{45 pt} + \dfrac14 \left(3 e_A^i C_{0i} + 2 n^i D_A C_{0i} + D_A C_{00} + 3 n^i e_A^j C_{ij} + n^i n^j D_A C_{ij} \right) \right).
\end{aligned}    
\end{equation}

\subsection{Graviton drag contribution at future null infinity}
In the harmonic gauge, we determined that the relevant part of the non-linear gothic metric perturbation at $\scri^+_+$ takes the form \eqref{Eq:GothicTailTower} in terms of the scattering data:
	\hspace{-10 pt}\begin{align}
			\left.\mathfrak{h}_{(2), \text{grav.}}^{\alpha \beta}\right|_{\scri^+_+} &= -8\sum_{j=1}^{n+m} \eta_j m_j{v}_j^{\alpha} {v}_j^{\beta}\left\{O\left( r^{-1}\right) +  \dfrac{\log u}{r^2} \left(\dfrac{(P \cdot v_j)}{(q \cdot v_j)^2} + \dfrac{(P \cdot q)}{(q \cdot v_j)^3}\right) \right.\\
			&\left. \hspace{-68 pt}+ \dfrac{u \log u}{r^3} \left(3 \dfrac{(P \cdot q)}{(q \cdot v_j)^5} +3 \dfrac{(P \cdot v_j)}{(q \cdot v_j)^4} -  \dfrac{P^0}{(q \cdot v_j)^3} + 3 \dfrac{v_j^0(P \cdot q)}{(q \cdot v_j)^4}+ 2 \dfrac{v_j^0(P \cdot v_j)}{(q \cdot v_j)^3}\right)  +O\left( r^{-4}\right)   \right\}.\nn
	\end{align}
Then, the covariant metric perturbation is simply
\begin{equation}
	\hspace{-21 pt}\begin{aligned}
			\left.h_{(2), \text{grav.}}^{\alpha \beta}\right|_{\scri^+_+}&= 8\sum_{j=1}^{n+m} \eta_j m_j\left({v}_j^{\alpha} {v}_j^{\beta}+ \dfrac12 \eta^{\alpha \beta}\right)\left\{O\left(r^{-1}\right) +  \dfrac{\log u}{r^2} \left(\dfrac{(P \cdot v_j)}{(q \cdot v_j)^2} + \dfrac{(P \cdot q)}{(q \cdot v_j)^3}\right) \right.\\
			&\left. \hspace{-60 pt}+ \dfrac{u \log u}{r^3} \left(3 \dfrac{(P \cdot q)}{(q \cdot v_j)^5} +3 \dfrac{(P \cdot v_j)}{(q \cdot v_j)^4} -  \dfrac{P^0}{(q \cdot v_j)^3} + 3 \dfrac{v_j^0(P \cdot q)}{(q \cdot v_j)^4}+ 2 \dfrac{v_j^0(P \cdot v_j)}{(q \cdot v_j)^3}\right)  +O\left( r^{-4}\right)   \right\},
	\end{aligned}
\end{equation} 
that is, in the notations of the previous section,
\begin{equation}
    B_{\alpha \beta}= 8\sum_{j=1}^{n+m} \eta_j m_j \left(v_{\alpha,j}v_{\beta,j}+ \dfrac12 \eta_{\alpha \beta}\right)\left(\dfrac{(P \cdot v_j)}{(q \cdot v_j)^2} + \dfrac{(P \cdot q)}{(q \cdot v_j)^3}\right)
\end{equation}
and 
\begin{equation}
\begin{aligned}
    C_{\alpha \beta}&= 8\sum_{j=1}^{n+m} \eta_j m_j \left(v_{\alpha,j}v_{\beta,j}+ \dfrac12 \eta_{\alpha \beta}\right)\left(3 \dfrac{(P \cdot q)}{(q \cdot v_j)^5} +3 \dfrac{(P \cdot v_j)}{(q \cdot v_j)^4}  \right.\\
    & \left. \hspace{170 pt} -  \dfrac{P^0}{(q \cdot v_j)^3}+ 3 \dfrac{v_j^0(P \cdot q)}{(q \cdot v_j)^4}+ 2 \dfrac{v_j^0(P \cdot v_j)}{(q \cdot v_j)^3}\right).
\end{aligned}
\end{equation}
We then get 
\begin{equation}
   \hspace{-10 pt} D_A B_{00}=  8\sum_{j=1}^{n+m} \eta_j m_j \left((v^0_{j})^2- \dfrac12 \right)\left( \dfrac{(e_A \cdot P)}{(q\cdot v_j)^3}-(e_A \cdot v_j)\left(2 \dfrac{(P\cdot v_j)}{(q\cdot v_j)^3}  + 3 \dfrac{(P\cdot q)}{(q\cdot v_j)^4} \right) \right),
\end{equation}
\begin{equation}
   \hspace{-7 pt} n^i D_A B_{0i}=  -8\sum_{j=1}^{n+m} \eta_j m_j v^0_j n^i v^i_j\left( \dfrac{(e_A \cdot P)}{(q\cdot v_j)^3}-(e_A \cdot v_j)\left(2 \dfrac{(P\cdot v_j)}{(q\cdot v_j)^3}  + 3 \dfrac{(P\cdot q)}{(q\cdot v_j)^4} \right) \right),
\end{equation}
and 
\begin{equation}
   2 e_A^i B_{0i}= -16\sum_{j=1}^{n+m} \eta_j m_j v^0_j (e_A \cdot v_j) \left(\dfrac{(P \cdot v_j)}{(q \cdot v_j)^2} + \dfrac{(P \cdot q)}{(q \cdot v_j)^3}\right).
\end{equation}
Summing these three contributions yields 
\begin{equation}
\begin{aligned}
    (\!\overset{2, \text{log.}}{\Psi}_{\hspace{-6 pt}1,B})^{(2), \text{grav.}}&= \sum_{j=1}^{n+m} \theta^A\eta_j m_j \times \dfrac34 \left( -8v^0_j \dfrac{(e_A \cdot P)}{(q \cdot v_j)^2} + 8 v^0_j (e_A \cdot v_j) \dfrac{(P\cdot q)}{(q \cdot v_j)^3}\right.\\
    &\left. \hspace{55 pt} + 8(e_A \cdot v_j)\dfrac{(P\cdot v_j)}{(q \cdot v_j)^3} - 4 \dfrac{(e_A \cdot P)}{(q \cdot v_j)^3} + 12 (e_A \cdot v_j) \dfrac{(P \cdot q)}{(q \cdot v_j)^4}   \right).
\end{aligned}
\end{equation}
Next, we have 
\begin{equation}
    \begin{aligned}
        3 e_A^i C_{0i}&= -24\sum_{j=1}^{n+m} \eta_j m_j v^0_j (e_A \cdot v_j) \left( 3 \dfrac{(P\cdot q)}{(q\cdot v_j)^5} + 3 \dfrac{(P\cdot v_j)}{(q \cdot v_j)^4}    \right.\\
        & \left.\hspace{140 pt} - \dfrac{P^0}{(q\cdot v_j)^3}+3 v^0_j \dfrac{(P\cdot q)}{(q\cdot v_j)^4} + 2 v^0_j \dfrac{(P\cdot v_j)}{(q \cdot v_j)^3} \right),
    \end{aligned}
\end{equation}
\begin{equation}
     \begin{aligned}
        2 n^i D_A C_{0i}&= -16\sum_{j=1}^{n+m} \eta_j m_j v^0_j n^i v^i_j  \left( 3 \dfrac{(e_A \cdot P)}{(q \cdot v_j)^5}- 15 (e_A \cdot v_j)\dfrac{(P\cdot q) }{(q \cdot v_j)^6}   \right.\\
        & \left. \hspace{30 pt}- 12 (e_A \cdot v_j) \dfrac{(P\cdot v_j)}{(q\cdot v_j)^5}+ 3 (e_A \cdot v_j) \dfrac{P^0}{(q \cdot v_j)^4}  + 3 (e_A \cdot P) \dfrac{v_j^0}{(q \cdot v_j)^4} \right. \\
        & \left. \hspace{110 pt}- 12 (e_A \cdot v_j)v^0_j \dfrac{(P \cdot q)}{(q \cdot v_j)^5}  - 6(e_A \cdot v_j)v_j^0 \dfrac{(P \cdot v_j)}{(q \cdot v_j)^4}\right),
    \end{aligned}
\end{equation}
\begin{equation}
\begin{aligned}
    3 n^i e_A^j C_{ij}&= 24\sum_{j=1}^{n+m} \eta_j m_j (e_A \cdot v_j) n^i v^i_j \left( 3 \dfrac{(P\cdot q)}{(q\cdot v_j)^5} + 3 \dfrac{(P\cdot v_j)}{(q \cdot v_j)^4}    \right.\\
        & \left.\hspace{140 pt} - \dfrac{P^0}{(q\cdot v_j)^3}+3 v^0_j \dfrac{(P\cdot q)}{(q\cdot v_j)^4} + 2 v^0_j \dfrac{(P\cdot v_j)}{(q \cdot v_j)^3} \right),
        \end{aligned}
\end{equation}
\begin{equation}
   \hspace{-10 pt} \begin{aligned}
     n^i n^j D_AC_{ij}&= 8\sum_{j=1}^{n+m} \eta_j m_j  \left((n^i v^i_j)^2+\dfrac12 \right) \left( 3 \dfrac{(e_A \cdot P)}{(q \cdot v_j)^5}- 15 (e_A \cdot v_j)\dfrac{(P\cdot q) }{(q \cdot v_j)^6}   \right.\\
        & \left. \hspace{30 pt}- 12 (e_A \cdot v_j) \dfrac{(P\cdot v_j)}{(q\cdot v_j)^5}+ 3 (e_A \cdot v_j) \dfrac{P^0}{(q \cdot v_j)^4}  + 3 (e_A \cdot P) \dfrac{v_j^0}{(q \cdot v_j)^4} \right. \\
        & \left. \hspace{110 pt}- 12 (e_A \cdot v_j)v^0_j \dfrac{(P \cdot q)}{(q \cdot v_j)^5}  - 6(e_A \cdot v_j)v_j^0 \dfrac{(P \cdot v_j)}{(q \cdot v_j)^4}\right),
        \end{aligned}
\end{equation}
and 
\begin{equation}
   \hspace{-10 pt} \begin{aligned}
      D_A C_{00}&= 8\sum_{j=1}^{n+m} \eta_j m_j  \left((v_j^0)^2-\dfrac12 \right) \left( 3 \dfrac{(e_A \cdot P)}{(q \cdot v_j)^5}- 15 (e_A \cdot v_j)\dfrac{(P\cdot q) }{(q \cdot v_j)^6}   \right.\\
        & \left. \hspace{30 pt}- 12 (e_A \cdot v_j) \dfrac{(P\cdot v_j)}{(q\cdot v_j)^5}+ 3 (e_A \cdot v_j) \dfrac{P^0}{(q \cdot v_j)^4}  + 3 (e_A \cdot P) \dfrac{v_j^0}{(q \cdot v_j)^4} \right. \\
        & \left. \hspace{110 pt}- 12 (e_A \cdot v_j)v^0_j \dfrac{(P \cdot q)}{(q \cdot v_j)^5}  - 6(e_A \cdot v_j)v_j^0 \dfrac{(P \cdot v_j)}{(q \cdot v_j)^4}\right).
        \end{aligned}
\end{equation}
Summing all these contributions yields
\begin{equation}
\begin{aligned}
    \overset{2, \text{log.}}{\Psi}_{\hspace{-6 pt}1,C}&= \sum_{j=1}^{n+m} \theta^A\eta_j m_j \times \left( -12(e_A \cdot v_j) \dfrac{(P\cdot q)}{(q \cdot v_j)^4} - 6 (e_A \cdot v_j) \dfrac{(P\cdot v_j)}{(q \cdot v_j)^3}  \right.\\
    &\left.\hspace{70 pt} - 6 v^0_j (e_A \cdot v_j) \dfrac{(P\cdot q)}{(q \cdot v_j)^3} + 6  \dfrac{(e_A \cdot P)}{(q\cdot v_j)^3} + 6 v^0_j \dfrac{(e_A \cdot P)}{(q \cdot v_j)^2} \right),
\end{aligned}
\end{equation}
which leads us to 
\begin{equation}
\begin{aligned}
     \overset{2, \text{log.}}{\Psi}_{\hspace{-6 pt}1}&= \overset{2, \text{log.}}{\Psi}_{\hspace{-6 pt}1,B}+ \overset{2, \text{log.}}{\Psi}_{\hspace{-6 pt}1,C}= \sum_{j=1}^{n+m} \theta^A\eta_j m_j \dfrac{3}{(q\cdot v_j)^4}\left((e_A \cdot P)(q \cdot v_j)-(P\cdot q)(e_A \cdot v_j)  \right)\\
     &= \sum_{a\in \pm} \dfrac{3m_a^4}{(q\cdot p_a)^4} \theta^A \partial_A q^\mu(p^a_{ \nu} P_\mu- p^a_{\mu} P_\nu )q^\nu.
\end{aligned}
\end{equation}


\begin{thebibliography}{10}

\bibitem{Weinberg:1965nx}
S.~Weinberg, \emph{{Infrared photons and gravitons}},
  \href{https://doi.org/10.1103/PhysRev.140.B516}{\emph{Phys. Rev.} {\bfseries
  140} (1965) B516}.

\bibitem{Cachazo:2014fwa}
F.~Cachazo and A.~Strominger, \emph{{Evidence for a New Soft Graviton
  Theorem}},  \href{https://arxiv.org/abs/1404.4091}{{\ttfamily 1404.4091}}.

\bibitem{1968PhRv..166.1287G}
D.J.~{Gross} and R.~{Jackiw}, \emph{{Low-Energy Theorem for Graviton
  Scattering}}, \href{https://doi.org/10.1103/PhysRev.166.1287}{\emph{Physical
  Review} {\bfseries 166} (1968) 1287}.

\bibitem{White:2011yy}
C.D.~White, \emph{{Factorization Properties of Soft Graviton Amplitudes}},
  \href{https://doi.org/10.1007/JHEP05(2011)060}{\emph{JHEP} {\bfseries 05}
  (2011) 060} [\href{https://arxiv.org/abs/1103.2981}{{\ttfamily 1103.2981}}].

\bibitem{Bern:2014oka}
Z.~Bern, S.~Davies and J.~Nohle, \emph{{On Loop Corrections to Subleading Soft
  Behavior of Gluons and Gravitons}},
  \href{https://doi.org/10.1103/PhysRevD.90.085015}{\emph{Phys. Rev. D}
  {\bfseries 90} (2014) 085015}
  [\href{https://arxiv.org/abs/1405.1015}{{\ttfamily 1405.1015}}].

\bibitem{He:2014bga}
S.~He, Y.-t.~Huang and C.~Wen, \emph{{Loop Corrections to Soft Theorems in
  Gauge Theories and Gravity}},
  \href{https://doi.org/10.1007/JHEP12(2014)115}{\emph{JHEP} {\bfseries 12}
  (2014) 115} [\href{https://arxiv.org/abs/1405.1410}{{\ttfamily 1405.1410}}].

\bibitem{Broedel:2014fsa}
J.~Broedel, M.~de~Leeuw, J.~Plefka and M.~Rosso, \emph{{Constraining subleading
  soft gluon and graviton theorems}},
  \href{https://doi.org/10.1103/PhysRevD.90.065024}{\emph{Phys. Rev. D}
  {\bfseries 90} (2014) 065024}
  [\href{https://arxiv.org/abs/1406.6574}{{\ttfamily 1406.6574}}].

\bibitem{Sahoo:2018lxl}
B.~Sahoo and A.~Sen, \emph{{Classical and Quantum Results on Logarithmic Terms
  in the Soft Theorem in Four Dimensions}},
  \href{https://doi.org/10.1007/JHEP02(2019)086}{\emph{JHEP} {\bfseries 02}
  (2019) 086} [\href{https://arxiv.org/abs/1808.03288}{{\ttfamily
  1808.03288}}].

\bibitem{Laddha:2018myi}
A.~Laddha and A.~Sen, \emph{{Logarithmic Terms in the Soft Expansion in Four
  Dimensions}}, \href{https://doi.org/10.1007/JHEP10(2018)056}{\emph{JHEP}
  {\bfseries 10} (2018) 056}
  [\href{https://arxiv.org/abs/1804.09193}{{\ttfamily 1804.09193}}].

\bibitem{Donnay:2022hkf}
L.~Donnay, K.~Nguyen and R.~Ruzziconi, \emph{{Loop-corrected subleading soft
  theorem and the celestial stress tensor}},
  \href{https://doi.org/10.1007/JHEP09(2022)063}{\emph{JHEP} {\bfseries 09}
  (2022) 063} [\href{https://arxiv.org/abs/2205.11477}{{\ttfamily
  2205.11477}}].

\bibitem{Agrawal:2023zea}
S.~Agrawal, L.~Donnay, K.~Nguyen and R.~Ruzziconi, \emph{{Logarithmic soft
  graviton theorems from superrotation Ward identities}},
  \href{https://doi.org/10.1007/JHEP02(2024)120}{\emph{JHEP} {\bfseries 02}
  (2024) 120} [\href{https://arxiv.org/abs/2309.11220}{{\ttfamily
  2309.11220}}].

\bibitem{Compere:2026jmk}
G.~Comp{\`e}re and S.~Robert, \emph{{Proof of Conservation Laws in
  Gravitational Scattering: Tails and Breaking of Peeling}},
  \href{https://doi.org/10.1103/hlx7-frmx}{\emph{Phys. Rev. Lett.} {\bfseries
  137} (2026) 131401} [\href{https://arxiv.org/abs/2603.08705}{{\ttfamily
  2603.08705}}].

\bibitem{Boschetti:2026ogm}
G.~Boschetti and M.~Campiglia, \emph{{Peeling-violating coefficients in
  classical gravitational scattering}},
  \href{https://doi.org/10.1103/k6ds-1c9l}{\emph{Phys. Rev. D} {\bfseries 113}
  (2026) 104064} [\href{https://arxiv.org/abs/2603.22681}{{\ttfamily
  2603.22681}}].

\bibitem{Boschetti:2026gfd}
G.~Boschetti and M.~Campiglia, \emph{{An asymptotic proof of the classical log
  soft graviton theorem}},
  \href{https://doi.org/10.1007/JHEP06(2026)222}{\emph{JHEP} {\bfseries 06}
  (2026) 222} [\href{https://arxiv.org/abs/2603.09844}{{\ttfamily
  2603.09844}}].

\bibitem{PhysRevD.19.3495}
M.~Walker and C.M.~Will, \emph{{Relativistic Kepler problem. II. Asymptotic
  behavior of the field in the infinite past}},
  \href{https://doi.org/10.1103/PhysRevD.19.3495}{\emph{Phys. Rev. D}
  {\bfseries 19} (1979) 3495}.

\bibitem{Damour:1985cm}
T.~Damour, \emph{{Analytical Calculations of Gravitational Radiation}},  in
  \emph{{4th Marcel Grossmann Meeting on the Recent Developments of General
  Relativity}}, 6, 1985.

\bibitem{1985FoPh...15..605W}
J.~{Winicour}, \emph{{Logarithmic asymptotic flatness}},
  \href{https://doi.org/10.1007/BF01882485}{\emph{Foundations of Physics}
  {\bfseries 15} (1985) 605}.

\bibitem{christodoulou2002global}
D.~Christodoulou, \emph{The global initial value problem in general
  relativity},  in \emph{The Ninth Marcel Grossmann Meeting: On Recent
  Developments in Theoretical and Experimental General Relativity, Gravitation
  and Relativistic Field Theories (In 3 Volumes)}, pp.~44--54, World
  Scientific, 2002.

\bibitem{Kehrberger:2021uvf}
L.M.A.~Kehrberger, \emph{{The Case Against Smooth Null Infinity I: Heuristics
  and Counter-Examples}},
  \href{https://doi.org/10.1007/s00023-021-01108-2}{\emph{Annales Henri
  Poincare} {\bfseries 23} (2022) 829}
  [\href{https://arxiv.org/abs/2105.08079}{{\ttfamily 2105.08079}}].

\bibitem{Kehrberger:2021vhp}
L.M.A.~Kehrberger, \emph{{The case against smooth null infinity II: A
  logarithmically modified Price{\textquoteright}s Law}},
  \href{https://doi.org/10.4310/atmp.2022.v26.n10.a6}{\emph{Adv. Theor. Math.
  Phys.} {\bfseries 26} (2024) 3633}
  [\href{https://arxiv.org/abs/2105.08084}{{\ttfamily 2105.08084}}].

\bibitem{Sahoo:2021ctw}
B.~Sahoo and A.~Sen, \emph{{Classical soft graviton theorem rewritten}},
  \href{https://doi.org/10.1007/JHEP01(2022)077}{\emph{JHEP} {\bfseries 01}
  (2022) 077} [\href{https://arxiv.org/abs/2105.08739}{{\ttfamily
  2105.08739}}].

\bibitem{Saha:2019tub}
A.P.~Saha, B.~Sahoo and A.~Sen, \emph{{Proof of the classical soft graviton
  theorem in $D$ = 4}},
  \href{https://doi.org/10.1007/JHEP06(2020)153}{\emph{JHEP} {\bfseries 06}
  (2020) 153} [\href{https://arxiv.org/abs/1912.06413}{{\ttfamily
  1912.06413}}].

\bibitem{Sen:2024qzb}
A.~Sen, \emph{{Gravitational wave tails from soft theorem: a short review}},
  \href{https://doi.org/10.1088/1361-6382/adec33}{\emph{Class. Quant. Grav.}
  {\bfseries 42} (2025) 143002}
  [\href{https://arxiv.org/abs/2408.08851}{{\ttfamily 2408.08851}}].

\bibitem{Strominger:2017zoo}
A.~Strominger, \emph{{Lectures on the Infrared Structure of Gravity and Gauge
  Theory}}, Princeton University Press (2018),
  [\href{https://arxiv.org/abs/1703.05448}{{\ttfamily 1703.05448}}].

\bibitem{McLoughlin:2022ljp}
T.~McLoughlin, A.~Puhm and A.-M.~Raclariu, \emph{{The SAGEX review on
  scattering amplitudes chapter 11: soft theorems and celestial amplitudes}},
  \href{https://doi.org/10.1088/1751-8121/ac9a40}{\emph{J. Phys. A} {\bfseries
  55} (2022) 443012} [\href{https://arxiv.org/abs/2203.13022}{{\ttfamily
  2203.13022}}].

\bibitem{Blanchet:2013haa}
L.~Blanchet, \emph{{Post-Newtonian Theory for Gravitational Waves}},
  \href{https://doi.org/10.1007/s41114-024-00050-z}{\emph{Living Rev. Rel.}
  {\bfseries 27} (2024) 4} [\href{https://arxiv.org/abs/1310.1528}{{\ttfamily
  1310.1528}}].

\bibitem{Compere:2025tzr}
G.~Comp{\`e}re, D.~Fontaine and K.~Nguyen, \emph{{Electromagnetic multipole
  expansions and the logarithmic soft photon theorem}},
  \href{https://doi.org/10.21468/SciPostPhysCore.8.4.066}{\emph{SciPost Phys.
  Core} {\bfseries 8} (2025) 066}
  [\href{https://arxiv.org/abs/2503.23937}{{\ttfamily 2503.23937}}].

\bibitem{Campiglia:2018dyi}
M.~Campiglia and A.~Laddha, \emph{{Asymptotic charges in massless QED
  revisited: A view from Spatial Infinity}},
  \href{https://doi.org/10.1007/JHEP05(2019)207}{\emph{JHEP} {\bfseries 05}
  (2019) 207} [\href{https://arxiv.org/abs/1810.04619}{{\ttfamily
  1810.04619}}].

\bibitem{AtulBhatkar:2020hqz}
S.~Atul~Bhatkar, \emph{{New asymptotic conservation laws for
  electromagnetism}},
  \href{https://doi.org/10.1007/JHEP02(2021)082}{\emph{JHEP} {\bfseries 02}
  (2021) 082} [\href{https://arxiv.org/abs/2007.03627}{{\ttfamily
  2007.03627}}].

\bibitem{Damour:1990gj}
T.~Damour and B.R.~Iyer, \emph{{Multipole analysis for electromagnetism and
  linearized gravity with irreducible cartesian tensors}},
  \href{https://doi.org/10.1103/PhysRevD.43.3259}{\emph{Phys. Rev. D}
  {\bfseries 43} (1991) 3259}.

\bibitem{PhysRevD.15.2156}
W.B.~Campbell, J.~Macek and T.A.~Morgan, \emph{Relativistic time-dependent
  multipole analysis for scalar, electromagnetic, and gravitational fields},
  \href{https://doi.org/10.1103/PhysRevD.15.2156}{\emph{Phys. Rev. D}
  {\bfseries 15} (1977) 2156}.

\bibitem{Thorne:1980ru}
K.S.~Thorne, \emph{{Multipole Expansions of Gravitational Radiation}},
  \href{https://doi.org/10.1103/RevModPhys.52.299}{\emph{Rev. Mod. Phys.}
  {\bfseries 52} (1980) 299}.

\bibitem{Blanchet:1985sp}
L.~Blanchet and T.~Damour, \emph{{Radiative gravitational fields in general
  relativity I. general structure of the field outside the source}},
  \href{https://doi.org/10.1098/rsta.1986.0125}{\emph{Phil. Trans. Roy. Soc.
  Lond. A} {\bfseries 320} (1986) 379}.

\bibitem{1989AIHPA..50..377B}
L.~{Blanchet} and T.~{Damour}, \emph{{Post-Newtonian generation of
  gravitational waves.}}, {\emph{Annales de L'Institut Henri Poincare Section
  (A) Physique Theorique} {\bfseries 50} (1989) 377}.

\bibitem{Geiller:2024ryw}
M.~Geiller, A.~Laddha and C.~Zwikel, \emph{{Symmetries of the gravitational
  scattering in the absence of peeling}},
  \href{https://doi.org/10.1007/JHEP12(2024)081}{\emph{JHEP} {\bfseries 12}
  (2024) 081} [\href{https://arxiv.org/abs/2407.07978}{{\ttfamily
  2407.07978}}].

\bibitem{Geiller:2022vto}
M.~Geiller and C.~Zwikel, \emph{{The partial Bondi gauge: Further enlarging the
  asymptotic structure of gravity}},
  \href{https://doi.org/10.21468/SciPostPhys.13.5.108}{\emph{SciPost Phys.}
  {\bfseries 13} (2022) 108}
  [\href{https://arxiv.org/abs/2205.11401}{{\ttfamily 2205.11401}}].

\bibitem{Blanchet:2020ngx}
L.~Blanchet, G.~Comp\`ere, G.~Faye, R.~Oliveri and A.~Seraj, \emph{{Multipole
  expansion of gravitational waves: from harmonic to Bondi coordinates}},
  \href{https://doi.org/10.1007/JHEP02(2021)029}{\emph{JHEP} {\bfseries 02}
  (2021) 029} [\href{https://arxiv.org/abs/2011.10000}{{\ttfamily
  2011.10000}}].

\bibitem{Chrusciel:1993hx}
P.T.~Chrusciel, M.A.H.~MacCallum and D.B.~Singleton, \emph{{Gravitational waves
  in general relativity: 14. Bondi expansions and the polyhomogeneity of
  Scri}},  \href{https://arxiv.org/abs/gr-qc/9305021}{{\ttfamily
  gr-qc/9305021}}.

\bibitem{Hintz:2017xxu}
P.~Hintz and A.~Vasy, \emph{{Stability of Minkowski space and polyhomogeneity
  of the metric}},  \href{https://arxiv.org/abs/1711.00195}{{\ttfamily
  1711.00195}}.

\bibitem{Kadar:2025xmo}
I.~Kadar and L.~Kehrberger, \emph{{Scattering, Polyhomogeneity and Asymptotics
  for Quasilinear Wave Equations From Past to Future Null Infinity}},
  \href{https://arxiv.org/abs/2501.09814}{{\ttfamily 2501.09814}}.

\bibitem{Blanchet:2018yqa}
L.~Blanchet and G.~Faye, \emph{{Flux-balance equations for linear momentum and
  center-of-mass position of self-gravitating post-Newtonian systems}},
  \href{https://doi.org/10.1088/1361-6382/ab0d4f}{\emph{Class. Quant. Grav.}
  {\bfseries 36} (2019) 085003}
  [\href{https://arxiv.org/abs/1811.08966}{{\ttfamily 1811.08966}}].

\bibitem{Compere:2019gft}
G.~Comp\`ere, R.~Oliveri and A.~Seraj, \emph{{The Poincar\'e and BMS
  flux-balance laws with application to binary systems}},
  \href{https://doi.org/10.1007/JHEP10(2020)116}{\emph{JHEP} {\bfseries 10}
  (2020) 116} [\href{https://arxiv.org/abs/1912.03164}{{\ttfamily
  1912.03164}}].

\bibitem{Compere:2022zdz}
G.~Comp{\`e}re, R.~Oliveri and A.~Seraj, \emph{{Metric reconstruction from
  celestial multipoles}},
  \href{https://doi.org/10.1007/JHEP11(2022)001}{\emph{JHEP} {\bfseries 11}
  (2022) 001} [\href{https://arxiv.org/abs/2206.12597}{{\ttfamily
  2206.12597}}].

\bibitem{Choi:2024ajz}
S.~Choi, A.~Laddha and A.~Puhm, \emph{{The classical super-rotation infrared
  triangle. Classical logarithmic soft theorem as conservation law in
  gravity}}, \href{https://doi.org/10.1007/JHEP04(2025)138}{\emph{JHEP}
  {\bfseries 04} (2025) 138}
  [\href{https://arxiv.org/abs/2412.16142}{{\ttfamily 2412.16142}}].

\bibitem{Compere:2026mdj}
G.~Comp{\`e}re and S.~Robert, \emph{{Conservation law of super-Lorentz charges
  across spatial infinity}},
  \href{https://arxiv.org/abs/2606.26848}{{\ttfamily 2606.26848}}.

\bibitem{Laddha:2018vbn}
A.~Laddha and A.~Sen, \emph{{Observational Signature of the Logarithmic Terms
  in the Soft Graviton Theorem}},
  \href{https://doi.org/10.1103/PhysRevD.100.024009}{\emph{Phys. Rev. D}
  {\bfseries 100} (2019) 024009}
  [\href{https://arxiv.org/abs/1806.01872}{{\ttfamily 1806.01872}}].

\bibitem{Compere:2023qoa}
G.~Comp{\`e}re, S.E.~Gralla and H.~Wei, \emph{{An asymptotic framework for
  gravitational scattering}},
  \href{https://doi.org/10.1088/1361-6382/acf5c1}{\emph{Class. Quant. Grav.}
  {\bfseries 40} (2023) 205018}
  [\href{https://arxiv.org/abs/2303.17124}{{\ttfamily 2303.17124}}].

\bibitem{Sonego_1998}
S.~Sonego and M.~Bruni, \emph{Gauge dependence in the theory of non-linear
  spacetime perturbations},
  \href{https://doi.org/10.1007/s002200050325}{\emph{Communications in
  Mathematical Physics} {\bfseries 193} (1998) 209–218}.

\bibitem{Bonetto:2021exn}
R.~Bonetto, A.~Pound and Z.~Sam, \emph{{Deformed Schwarzschild horizons in
  second-order perturbation theory: Mass, geometry, and teleology}},
  \href{https://doi.org/10.1103/PhysRevD.105.024048}{\emph{Phys. Rev. D}
  {\bfseries 105} (2022) 024048}
  [\href{https://arxiv.org/abs/2109.09514}{{\ttfamily 2109.09514}}].

\bibitem{Strominger:2013jfa}
A.~Strominger, \emph{{On BMS Invariance of Gravitational Scattering}},
  \href{https://doi.org/10.1007/JHEP07(2014)152}{\emph{JHEP} {\bfseries 07}
  (2014) 152} [\href{https://arxiv.org/abs/1312.2229}{{\ttfamily 1312.2229}}].

\bibitem{He:2014laa}
T.~He, V.~Lysov, P.~Mitra and A.~Strominger, \emph{{BMS supertranslations and
  Weinberg{\textquoteright}s soft graviton theorem}},
  \href{https://doi.org/10.1007/JHEP05(2015)151}{\emph{JHEP} {\bfseries 05}
  (2015) 151} [\href{https://arxiv.org/abs/1401.7026}{{\ttfamily 1401.7026}}].

\bibitem{Campiglia:2015kxa}
M.~Campiglia and A.~Laddha, \emph{{Asymptotic symmetries of gravity and soft
  theorems for massive particles}},
  \href{https://doi.org/10.1007/JHEP12(2015)094}{\emph{JHEP} {\bfseries 12}
  (2015) 094} [\href{https://arxiv.org/abs/1509.01406}{{\ttfamily
  1509.01406}}].

\bibitem{Ashtekar:1978zz}
A.~Ashtekar and R.O.~Hansen, \emph{{A unified treatment of null and spatial
  infinity in general relativity. I - Universal structure, asymptotic
  symmetries, and conserved quantities at spatial infinity}},
  \href{https://doi.org/10.1063/1.523863}{\emph{J. Math. Phys.} {\bfseries 19}
  (1978) 1542}.

\bibitem{PhysRevLett.43.649}
A.~Ashtekar and A.~Magnon-Ashtekar, \emph{Energy-momentum in general
  relativity.}, \href{https://doi.org/10.1103/PhysRevLett.43.649}{\emph{Phys.
  Rev. Lett.} {\bfseries 43} (1979) 649}.

\bibitem{1979JMP....20.1362A}
A.~{Ashtekar} and M.~{Streubel}, \emph{{On angular momentum of stationary
  gravitating systems}}, \href{https://doi.org/10.1063/1.524242}{\emph{Journal
  of Mathematical Physics} {\bfseries 20} (1979) 1362}.

\bibitem{Herberthson:1992gcz}
M.~Herberthson and M.~Ludvigsen, \emph{{A relationship between future and past
  null infinity}}, \href{https://doi.org/10.1007/BF00756992}{\emph{Gen. Rel.
  Grav.} {\bfseries 24} (1992) 1185}.

\bibitem{Friedrich:1998xmu}
H.~Friedrich, \emph{{Gravitational fields near space-like and null infinity}},
  \href{https://doi.org/10.1016/s0393-0440(97)82168-7}{\emph{J. Geom. Phys.}
  {\bfseries 24} (1998) 83}.

\bibitem{Friedrich:1999ax}
H.~Friedrich and J.~Kannar, \emph{{Calculating asymptotic quantities near space
  - like and null infinity from Cauchy data}},
  \href{https://doi.org/10.1002/(SICI)1521-3889(200005)9:3/5<321::AID-ANDP321>3.0.CO}{\emph{Annalen
  Phys.} {\bfseries 9} (2000) 321}
  [\href{https://arxiv.org/abs/gr-qc/9911103}{{\ttfamily gr-qc/9911103}}].

\bibitem{Troessaert:2017jcm}
C.~Troessaert, \emph{{The BMS4 algebra at spatial infinity}},
  \href{https://doi.org/10.1088/1361-6382/aaae22}{\emph{Class. Quant. Grav.}
  {\bfseries 35} (2018) 074003}
  [\href{https://arxiv.org/abs/1704.06223}{{\ttfamily 1704.06223}}].

\bibitem{Prabhu:2021cgk}
K.~Prabhu and I.~Shehzad, \emph{{Conservation of asymptotic charges from past
  to future null infinity: Lorentz charges in general relativity}},
  \href{https://doi.org/10.1007/JHEP08(2022)029}{\emph{JHEP} {\bfseries 08}
  (2022) 029} [\href{https://arxiv.org/abs/2110.04900}{{\ttfamily
  2110.04900}}].

\bibitem{Capone:2022gme}
F.~Capone, K.~Nguyen and E.~Parisini, \emph{{Charge and antipodal matching
  across spatial infinity}},
  \href{https://doi.org/10.21468/SciPostPhys.14.2.014}{\emph{SciPost Phys.}
  {\bfseries 14} (2023) 014}
  [\href{https://arxiv.org/abs/2204.06571}{{\ttfamily 2204.06571}}].

\bibitem{Compere:2018aar}
G.~Comp{\`e}re and A.~Fiorucci, \emph{{Advanced Lectures on General
  Relativity}},  \href{https://arxiv.org/abs/1801.07064}{{\ttfamily
  1801.07064}}.

\bibitem{Campiglia:2015lxa}
M.~Campiglia, \emph{{Null to time-like infinity Green{\textquoteright}s
  functions for asymptotic symmetries in Minkowski spacetime}},
  \href{https://doi.org/10.1007/JHEP11(2015)160}{\emph{JHEP} {\bfseries 11}
  (2015) 160} [\href{https://arxiv.org/abs/1509.01408}{{\ttfamily
  1509.01408}}].

\bibitem{Alessio:2024onn}
F.~Alessio, P.~Di~Vecchia and C.~Heissenberg, \emph{{Logarithmic soft theorems
  and soft spectra}},
  \href{https://doi.org/10.1007/JHEP11(2024)124}{\emph{JHEP} {\bfseries 11}
  (2024) 124} [\href{https://arxiv.org/abs/2407.04128}{{\ttfamily
  2407.04128}}].

\bibitem{Fucito:2024wlg}
F.~Fucito, J.F.~Morales and R.~Russo, \emph{{Gravitational wave forms for
  extreme mass ratio collisions from supersymmetric gauge theories}},
  \href{https://doi.org/10.1103/PhysRevD.111.044054}{\emph{Phys. Rev. D}
  {\bfseries 111} (2025) 044054}
  [\href{https://arxiv.org/abs/2408.07329}{{\ttfamily 2408.07329}}].

\bibitem{Karan:2025ndk}
D.~Karan, B.~Khatun, B.~Sahoo and A.~Sen, \emph{{All order classical
  electromagnetic soft theorems}},
  \href{https://doi.org/10.1007/JHEP11(2025)025}{\emph{JHEP} {\bfseries 11}
  (2025) 025} [\href{https://arxiv.org/abs/2501.07328}{{\ttfamily
  2501.07328}}].

\bibitem{Strominger:2021mtt}
A.~Strominger, \emph{{$w_{1+\infty}$ Algebra and the Celestial Sphere: Infinite
  Towers of Soft Graviton, Photon, and Gluon Symmetries}},
  \href{https://doi.org/10.1103/PhysRevLett.127.221601}{\emph{Phys. Rev. Lett.}
  {\bfseries 127} (2021) 221601}
  [\href{https://arxiv.org/abs/2105.14346}{{\ttfamily 2105.14346}}].

\bibitem{Freidel:2021ytz}
L.~Freidel, D.~Pranzetti and A.-M.~Raclariu, \emph{{Higher spin dynamics in
  gravity and w1+{\ensuremath{\infty}} celestial symmetries}},
  \href{https://doi.org/10.1103/PhysRevD.106.086013}{\emph{Phys. Rev. D}
  {\bfseries 106} (2022) 086013}
  [\href{https://arxiv.org/abs/2112.15573}{{\ttfamily 2112.15573}}].

\bibitem{Geiller:2024bgf}
M.~Geiller, \emph{{Celestial $w_{1+\infty}$ charges and the subleading
  structure of asymptotically-flat spacetimes}},
  \href{https://doi.org/10.21468/SciPostPhys.18.1.023}{\emph{SciPost Phys.}
  {\bfseries 18} (2025) 023}
  [\href{https://arxiv.org/abs/2403.05195}{{\ttfamily 2403.05195}}].

\bibitem{Akhtar:2026pkm}
S.~Akhtar and R.~Sturani, \emph{{Classical soft theorems meet the
  post-Newtonian expansion}},
  \href{https://arxiv.org/abs/2609.29895}{{\ttfamily 2609.29895}}.

\bibitem{Sahoo:2020ryf}
B.~Sahoo, \emph{{Classical Sub-subleading Soft Photon and Soft Graviton
  Theorems in Four Spacetime Dimensions}},
  \href{https://doi.org/10.1007/JHEP12(2020)070}{\emph{JHEP} {\bfseries 12}
  (2020) 070} [\href{https://arxiv.org/abs/2008.04376}{{\ttfamily
  2008.04376}}].

\bibitem{AtulBhatkar:2021txo}
S.~Atul~Bhatkar, \emph{{Asymptotic conservation law with Feynman boundary
  condition}}, \href{https://doi.org/10.1103/PhysRevD.103.125026}{\emph{Phys.
  Rev. D} {\bfseries 103} (2021) 125026}
  [\href{https://arxiv.org/abs/2101.09734}{{\ttfamily 2101.09734}}].

\end{thebibliography}

\providecommand{\href}[2]{#2}\begingroup\raggedright\endgroup

\end{document}